\documentclass[a4paper,11pt]{article}
\pdfoutput=1

\usepackage{jcappub}

\usepackage[T1]{fontenc}

\usepackage{bm}
\usepackage{arydshln}

\def\setsymbol#1#2{\expandafter\def\csname #1\endcsname{#2}}
\def\getsymbol#1{\csname #1\endcsname}

\def\Planck{\textit{Planck}}

\newbox\tablebox    \newdimen\tablewidth
\def\leaderfil{\leaders\hbox to 5pt{\hss.\hss}\hfil}
\def\tablenote#1 #2\par{\begingroup \parindent=0.8em
    \abovedisplayshortskip=0pt\belowdisplayshortskip=0pt
    \noindent
    $$\hss\vbox{\hsize\tablewidth \hangindent=\parindent \hangafter=1 \noindent
    \hbox to \parindent{$^#1$\hss}\strut#2\strut\par}\hss$$
    \endgroup}

\def\L2{\ifmmode L_2\else $L_2$\fi}

\def\DeltaT{\ifmmode \Delta T\else $\Delta T$\fi}
\def\deltat{\ifmmode \Delta t\else $\Delta t$\fi}
\def\fknee{\ifmmode f_{\rm knee}\else $f_{\rm knee}$\fi}
\def\Fmax{\ifmmode F_{\rm max}\else $F_{\rm max}$\fi}
\def\solar{\ifmmode{\rm M}_{\mathord\odot}\else${\rm M}_{\mathord\odot}$\fi}
\def\Msolar{\ifmmode{\rm M}_{\mathord\odot}\else${\rm M}_{\mathord\odot}$\fi}
\def\Lsolar{\ifmmode{\rm L}_{\mathord\odot}\else${\rm L}_{\mathord\odot}$\fi}
\def\inv{\ifmmode^{-1}\else$^{-1}$\fi}
\def\mo{\ifmmode^{-1}\else$^{-1}$\fi}
\def\sup#1{\ifmmode ^{\rm #1}\else $^{\rm #1}$\fi}
\def\expo#1{\ifmmode \times 10^{#1}\else $\times 10^{#1}$\fi}
\def\,{\thinspace}
\def\lsim{\mathrel{\raise .4ex\hbox{\rlap{$<$}\lower 1.2ex\hbox{$\sim$}}}}
\def\gsim{\mathrel{\raise .4ex\hbox{\rlap{$>$}\lower 1.2ex\hbox{$\sim$}}}}

\def\simprop{\mathrel{\raise .4ex\hbox{\rlap{$\propto$}\lower 1.2ex\hbox{$\sim$}}}}
\def\deg{\ifmmode^\circ\else$^\circ$\fi}
\def\pdeg{\ifmmode $\setbox0=\hbox{$^{\circ}$}\rlap{\hskip.11\wd0 .}$^{\circ}
          \else \setbox0=\hbox{$^{\circ}$}\rlap{\hskip.11\wd0 .}$^{\circ}$\fi}
\def\arcs{\ifmmode {^{\scriptstyle\prime\prime}}
          \else $^{\scriptstyle\prime\prime}$\fi}
\def\arcm{\ifmmode {^{\scriptstyle\prime}}
          \else $^{\scriptstyle\prime}$\fi}
\newdimen\sa  \newdimen\sb
\def\parcs{\sa=.07em \sb=.03em
     \ifmmode \hbox{\rlap{.}}^{\scriptstyle\prime\kern -\sb\prime}\hbox{\kern -\sa}
     \else \rlap{.}$^{\scriptstyle\prime\kern -\sb\prime}$\kern -\sa\fi}
\def\parcm{\sa=.08em \sb=.03em
     \ifmmode \hbox{\rlap{.}\kern\sa}^{\scriptstyle\prime}\hbox{\kern-\sb}
     \else \rlap{.}\kern\sa$^{\scriptstyle\prime}$\kern-\sb\fi}
\def\ra[#1 #2 #3.#4]{#1\sup{h}#2\sup{m}#3\sup{s}\llap.#4}
\def\dec[#1 #2 #3.#4]{#1\deg#2\arcm#3\arcs\llap.#4}
\def\deco[#1 #2 #3]{#1\deg#2\arcm#3\arcs}
\def\rra[#1 #2]{#1\sup{h}#2\sup{m}}

\def\dots{\relax\ifmmode \ldots\else $\ldots$\fi}
\def\WHzsr{\ifmmode $W\,Hz\mo\,sr\mo$\else W\,Hz\mo\,sr\mo\fi}
\def\mHz{\ifmmode $\,mHz$\else \,mHz\fi}
\def\GHz{\ifmmode $\,GHz$\else \,GHz\fi}
\def\mKs{\ifmmode $\,mK\,s$^{1/2}\else \,mK\,s$^{1/2}$\fi}
\def\muKs{\ifmmode \,\mu$K\,s$^{1/2}\else \,$\mu$K\,s$^{1/2}$\fi}
\def\muKRJs{\ifmmode \,\mu$K$_{\rm RJ}$\,s$^{1/2}\else \,$\mu$K$_{\rm RJ}$\,s$^{1/2}$\fi}
\def\muKHz{\ifmmode \,\mu$K\,Hz$^{-1/2}\else \,$\mu$K\,Hz$^{-1/2}$\fi}
\def\MJysr{\ifmmode \,$MJy\,sr\mo$\else \,MJy\,sr\mo\fi}
\def\MJysrmK{\ifmmode \,$MJy\,sr\mo$\,mK$_{\rm CMB}\mo\else \,MJy\,sr\mo\,mK$_{\rm CMB}\mo$\fi}
\def\microns{\ifmmode \,\mu$m$\else \,$\mu$m\fi}

\def\muK{\ifmmode \,\mu$K$\else \,$\mu$\hbox{K}\fi}
\def\microK{\ifmmode \,\mu$K$\else \,$\mu$\hbox{K}\fi}
\def\muW{\ifmmode \,\mu$W$\else \,$\mu$\hbox{W}\fi}
\def\kms{\ifmmode $\,km\,s$^{-1}\else \,km\,s$^{-1}$\fi}
\def\kmsMpc{\ifmmode $\,\kms\,Mpc\mo$\else \,\kms\,Mpc\mo\fi}
\providecommand{\sorthelp}[1]{}

\usepackage[dvipsnames]{xcolor}

\def\NHUNIT{\ifmmode {\rm \,cm^{-2}} \else $\rm \,cm^{-2}$ \fi} 

\def\muKcmb{\ifmmode \,\mu$K$_{\rm CMB}$\else \,$\mu$K$_{\rm CMB}$\fi}
\newcommand{\planck}{\Planck}

\newcommand{\OmegaM}{\ifmmode\Omega_{\rm M}\else $\Omega_{\rm M}$\fi}

\newcommand{\commander}{{\tt Commander}}

\newcommand{\nilc}{{\tt NILC}}   
\newcommand{\sevem}{{\tt SEVEM}} 
\newcommand{\smica}{{\tt SMICA}}

\providecommand{\Planck}{\textit{Planck}}
\providecommand{\planck}{\Planck}

\providecommand{\text}[1]{\rm{#1}}

\providecommand{\muK}{\mu\rm{K}}

\newcommand{\begm}{\begin{pmatrix}}
\newcommand{\enm}{\end{pmatrix}}

\def\pmb#1{\setbox0=\hbox{#1}%
    \kern-.025em\copy0\kern-\wd0
    \kern.05em\copy0\kern-\wd0
    \kern-.025em\raise.0433em\box0}

\def\p2Y{\;_2Y}
\def\m2Y{\;_{-2}Y}

\newcommand{\mksym}[1]{\ifmmode {\rm #1}\else #1\fi}

\providecommand{\text}[1]{\rm{#1}}

\providecommand{\muK}{\mu\rm{K}}

\providecommand{\healpix}{\texttt{HEALPix}}

\newcommand\ba{\begin{eqnarray}}
\newcommand\ea{\end{eqnarray}}
\newcommand\bea{\begin{eqnarray}}
\newcommand\eea{\end{eqnarray}}

\newcommand\be{\begin{equation}}
\newcommand\ee{\end{equation}}

\title{\boldmath {Planck constraints on cross-correlations between anisotropic cosmic birefringence and CMB polarization}}

\author[a,b]{M. Bortolami,}
\author[c,d,e,f]{M. Billi,}
\author[d,e,a]{A. Gruppuso,}
\author[a,b]{P. Natoli}
\author[a,b,g]{and L. Pagano}

\affiliation[a]{Dipartimento di Fisica e Scienze della Terra, Università degli Studi di Ferrara, via Saragat 1, I-44122 Ferrara, Italy}
\affiliation[b]{Istituto Nazionale di Fisica Nucleare, Sezione di Ferrara, via Saragat 1, I-44122 Ferrara, Italy}
\affiliation[c]{Dipartimento di Fisica e Astronomia, Alma Mater Studiorum Università di Bologna, Via Gobetti 93/2, I-40129 Bologna, Italy}
\affiliation[d]{Istituto Nazionale di Astrofisica - Osservatorio di Astrofisica e Scienza dello Spazio di Bologna, via Gobetti 101, I-40129 Bologna, Italy}
\affiliation[e]{Istituto Nazionale di Fisica Nucleare, Sezione di Bologna, viale Berti Pichat 6/2, I-40127 Bologna, Italy}
\affiliation[f]{Instituto de Física de Cantabria (IFCA), CSIC-UC, Avenida de Los Castros s/n, 39005 Santander, Spain}
\affiliation[g]{Institut d'Astrophysique Spatiale, CNRS, Univ. Paris-Sud, Universit\'{e} Paris-Saclay, B\^{a}t. 121, 91405 Orsay cedex, France}

\emailAdd{marco.bortolami@unife.it}
\emailAdd{matteo.billi3@unibo.it}
\emailAdd{alessandro.gruppuso@inaf.it}
\emailAdd{paolo.natoli@unife.it}
\emailAdd{luca.pagano@unife.it}

\abstract{Cosmic Birefringence (CB) is the in-vacuo rotation of the linear polarization direction of photons during propagation, caused by parity-violating extensions of Maxwell electromagnetism. We build low resolution CB angle maps using \planck\ Legacy and NPIPE products and provide for the first time estimates of the cross-correlation spectra $C_L^{\alpha E}$ and $C_L^{\alpha B}$ between the CB and the CMB polarization fields. We also provide updated CB auto-correlation spectra $C_L^{\alpha\alpha}$ as well as the cross-correlation $C_L^{\alpha T}$ with the CMB temperature field. We report constraints by defining the scale-invariant amplitudes $A^{\alpha X} \equiv L(L + 1)C_L^{\alpha X}/2\pi$, where $X = \alpha, T, E, B$, finding no evidence of CB. In particular, we find $A^{\alpha E} = (-7.8 \pm 5.6)$~nK~deg and $A^{\alpha B} = (0.3 \pm 4.0)$~nK~deg at 68\% C.L..}

\begin{document}
\maketitle
\flushbottom

\section{Introduction}\label{sec:intro}
The standard model of particle physics predicts that the electromagnetic (EM) interaction is invariant under the parity operator. 
Nonetheless, models of parity-violating electromagnetism have been proposed, reflecting the phenomenology of weak interactions~\cite{Wu:1957my}.
A popular model extends Maxwell's Lagrangian with a parity-violating Chern-Simons term~\cite{Carroll:1989vb,Carroll:1991zs}:
\begin{equation}
    \mathcal{L} = \mathcal{L}_{EM} + \mathcal{L}_{CS} = -\dfrac{1}{4} F_{\mu\nu}F^{\mu\nu} - \dfrac{1}{2} p_\mu A_\nu \widetilde{F}^{\mu\nu},
\end{equation}
where $F_{\mu\nu}$ is the EM field tensor, $p_\mu$ is a constant kinematic four-vector with dimensions of mass, $A_\nu$ is the four-potential and $\widetilde{F}^{\mu\nu} = \frac{1}{2}\epsilon^{\mu\nu\alpha\beta}F_{\alpha\beta}$, $\epsilon^{\mu\nu\alpha\beta}$ being the four-dimensional Levi-Civita symbol. The additional parity-violating term $\mathcal{L}_{CS}$ is responsible for the Cosmic Birefringence (CB) effect, the in-vacuo rotation of the linear polarization plane of photons during propagation.\par
Hence, to observe such an effect one needs a source of linear polarized light.
The Cosmic Microwave Background (CMB) is linearly polarized at the level of 1\% to 10\% due to Thomson scattering off free electrons at the surface of last scattering and it is therefore a good observable to constrain this phenomenon \cite{Kostelecky:2007zz,Kostelecky:2008ts}. Moreover, the CMB is the farthest EM radiation available in nature and this increases the chances of detecting the CB effect, as the polarization plane had more time to rotate, since the effect accumulates during propagation.\par
The CB effect induces correlations between the temperature field T and the B-modes of the polarization field or between the E- and B-modes of the polarization field, that are null with standard EM as the parity symmetry prevents these correlations to be active. Thus, we can use TB and/or EB correlations to detect the CB effect. However, for the \planck\ data, which are employed in this work, the signal-to-noise ratio due to TB correlation is about a factor of two smaller than EB~\cite{planck2014-a23}. Hence, in this paper we focus only on the EB correlation.\par
If the parity-violating cross-correlations of the CMB fields do not depend on the direction of observation~\cite{Lue:1998mq,Feng:2006dp,Xia:2007qs,Finelli:2008jv,Li:2008tma,Gubitosi:2009eu,Pagano:2009kj,Gubitosi:2012rg,Gubitosi:2014cua,Gruppuso:2016nhj}, the CB effect is isotropic and it can be modelled with a single angle $\alpha$. Otherwise, the CB effect is anisotropic~\cite{Kamionkowski:2008fp,Gluscevic:2009mm,Gubitosi:2011ue,Gluscevic:2012me} and it has to be modelled with a set of angles $\alpha(\hat{n})$. Recent constraints on the CB effect can be found in~\cite{planck2014-a23,Gruppuso:2020kfy,Eskilt:2022wav,MinamiKomatsu2020,Diego-Palazuelos:2022dsq,Komatsu:2022nvu,Eskilt:2022cff,Abghari:2022bet} for the isotropic case\footnote{For a detailed summary of previous constraints see e.g. references in \cite{planck2014-a23}.} and in~\cite{Contreras:2017sgi,Gruppuso:2020kfy,Namikawa:2020ffr,SPT:2020cxx} for the anisotropic case. Isotropic and anisotropic CB can be originated by different physical models, see e.g.~\cite{PhysRevD.89.103518}. In this work, following a phenomenological approach, we focus on the anisotropic case but we also estimate, as consistency check with previous results, the CB monopole, i.e. the isotropic signal.\par
Parity violating electromagnetism Lagrangian also induce a non-null cross-correlation between the CB field and the CMB temperature and polarization maps, see e.g.~\cite{Greco:2022ufo} and references therein for a detailed treatment of cross-spectra and bi-spectra. Furthermore, in recent years, early dark energy models inspired by ultra-light axion fields~\cite{Karwal:2016vyq,Poulin:2018cxd} have been proposed to alleviate the $H_0$ tension~\cite{Schoneberg:2021qvd}. These models are known to produce CB signal and a non-zero cross-correlation between CB and the CMB fields, see e.g. \cite{Capparelli:2019rtn}.\par
In this paper we analyze CB extending what was done in Ref.~\cite{Gruppuso:2020kfy}. In particular, using the \planck\ Public Data Release 3 (PR3) products, we provide for the first time constraints on the cross-correlation spectra $C_L^{\alpha E}$ and $C_L^{\alpha B}$ between the CB angle maps and the CMB polarization maps at large angular scales. 
We also provide the CB auto-correlation spectra $C_L^{\alpha\alpha}$ and its cross-correlation spectra $C_L^{\alpha T}$ with the CMB temperature map up to $L=24$, exploiting the \planck\ Public Data Release 4 (NPIPE) data in addition to the \planck\ PR3 products.\par
This paper is organized as follows: in Section~\ref{sec:Dataset and Simulations} we review the features of the \planck\ datasets and simulations employed for this study; in Section~\ref{sec:Analysis pipeline} we describe the analysis pipeline; in Section~\ref{sec:Results} we present the results of this work; in Section~\ref{sec:Conclusions} we draw the conclusions; in Appendix~\ref{E and B modes purification for CMB APS extraction} we discuss the purification of E and B modes for the CMB Angular Power Spectra (APS) estimation; in Appendix~\ref{app:effective beam calculation} we explain the effective beam calculation used for the CMB spectra; in Appendix~\ref{app:results on simulations} we report the results of the analysis on simulations; in Appendix~\ref{app:Analysis robustness tests} we perform robustness tests for the analysis pipeline; in Appendix~\ref{app:2D plots} we set 2-dimensional (2D) joint constraints on $A^{\alpha \alpha}$ and $A^{\alpha T}$.

\section{Datasets and Simulations}\label{sec:Dataset and Simulations}
In this work we employ CMB maps provided by the \planck\ collaboration in the PR3 analysis \cite{planck2016-l01} and in the subsequent NPIPE re-analysis \cite{planck2020-LVII}. The PR3 maps have been cleaned from foregrounds by the four official \Planck\ component separation methods \commander, \nilc, \sevem\ and \smica\ \cite{planck2016-l04}, while NPIPE maps have been processed only by \commander\ and \sevem\ \cite{planck2020-LVII}. The corresponding simulations, provided by the \Planck\ collaboration, have been also employed. The resolution of the maps, following the \healpix\footnote{\href{http://healpix.sourceforge.net}{http://healpix.sourceforge.net}}~\cite{Gorski:2004by} pixelization scheme, is $N_{side} = 2048$. The NPIPE products have never been used to estimate the CB auto-correlation spectra $C_L^{\alpha\alpha}$ before this work.\par
The PR3 simulation set is composed of 1000 CMB Monte Carlo simulations generated using the \planck\ $\Lambda$CDM best-fit model, 300 noise simulations for the full-mission (FM) and 300 noise simulations for each of the two half-mission (HM) data splits, and for each of the component separation methods. We sum the first 300 CMB realizations with each set of 300 HM noise simulations that realistically describe the \planck\ 2018 data. We employ the two data splits in the construction of the $\alpha$ map. We also sum the first 300 CMB realizations with the 300 FM noise simulations in order to obtain 300 FM simulations. We use these maps for the cross-correlation spectra $C_L^{\alpha E}$ and $C_L^{\alpha B}$ estimation. The PR3 simulations contain beam leakage effects~\cite{planck2014-a14}, ADC non linearities, thermal fluctuations and other systematic effects~\cite{planck2016-l03}.\par
The NPIPE data and simulation sets are an evolution over PR3~\cite{planck2020-LVII}. One of the effects of these improvements is lower levels of noise and systematics in the CMB maps at essentially all angular scales~\cite{planck2020-LVII}. We employ the 400 simulations available for \commander\footnote{The simulations used in the analysis are actually 399 because one turned out corrupted.} and the 600 simulations available for \sevem. The NPIPE \commander\ simulations contain the PR3 CMB signal and the instrumental noise. To extract the noise-only maps, we subtract the signal-only PR3 simulations to the complete NPIPE simulations. The NPIPE \sevem\ CMB simulations are divided in noise-only and signal-only as the PR3 ones, which we sum for a set of noisy CMB simulations. The NPIPE data splits, where the systematics between the splits are expected to be uncorrelated, are called A and B (see again~\cite{planck2020-LVII} for a full description). We calculate the NPIPE CMB APS by cross-correlating those A and B maps. The NPIPE simulations employed in this work are the ``noise aligned'' ones, see again~\cite{planck2020-LVII} for details.

\section{Analysis pipeline}\label{sec:Analysis pipeline}
As first step of the analysis we need to define the processing masks. We divide the \healpix\ sky at $N_{side} = 2048$ in small regions following the pixelization scheme at $N_{side} = 8$. In the rest of the paper we refer to these regions as ``patches''. The sky fraction covered by one single patch is $f_{sky,patch} \simeq 0.13\%$ and the total number of available patches is 768. Working at resolution $N_{side}=8$ is an improvement with respect to~\cite{Gruppuso:2020kfy}, where they used $N_{side}=4$. We tried to increase the resolution to $N_{side}=16$, but the patch sky fraction was too low to get reliable CMB spectra.\par
\begin{figure}[h]
\centering
\includegraphics[width=.49\textwidth]{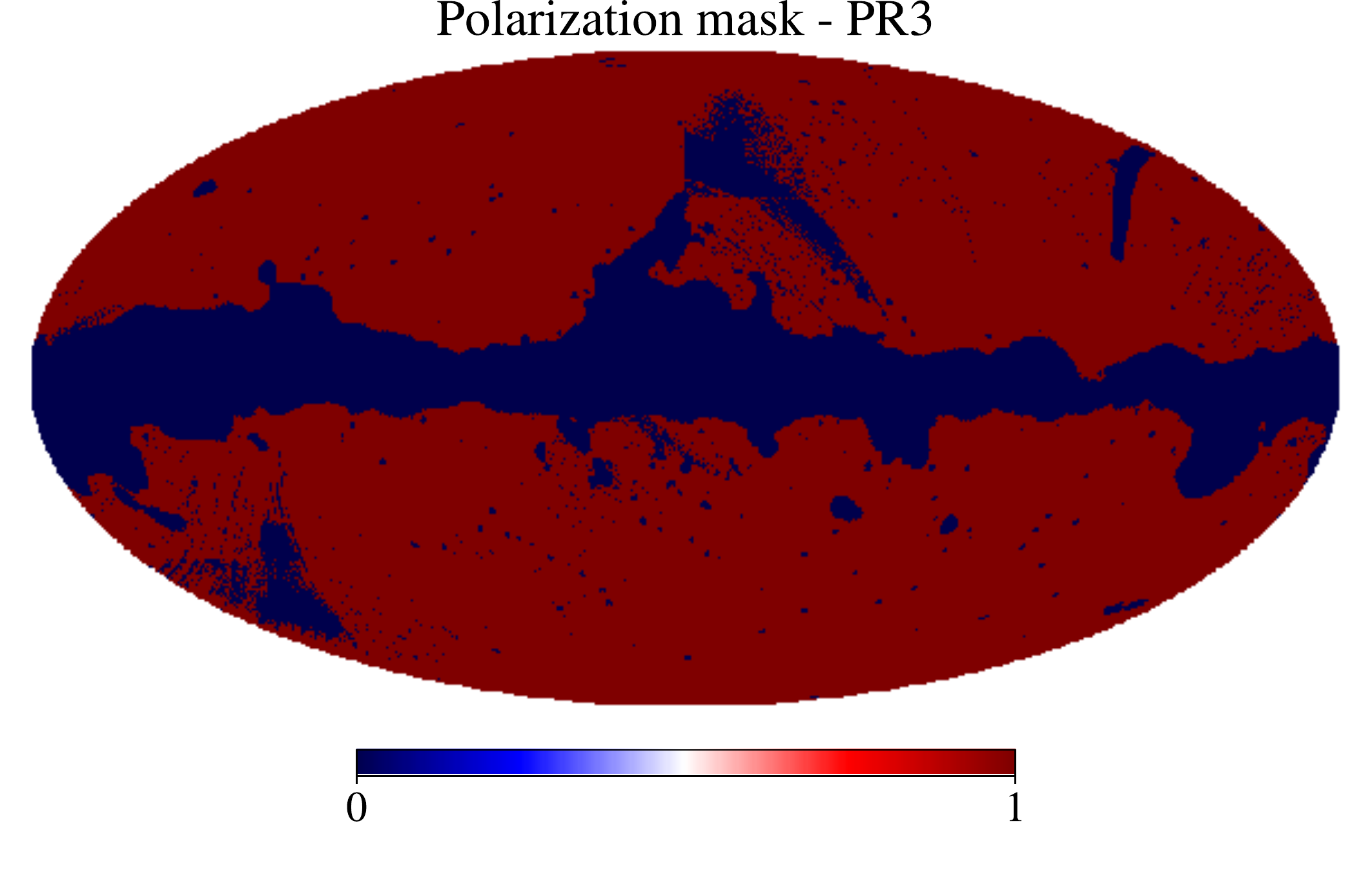}
\hfill
\includegraphics[width=.49\textwidth]{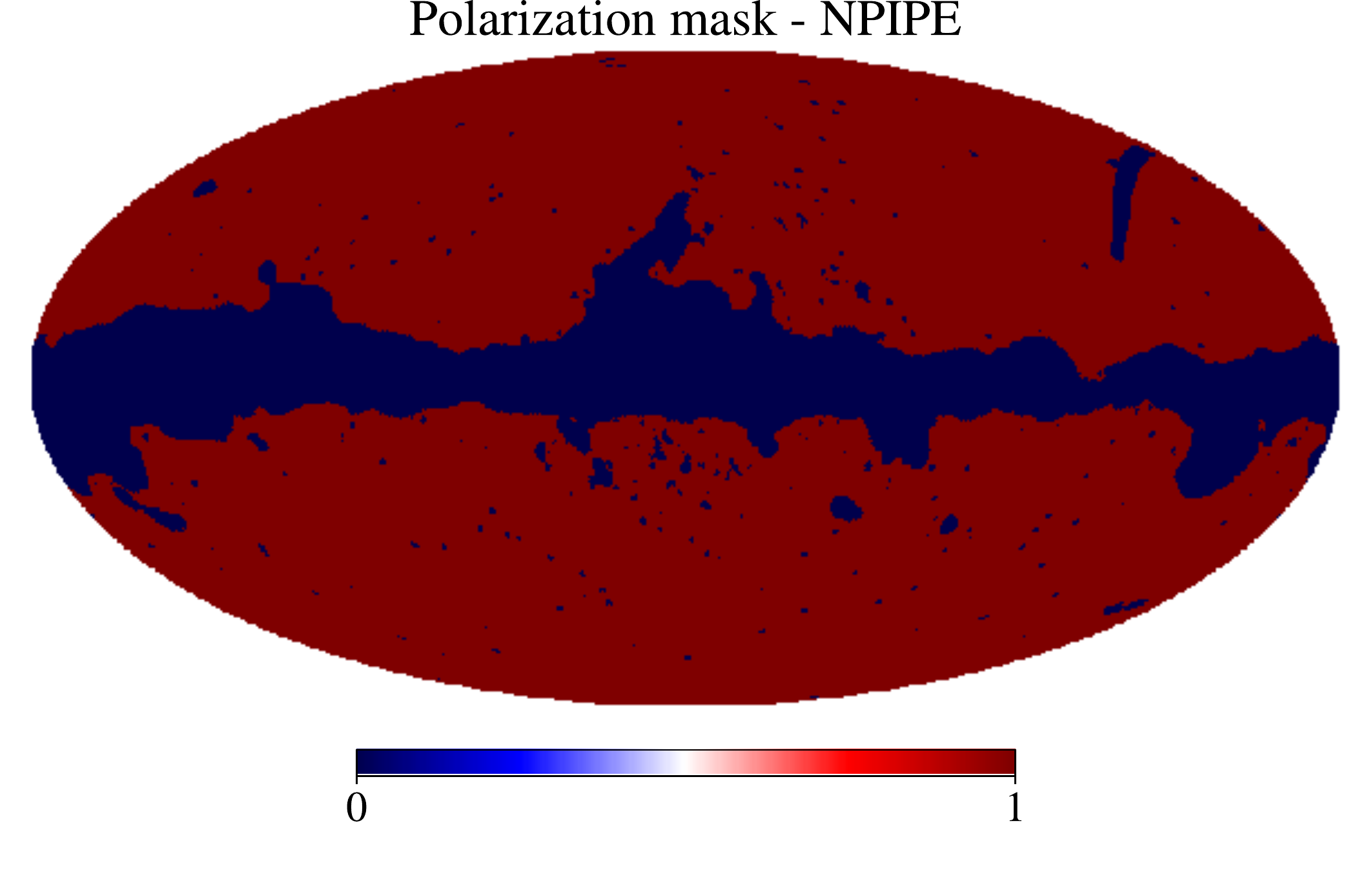}
\includegraphics[width=.49\textwidth]{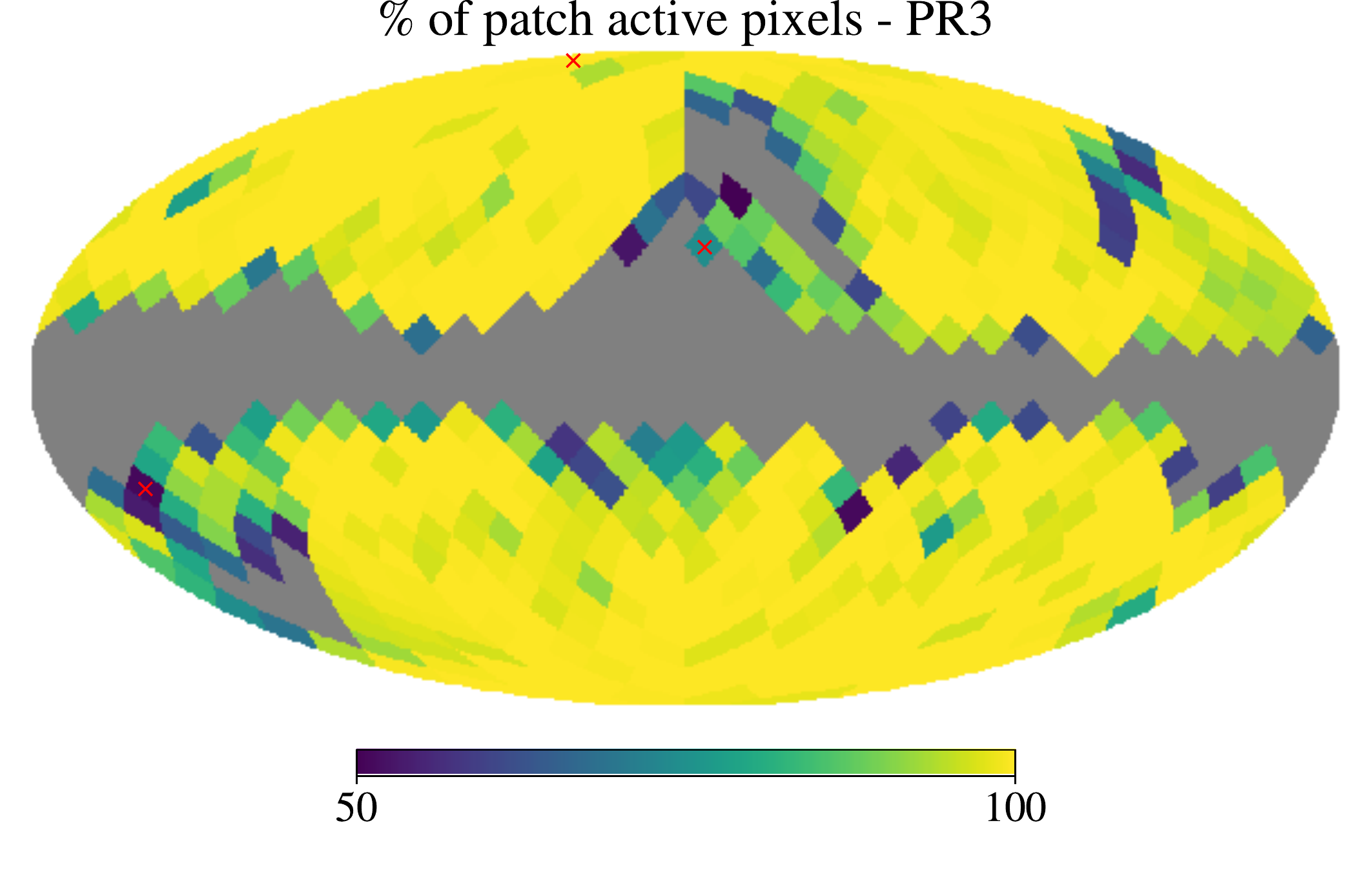}
\hfill
\includegraphics[width=.49\textwidth]{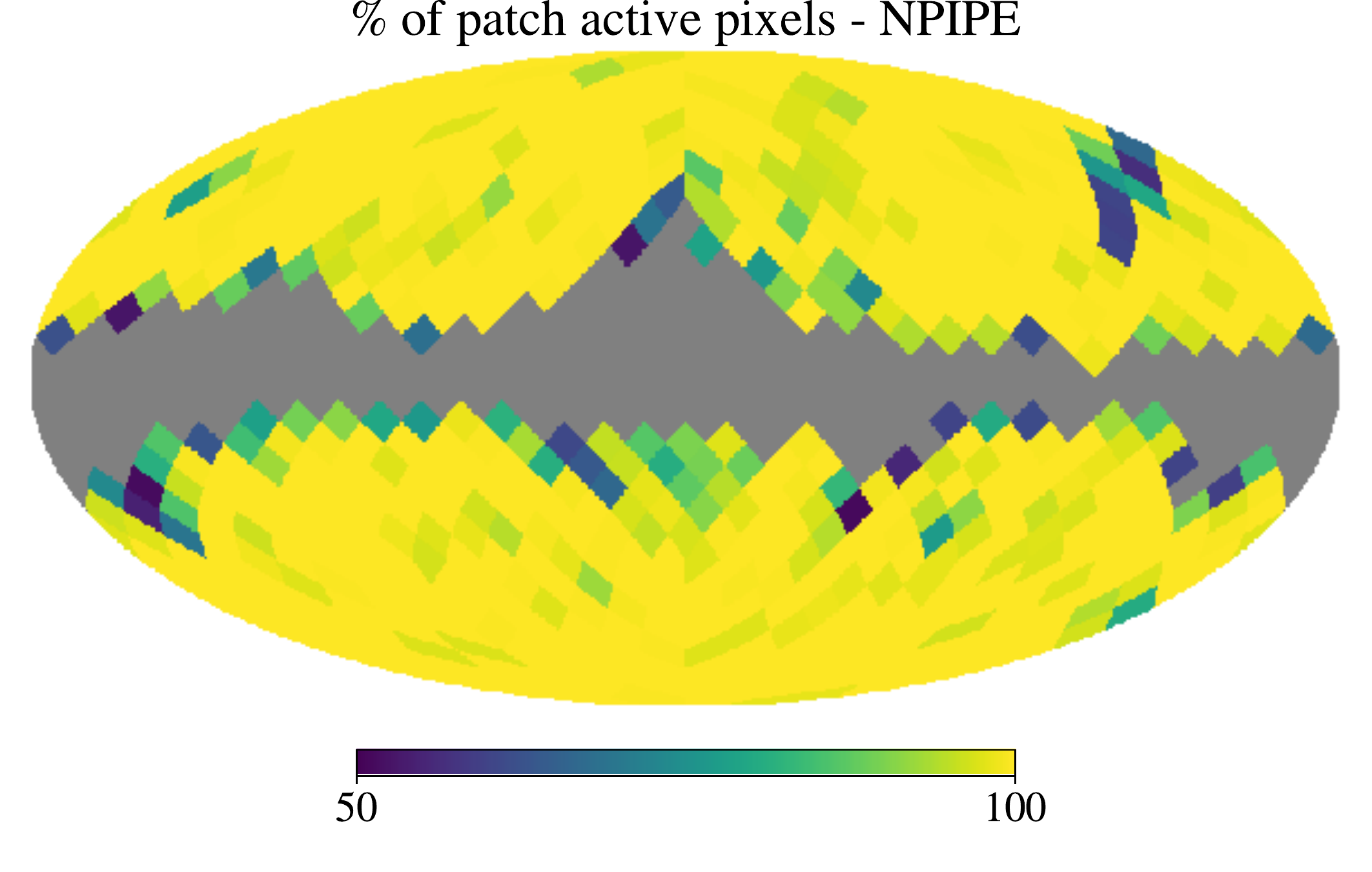}
\caption{\label{fig:mask} Top: polarization masks used in this work (left for PR3, right for NPIPE). Bottom: percentage of active pixels in each patch (left for PR3, right for NPIPE). The grey patches are not used in the analysis. The patches marked with a red x in the bottom left panel are the ones whose CMB spectra validation is shown in Fig.~\ref{fig:CMB_spectra_validation}, i.e. patches with \healpix\ indexes 1, 207, 542 from top to bottom.}
\end{figure}

We then combine the masks of these patches with a polarization mask that excludes the regions unobserved or heavily contaminated by foregrounds. For PR3 we combine the foreground mask (Galactic and point sources) with the HM missing pixels mask. For NPIPE only the foreground mask is employed. The polarization masks used are shown in the top panels of Fig.~\ref{fig:mask} (left for PR3, right for NPIPE).
\par
Some of the aforementioned patches contain masked pixels. A large fraction of masked pixels in a patch prevents the extraction of reliable CMB spectra. For this reason, we select and analyze only the patches that have at least 50\% of active (i.e. non masked) pixels. This corresponds to $f_{sky,patch} \gtrsim 0.065\%$. The maps showing the selected patches and the fraction of active pixels in each of them are shown in the bottom panels of Fig.~\ref{fig:mask} (left for PR3, right for NPIPE), where the grey patches are the excluded ones. As expected, among the selected patches, the ones close to the Galactic mask borders or the ones containing a large number of point sources have a lower percentage of active pixels. The number of selected patches is 571 for PR3 and 592 for NPIPE. The final masks are not apodized because the apodization process reduces the effective sky fraction covered by each patch and, for the sky fraction considered, it does not provide any benefit in the CMB computed APS, see also the discussion about polarization purification in the next paragraph.\par
For each selected patch, the CMB APS is calculated with the Python package Pymaster\footnote{\href{https://namaster.readthedocs.io}{https://namaster.readthedocs.io}}~\cite{Alonso:2018jzx} by cross-correlating the two splits both for data and simulations. We recall that the data splits are HM 1 and 2 for PR3, A and B for NPIPE, as explained in Sec.~\ref{sec:Dataset and Simulations}. We compute the APS in cross-mode to reduce systematic effects and noise mismatches~\cite{planck2016-l02,planck2016-l03}. When analysing a masked CMB map to produce the APS, E to B leakage can be generated~\cite{Grain:2009wq}. We have the possibility to purify the E and/or B modes in order to remove the misinterpreted modes, at a cost of information loss \cite{Grain:2009wq,Alonso:2018jzx}. 
We choose not to purify E and B modes: for the noise level of \Planck, the size of the patches considered and for the multipole range of interest, the purification does not provide any substantial advantage, and does increase the error bars (see App.~\ref{E and B modes purification for CMB APS extraction} for more details). We calculate the APS for multipoles from 2 to 2047. We deconvolve a non-Gaussian effective beam obtained from simulations, as explained in App.~\ref{app:effective beam calculation}. The APS is binned with $\Delta l = 60$ multipoles in order to reduce the errors and the correlations induced by the cut-sky~\cite{Hivon:2001jp}.\par
\begin{figure}[h]
    \centering
    \includegraphics[width=1.\textwidth]{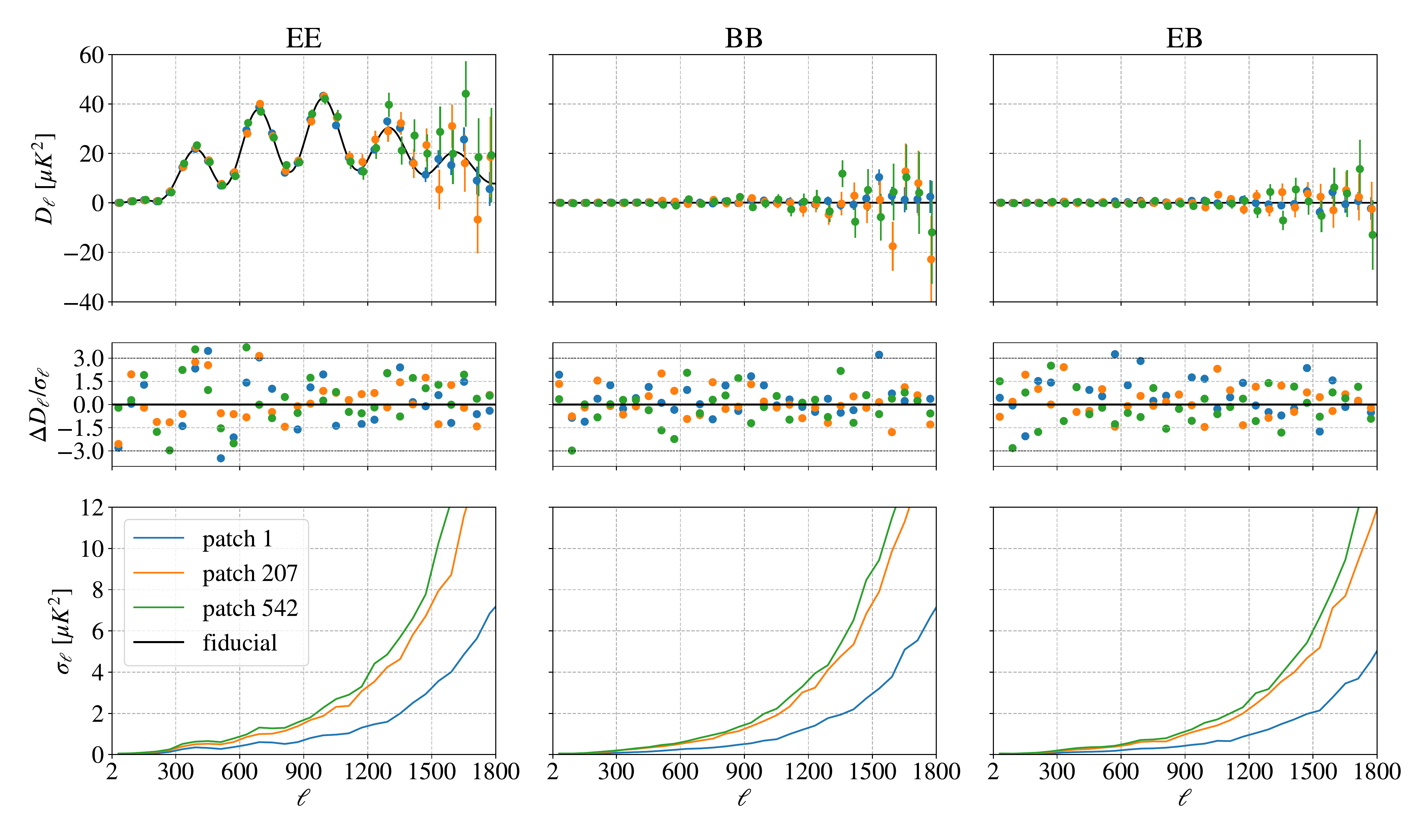}
    \caption{Top: EE (left), BB (center) and EB (right) CMB power spectra for three patches and for the \commander\ component separation method. The black line represents the fiducial, while the blue, orange and green points are the simulation average spectra for the patches with \healpix\ RING ordering indexes 1, 207 and 542, respectively (highlighted with a red cross in Fig.~\ref{fig:mask}). The error bars represent the error of the mean. Center: EE, BB and EB difference between the simulation average and the fiducial spectra, divided by the error of the mean. Bottom: EE, BB and EB errors of the mean.}
    \label{fig:CMB_spectra_validation}
\end{figure}

We show in the top panels of Fig.~\ref{fig:CMB_spectra_validation} the PR3 \commander\ validation plots for the EE, BB and EB power spectra for three patches, in terms of the bandpowers $D_\ell = \ell(\ell+1)C_\ell/2\pi$.
The three patches, following the \healpix\ RING ordering scheme, have indexes 1 (blue), 207 (orange) and 542 (green) and fractions of patch active pixels 100\%, 74.5\% and 51.1\%, respectively. The dots represent the mean of the simulations and the error bars show the uncertainty on the mean. The fiducial power spectra are represented by the black lines in the plots. These plots show that the simulations CMB spectra are validated. The scatter around the fiducial at the highest multipoles is large and, for this reason, we do consider the APS with $\ell \lesssim 1500$. We show in the central panels of Fig.~\ref{fig:CMB_spectra_validation} the residuals plots, i.e. the difference between the simulation average and the fiducial spectra, divided by the error of the mean. For all the cases considered the simulation average is less than $3.5\sigma_{\rm mean}$ away from the fiducial. The residuals do not show any systematic behaviour. We report in the bottom panels of Fig.~\ref{fig:CMB_spectra_validation} the error of the mean. As expected, the error increases for higher multipoles and for lower numbers of active pixels in the patch. We verified that the behaviour of other patches and other component separation methods is similar to the one shown in Fig.~\ref{fig:CMB_spectra_validation}. For the NPIPE case the simulations are validated with a similar procedure, the only difference is that the errors are slightly smaller. This is due to the lower level of noise with respect to the PR3 simulation set~\cite{planck2020-LVII}.\par
Given the low number of simulations, the CMB covariance matrices for each patch and for each component separation method are obtained via the analytical implementation of Pymaster~\cite{Alonso:2018jzx}. The module that produces the matrices needs the CMB signal APS, the noise APS and the smoothing function. The CMB signal APS used for the covariance matrix calculation is the \planck\ FFP10 fiducial~\cite{planck2016-l03}. The noise APS for each of the two data splits (HM1 and HM2 for PR3, A and B for NPIPE) is obtained by averaging the auto-correlation of the corresponding noise-only simulations. The smoothing function considered is obtained as discussed in App.~\ref{app:effective beam calculation}. We verified that these analytical covariance matrices are a good approximation of the ones obtained using the simulations. This is not valid for the first multipole bin, so we exclude the latter from the analysis and use the CMB spectra starting from $\ell_{min}=62$.\par
After obtaining the CMB APS and the corresponding covariance matrices, we estimate the CB angle in each patch assuming that the CB effect is isotropic inside each patch\footnote{Since the scales of the anisotropic birefringence angle we are interested in, are larger than the size of the aforementioned patches, we can consider constant the birefringence angle at the scale of the patches or smaller.}. The CMB APS is rotated by the isotropic CB effect in the following way~\cite{Gruppuso:2016nhj}:
\begin{align}
    C_\ell^{TT,obs} &= C_\ell^{TT},\\
    C_\ell^{TE,obs} &= C_\ell^{TE}\cos(2\alpha),\\
    C_\ell^{TB,obs} &= C_\ell^{TE}\sin(2\alpha),\\
    C_\ell^{EE,obs} &= C_\ell^{EE}\cos^2(2\alpha) + C_\ell^{BB}\sin^2(2\alpha),\\
    C_\ell^{BB,obs} &= C_\ell^{BB}\cos^2(2\alpha) + C_\ell^{EE}\sin^2(2\alpha),\\
    C_\ell^{EB,obs} &= \frac{1}{2} \left(C_\ell^{EE} - C_\ell^{BB}\right)\sin(4\alpha),
\end{align}
where $\alpha$ is the CB angle, the APS with the $obs$ tag is the rotated (observed) one and the APS without the $obs$ tag is the unrotated one. Without the CB effect, i.e. $\alpha = 0$, the TB and EB spectra are null, as expected.\par
Harmonic based estimators for $\alpha$ can be built exploiting the observed EB and/or TB spectra, see e.g.~\cite{Gruppuso:2016nhj}. Since the signal-to-noise ratio is smaller for the TB-based estimator as compared to EB one, we use the EB-only D-estimator defined in~\cite{Gruppuso:2016nhj}:
\begin{equation}
    D_\ell^{EB,obs} = C_\ell^{EB,obs} \cos(4\alpha) - \frac{1}{2} \left(C_\ell^{EE,obs} - C_\ell^{BB,obs}\right)\sin(4\alpha) \, .
    \label{eq:Dest}
\end{equation}\par
We obtain an estimate of the isotropic CB angle $\alpha$ in each patch by minimizing the following $\chi^2_{EB}$:
\begin{equation}
    \chi^2_{EB} = \sum_{\ell,\ell'}D_\ell^{EB,obs} M_{\ell\ell'}^{EB} D_{\ell'}^{EB,obs},
\end{equation}
where $D_\ell^{EB,obs}$ are the D-estimators defined in Eq.~(\ref{eq:Dest}) and $M_{\ell\ell'}^{EB} = \langle D_{\ell}^{EB,obs}D_{\ell'}^{EB,obs}\rangle^{-1}$ is their covariance matrix. The $M_{\ell\ell'}^{EB}$ matrix reduces to the EB covariance matrix, i.e. without the CB effect~\cite{Gruppuso:2016nhj}, when the noise on the E and B CMB fields is equivalent. Since the latter is valid for the Planck maps we are employing~\cite{planck2016-l02,planck2016-l03}, we use the EB covariance matrix in the $\chi^2_{EB}$ minimization. Only the CMB spectra and covariance matrix related to a specific patch are used in the $\chi^2_{EB}$ to estimate the CB angle in that patch.\par
For each CMB map we minimize the $\chi^2_{EB}$ in each patch obtaining a map of CB angles. From these maps we firstly estimate the CB monopole averaging the angle values in the pixels outside the masks shown in the bottom panels of Fig.~\ref{fig:mask}. Then, we calculate their auto-correlation spectra $C_L^{\alpha\alpha}$ and their cross-correlation spectra $C_L^{\alpha T}$, $C_L^{\alpha E}$ and $C_L^{\alpha B}$ with the CMB T, Q, U maps employing a Quadratic Maximum Likelihood (QML) estimator~\cite{Tegmark:2001zv,Gruppuso:2009ab,Vanneste:2018azc,Pagano:2019tci}, where $\alpha$ represents an extra scalar field. The maps and the associated covariance matrices used in the QML estimator are described in the next paragraphs.\par
The CB fields for data and simulations are obtained through the $\chi^2_{EB}$ minimization, described above. Following~\cite{Gruppuso:2020kfy}, the covariance matrices associated to the data are assumed to be diagonal, and they are computed from the variance of the simulations in each pixel. See App.~\ref{app:results on simulations} for more details about this choice. In the computation of the angular power spectra, we marginalize over monopole (for all the cases) and dipole (for the $\alpha T$, $\alpha E$ and $\alpha B$ cases) contributions following the procedure described in \cite{planck2014-a13,planck2016-l05}. Hence, note that we retain $L=1$ as the lowest multipole for $\alpha\alpha$ spectra.\par
For the CMB temperature field we always use the \commander\ map since it is the one used by the \planck\ collaboration for the large scale temperature cosmological analysis. We produce 300 signal-only CMB temperature simulations at \healpix\ resolution $N_{side}=8$ smoothed with a Gaussian beam with 880' FWHM and an analytic pixel window function, to which we sum isotropic white noise realizations with standard deviation of 500 nK for regularization. The \commander\ data map is down-sampled and regularized accordingly. The CMB temperature covariance matrix is analytically calculated from the FFP10 fiducial spectrum~\cite{planck2016-l03} and contains the marginalization of monopole and dipole and a diagonal term to take into account the 500 nK regularization noise.\par
For the CMB polarization fields, instead, we use the \smica\ polarization maps because it is the only component separation method providing harmonic weights\footnote{\href{https://wiki.cosmos.esa.int/planck-legacy-archive/index.php/SMICA_propagation_code}{https://wiki.cosmos.esa.int/planck-legacy-archive/index.php/SMICA\_propagation\_code}}~\cite{planck2016-l04,planck2016-ES} that allow to correctly compute the pixel space noise covariance matrix, from single frequency noise covariances. We down-sample the 300 PR3 polarization simulations and the data maps from $N_{side}=2048$ to $N_{side}=8$, applying, in addition to an analytic pixel window function, a cosine window function \cite{Benabed:2009af} defined as:
\begin{equation}
     b_{\ell} = \left\{\begin{matrix}
    1 & \hspace{-28pt} \mbox{for $\ell < N_{side}$}\\
   \frac{1}{2}\left(1+\sin\left(\frac{\pi}{2}\frac{\ell}{N_{side}}\right)\right) & \hspace{20pt} \mbox{for $N_{side} \leq \ell < 3\ N_{side}$}\\
    0 & \hspace{-44.5pt} \mbox{otherwise}\\
\end{matrix}\right. .
\end{equation}
We then sum a 20 nK regularization noise. The corresponding covariance matrix contains: the correlated noise term (computed using the \smica\ weights and the FFP8 \cite{planck2014-a14} noise covariance matrices), the signal term based on a fiducial power spectrum, and the regularization noise term. We limit the estimates of the $\alpha E$ and $\alpha B$ spectra at $L = 23$ by definition of the considered smoothing window.\par
\begin{figure}[h!]
\centering
\includegraphics[width=.49\textwidth]{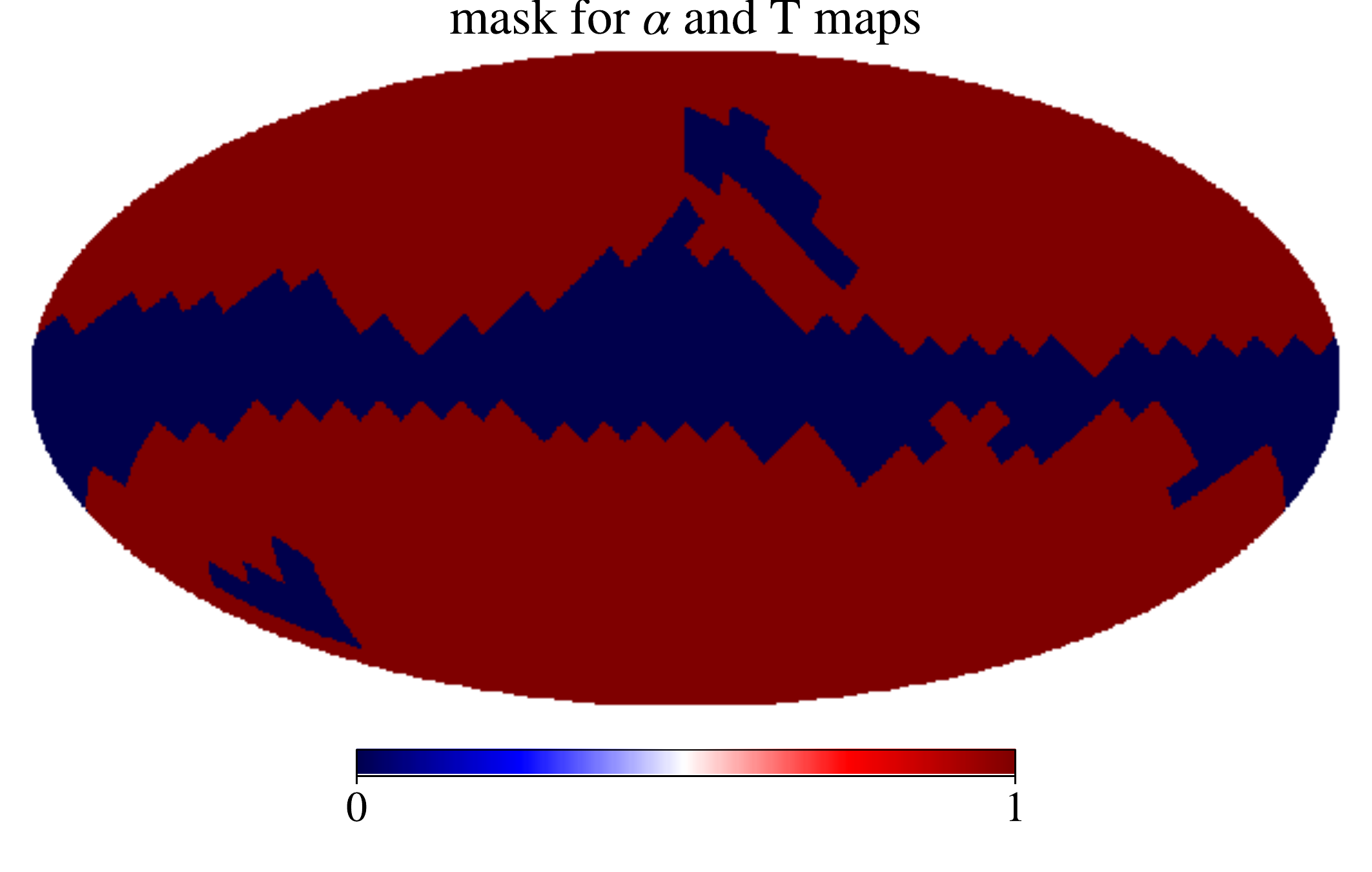}
\hfill
\includegraphics[width=.49\textwidth]{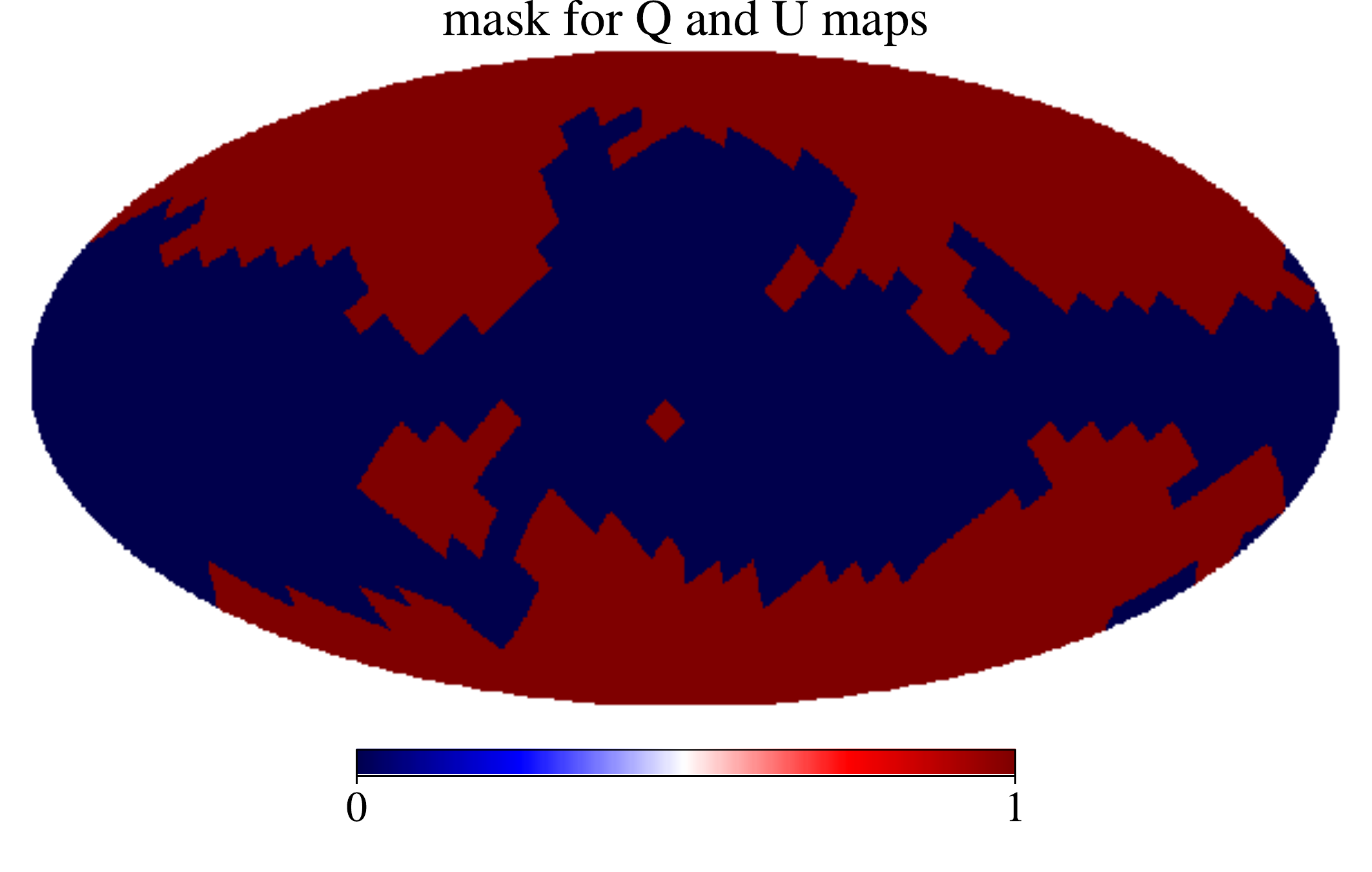}
\caption{\label{fig:CB_spectra_masks} Masks used for the CB spectra QML estimator for the $\alpha$ and $T$ (left), $Q$ and $U$ (right) maps.}
\end{figure}

We show in Fig.~\ref{fig:CB_spectra_masks} the masks used for the extraction of the spectra with the QML estimator. The mask employed with the $\alpha$ and $T$ maps has $\approx$ 74\% of non-masked pixels, while the fraction of active pixels for the $Q$ and $U$ mask is $\approx$ 47\%. The masks are obtained as the product of the PR3 $\alpha$ mask, that excludes the grey patches shown in the bottom left panel of Fig.~\ref{fig:mask}, and the CMB galactic mask for intensity ($\alpha$ and $T$ maps) or polarization ($Q$ and $U$ maps).
\par
After having obtained the $C_L^{\alpha X}$ spectra, where X can be $\alpha$, T, E or B, we set constraints in terms of the scale-invariant amplitude $A^{\alpha X}$. To do so, we minimize the following $\chi^2\left(A^{\alpha X}\right)$:
\begin{equation}\label{eq:chi2_A}
    \chi^2\left(A^{\alpha X}\right) = \sum_{L,L'}\left(\dfrac{L(L+1)}{2\pi}C_L^{\alpha X}-A^{\alpha X}\right) M_{LL'}^{-1} \left(\dfrac{L'(L'+1)}{2\pi}C_{L'}^{\alpha X}-A^{\alpha X}\right),
\end{equation}
where $C_L^{\alpha X}$ are the data spectra and $M_{LL'} = \left< \dfrac{L(L+1)}{2\pi} C_L^{\alpha X} \dfrac{L'(L'+1)}{2\pi}C_{L'}^{\alpha X} \right> $ is calculated from simulations.

\section{Results}\label{sec:Results}
In this section we show the results of our analysis, adopting the following color code: red for \commander, orange for \nilc, green for \sevem, and blue for \smica.\par
\begin{figure}[h!]
\centering
\includegraphics[width=.49\textwidth]{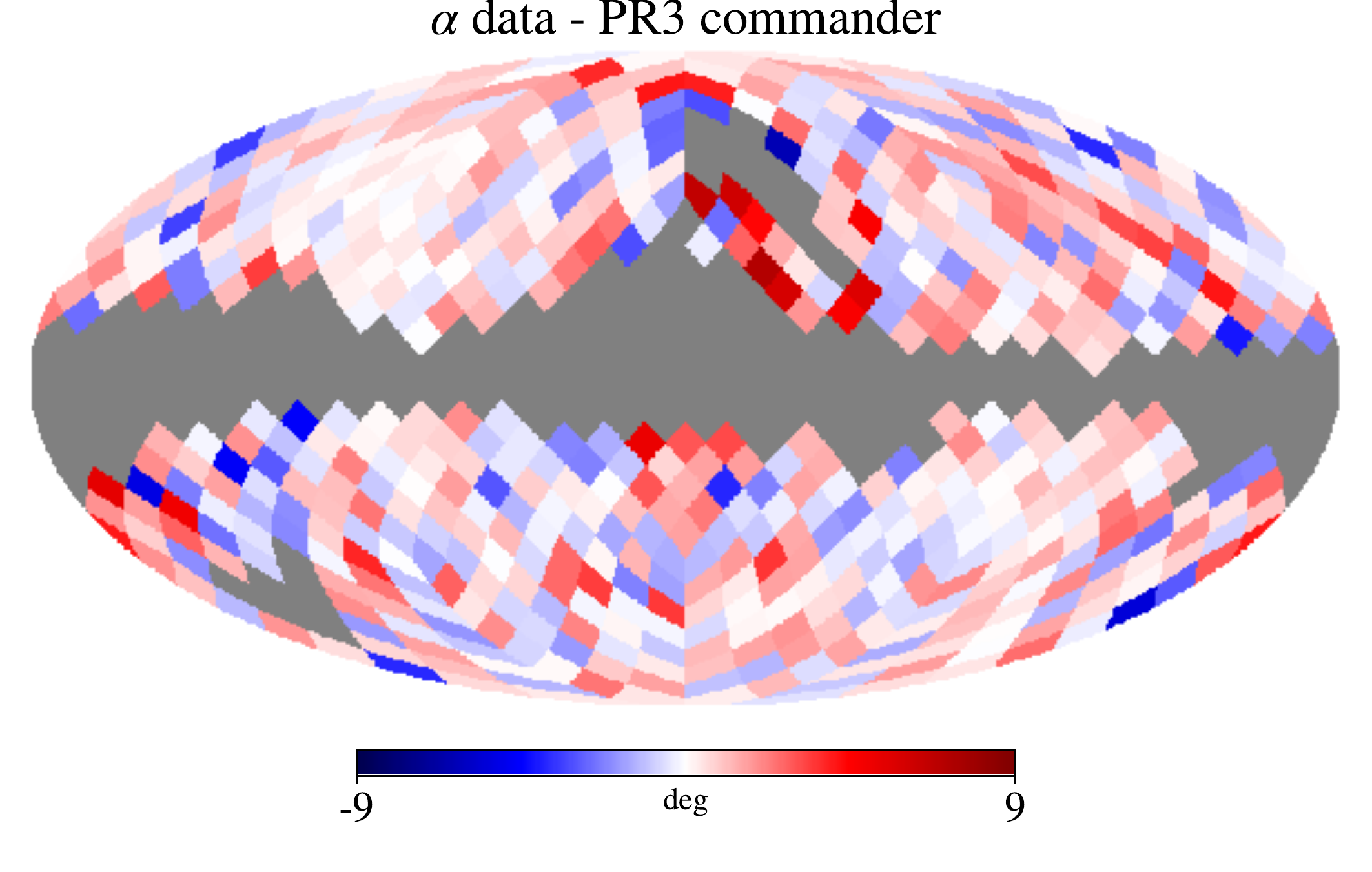}
\hfill
\includegraphics[width=.49\textwidth]{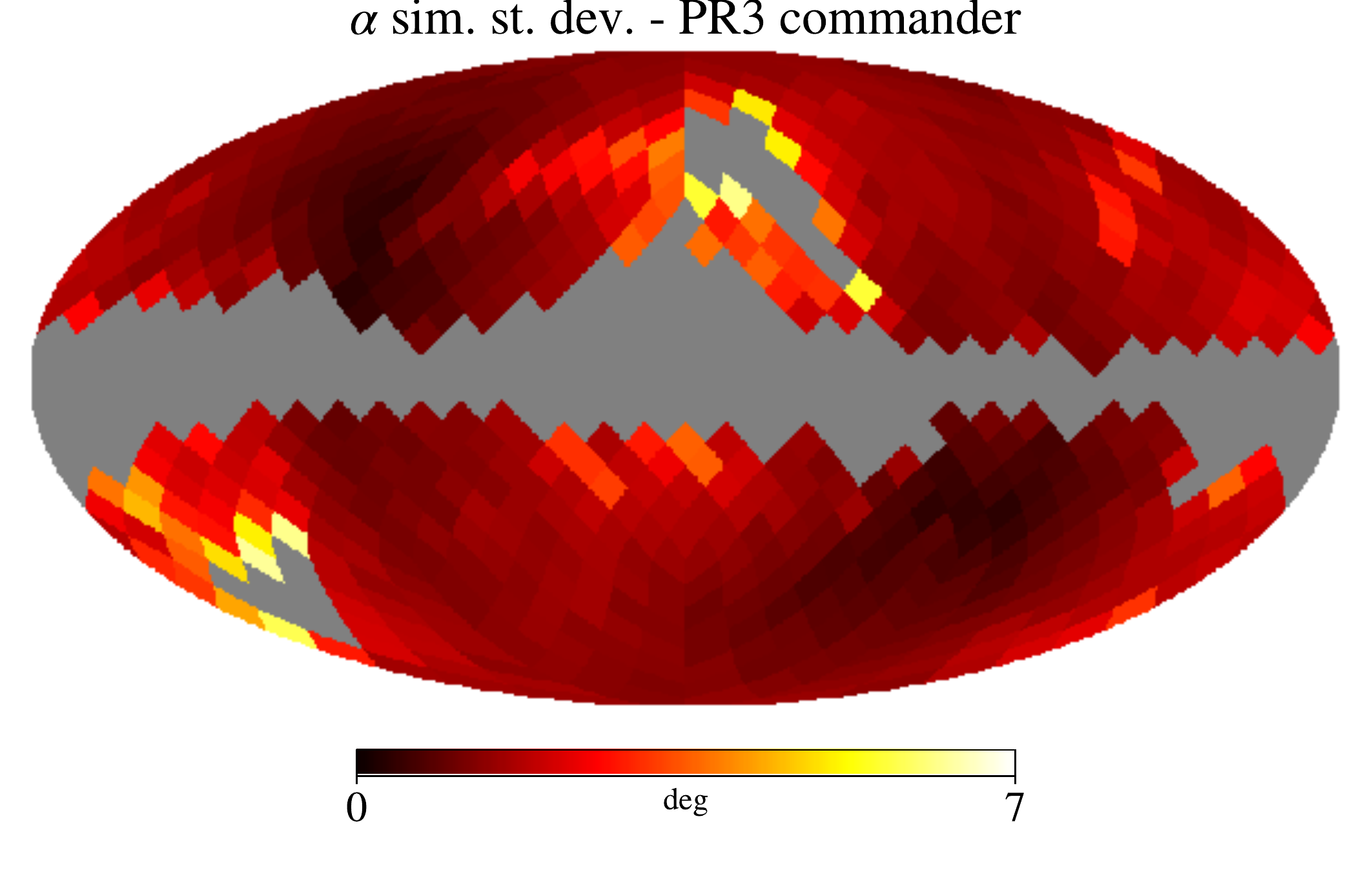}
\includegraphics[width=.49\textwidth]{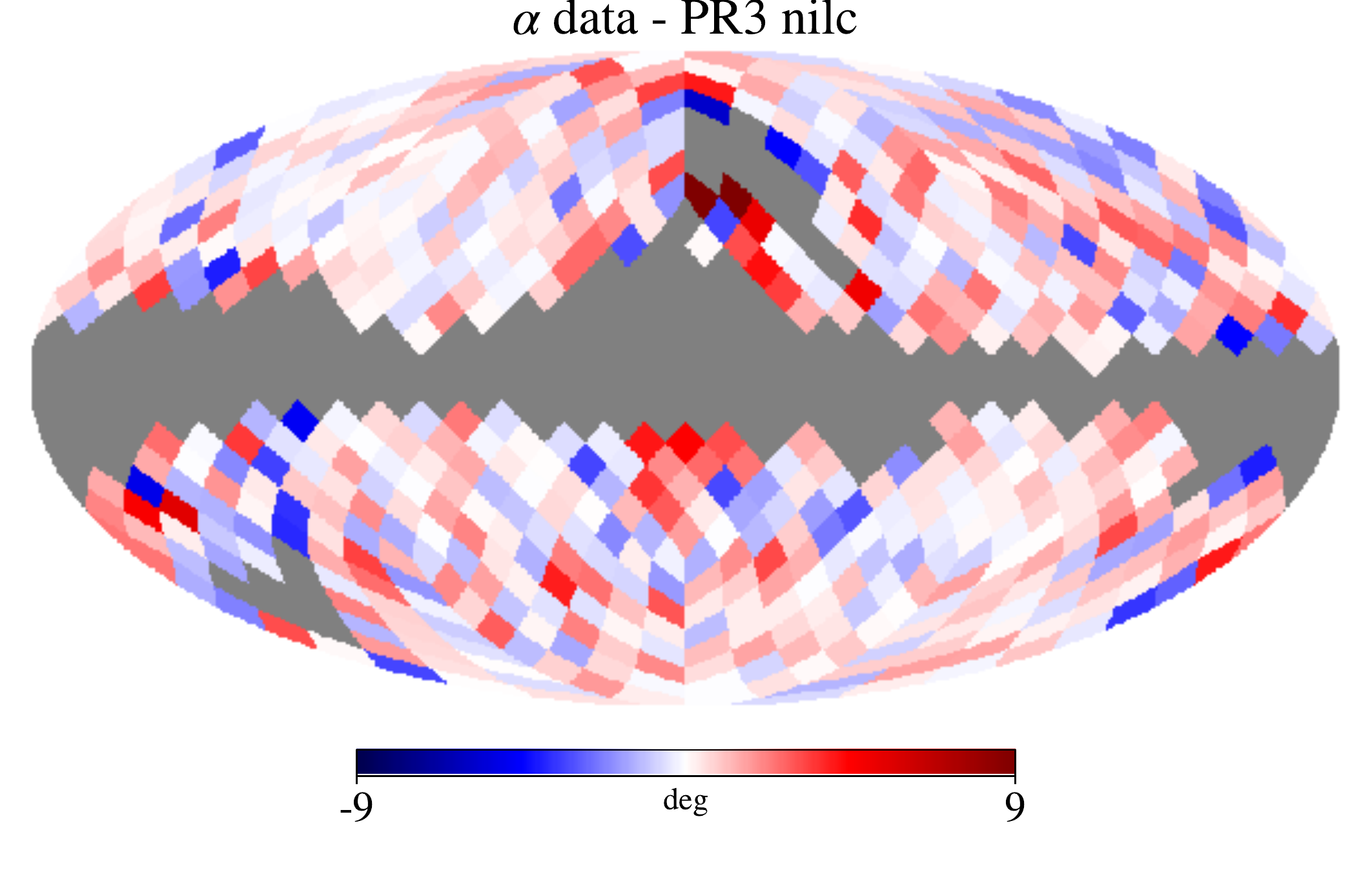}
\hfill
\includegraphics[width=.49\textwidth]{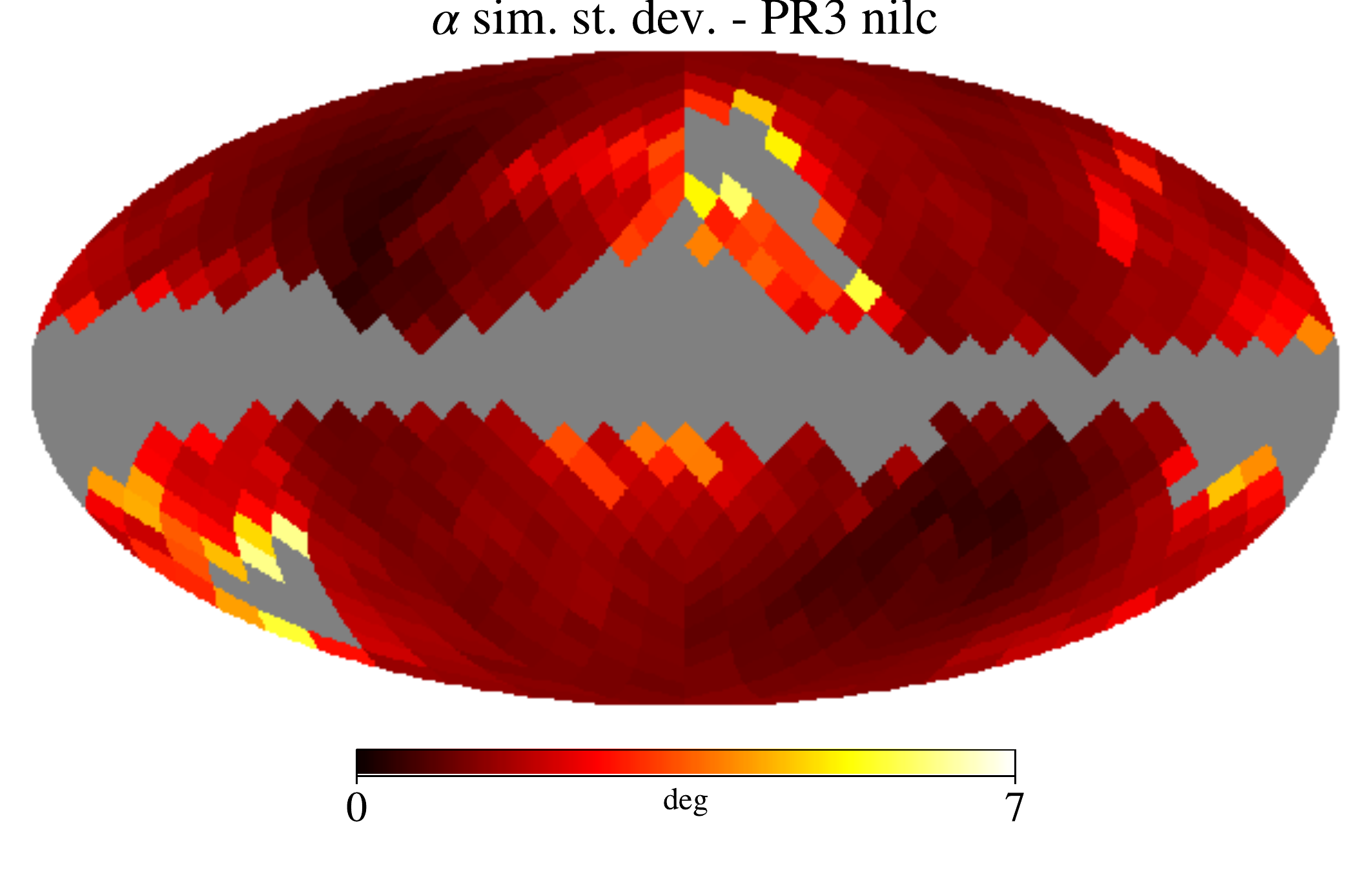}
\includegraphics[width=.49\textwidth]{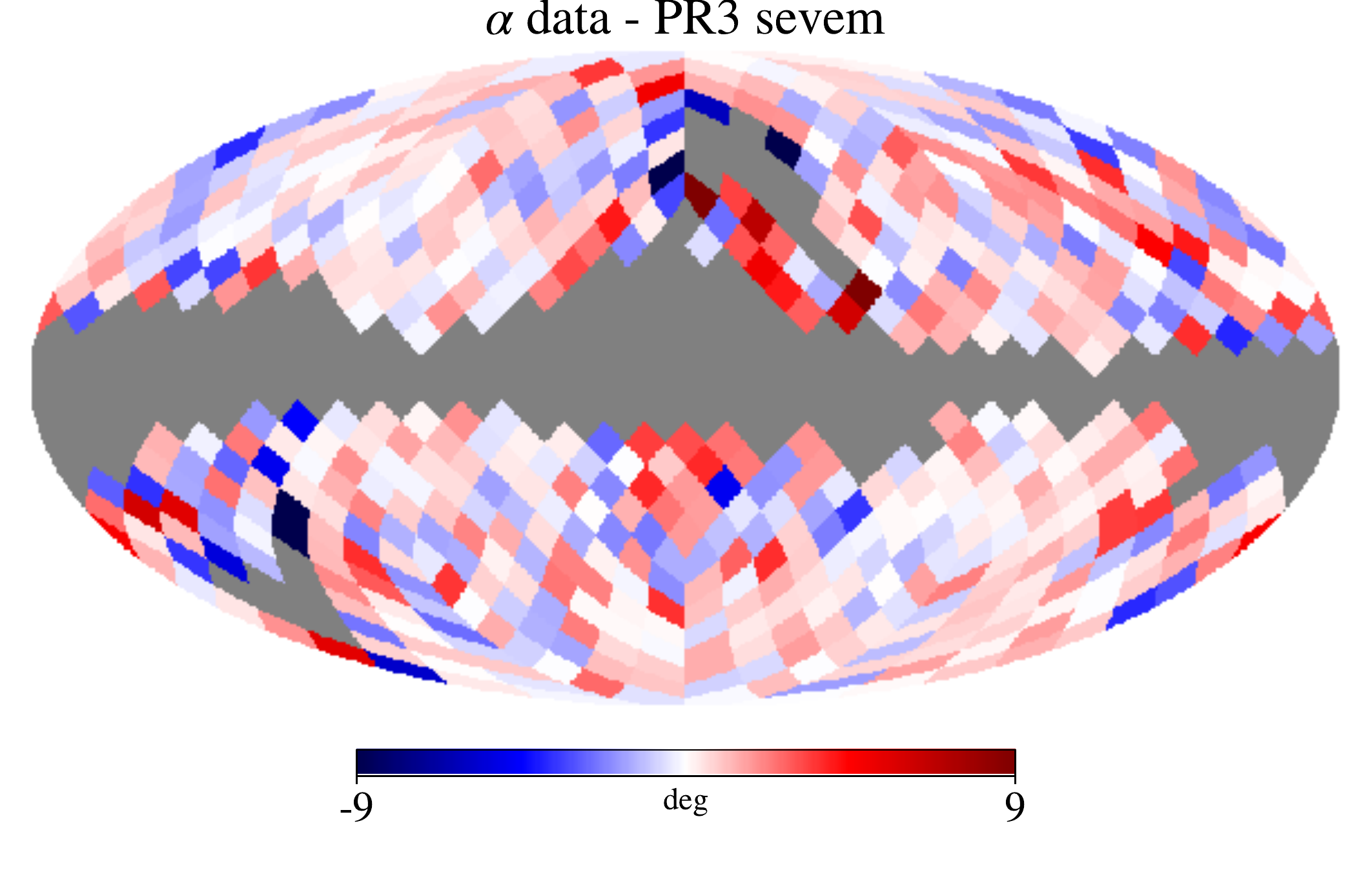}
\hfill
\includegraphics[width=.49\textwidth]{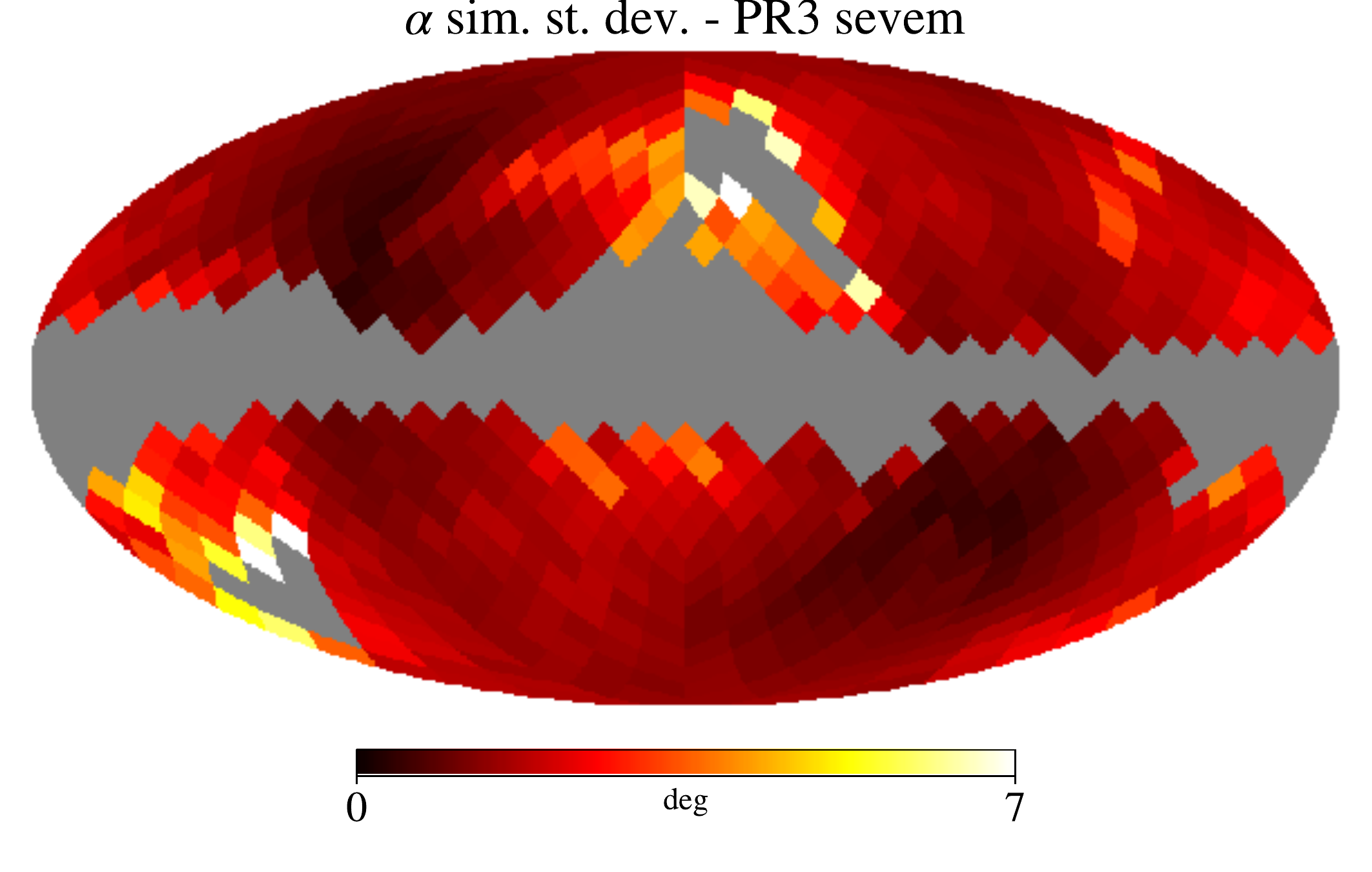}
\includegraphics[width=.49\textwidth]{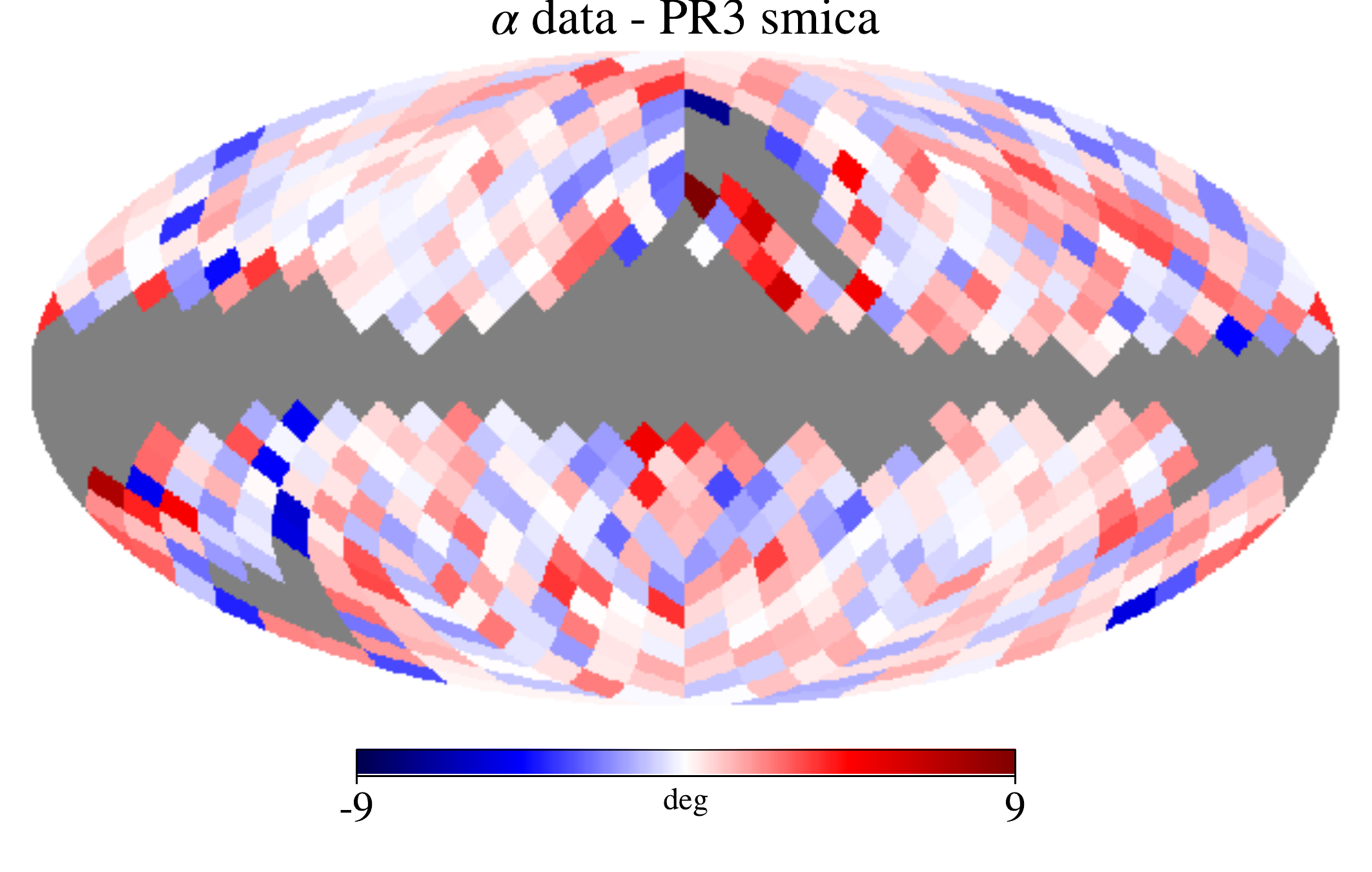}
\hfill
\includegraphics[width=.49\textwidth]{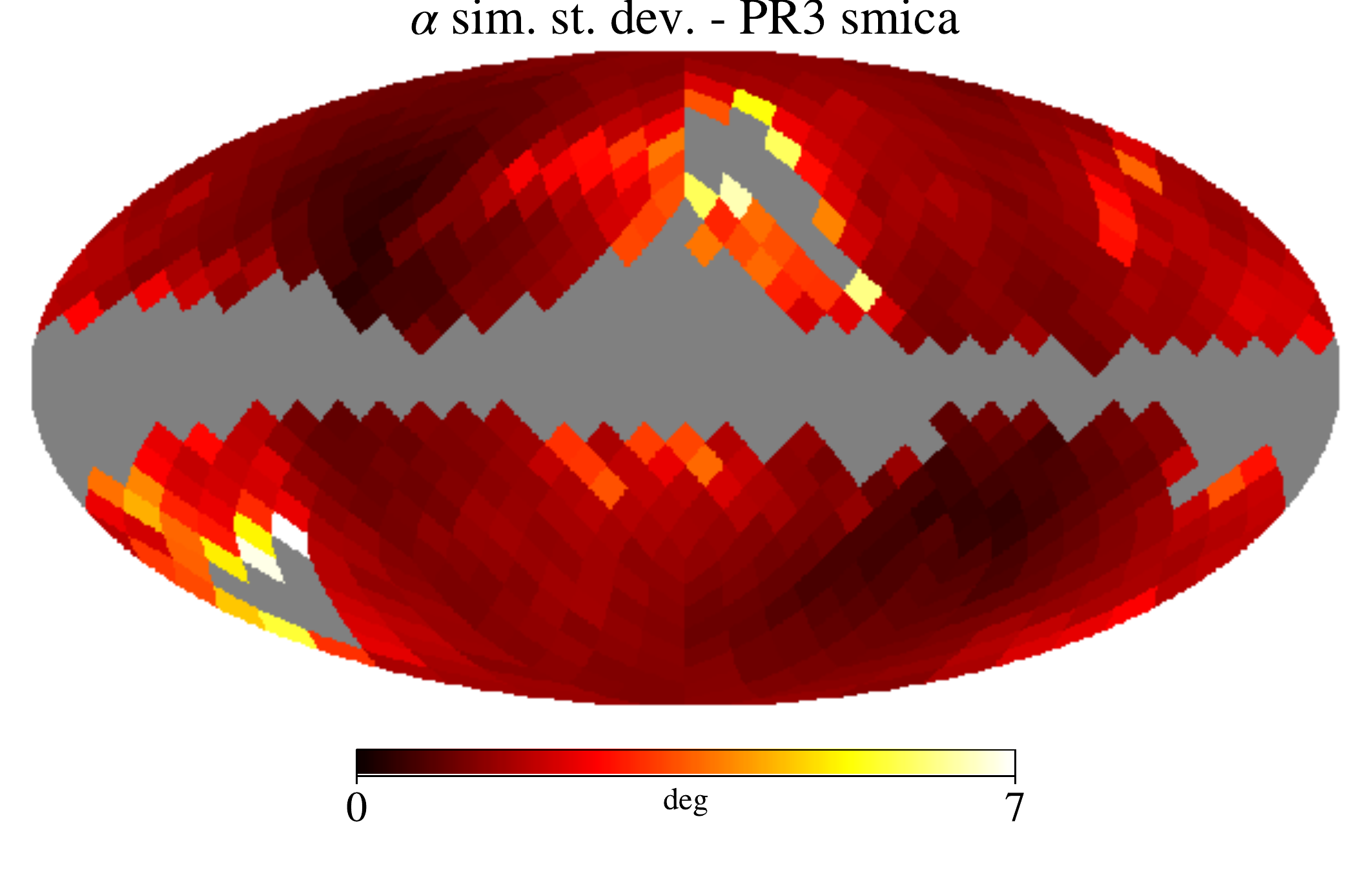}
\caption{Cosmic Birefringence angle maps obtained from the PR3 polarization maps for the \commander, \nilc, \sevem\ and \smica\ component separation methods (from top to bottom). Left: data. Right: standard deviation of simulations.\label{fig:beta_maps_PR3}}
\end{figure}
\begin{figure}[h!]
\centering
\includegraphics[width=.49\textwidth]{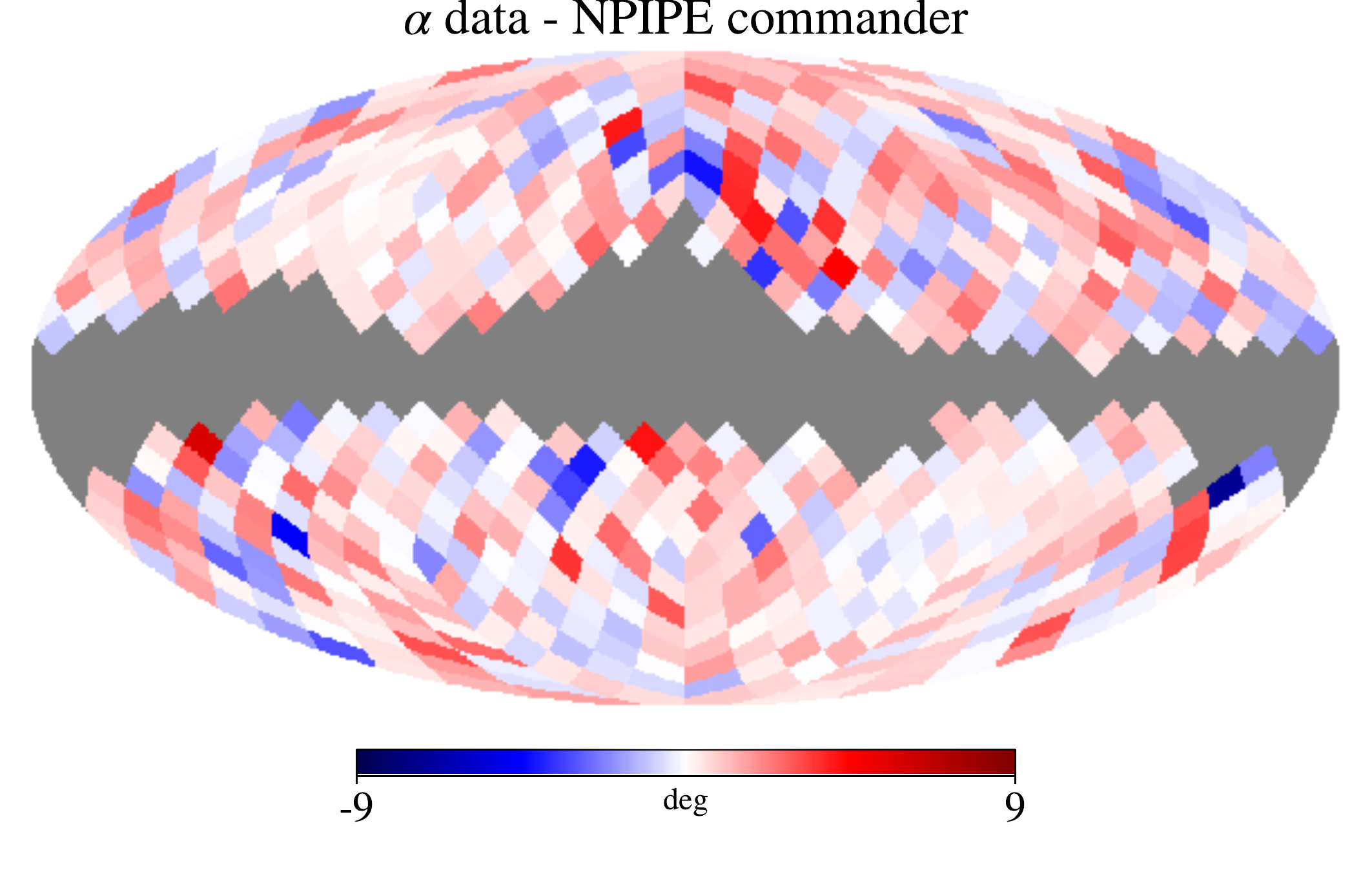}
\hfill
\includegraphics[width=.49\textwidth]{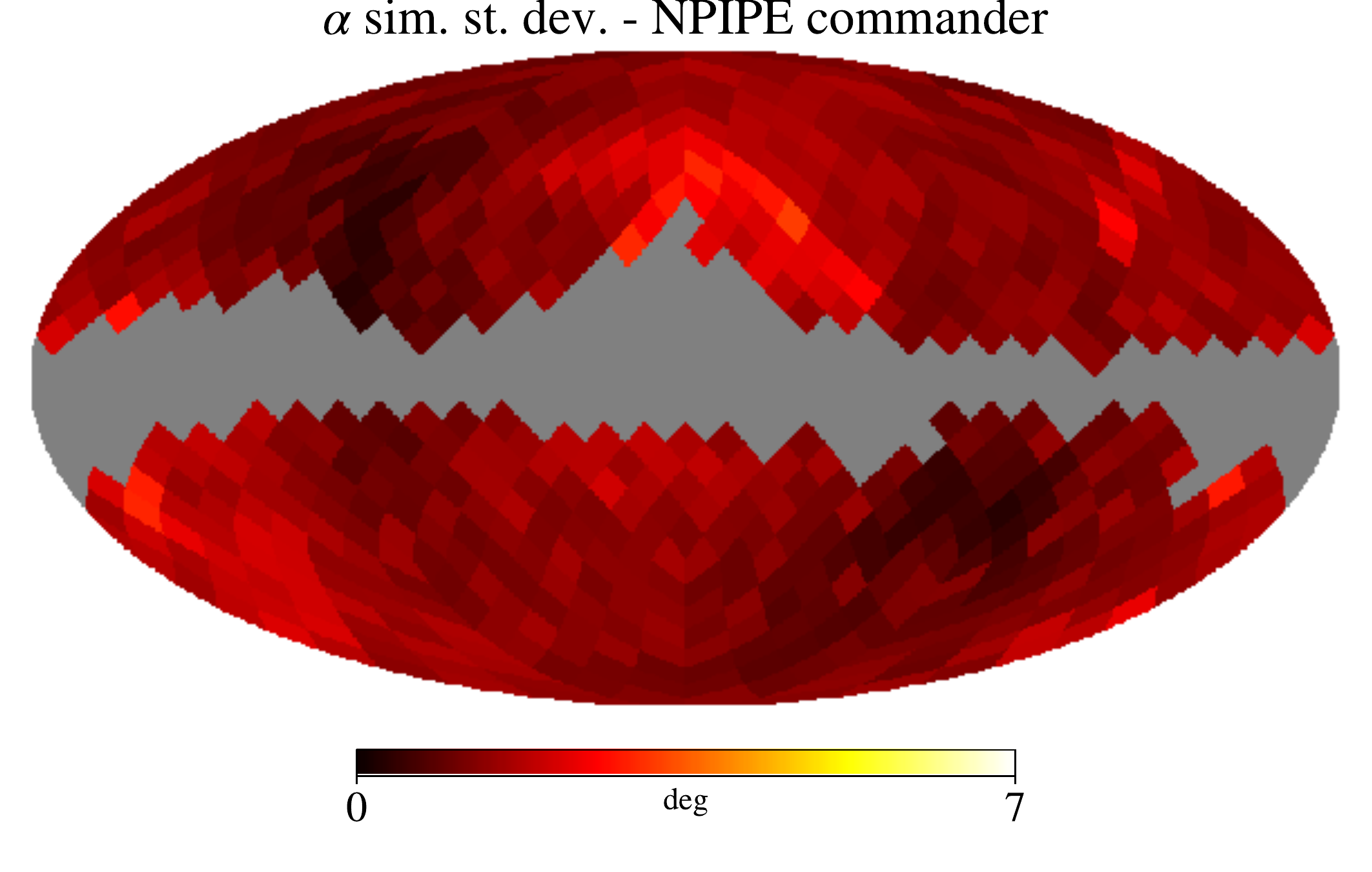}
\includegraphics[width=.49\textwidth]{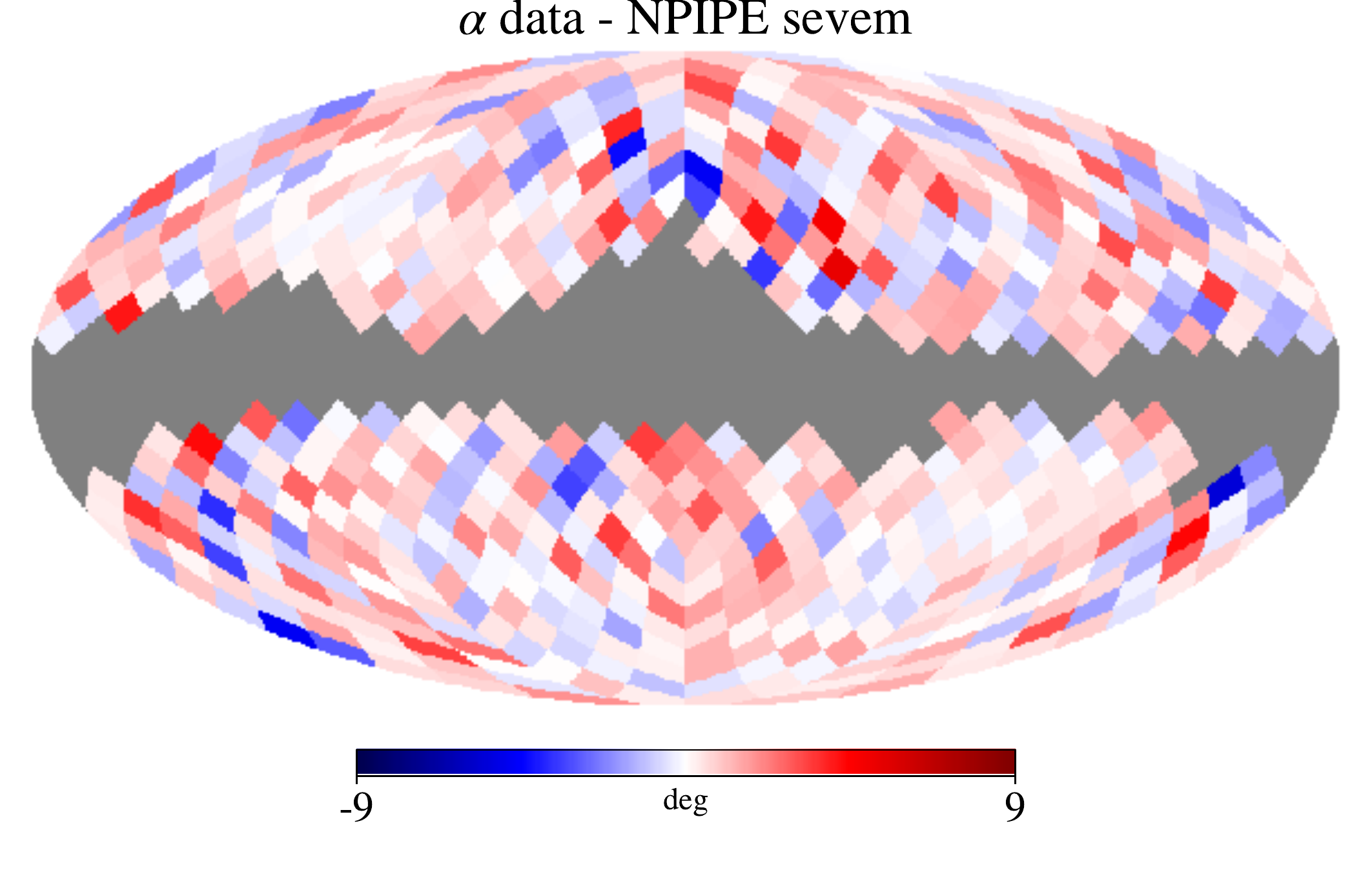}
\hfill
\includegraphics[width=.49\textwidth]{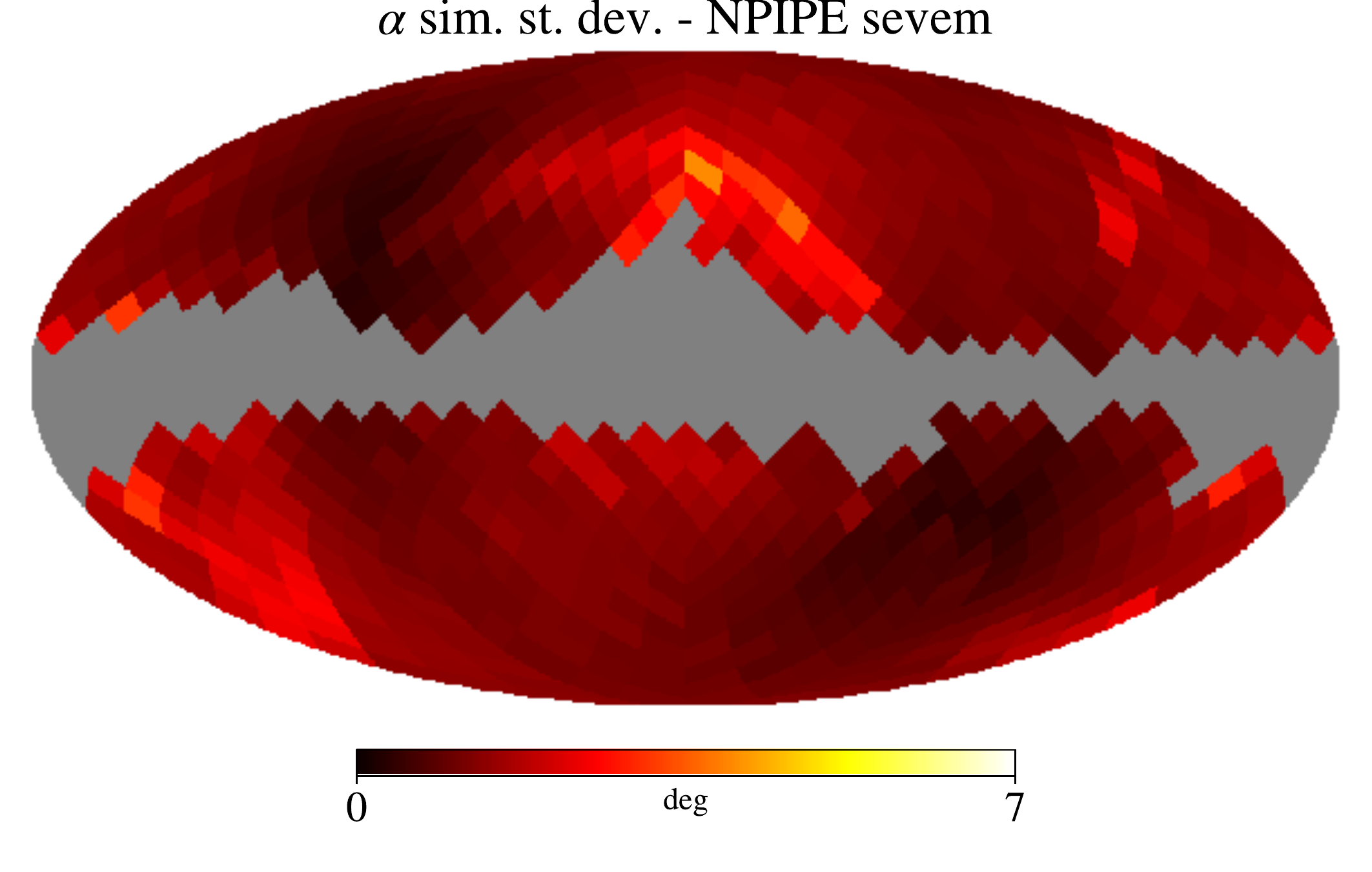}
\caption{Cosmic Birefringence angle maps obtained from the NPIPE polarization maps for the \commander\ and \sevem\ component separation methods (from top to bottom). Left: data. Right: standard deviation of simulations.\label{fig:beta_maps_NPIPE}}
\end{figure}

In Figure~\ref{fig:beta_maps_PR3} we show the CB maps (left panels) extracted from CMB data, and the associated error maps computed from simulations (right panels) for the PR3 dataset. In Figure~\ref{fig:beta_maps_NPIPE} the same quantities are displayed for NPIPE.
By visually comparing PR3 CB maps, we can recognize similar patterns for all the four component separation methods. The same applies to the NPIPE dataset. The errors associated to NPIPE are generally lower with respect to the PR3 ones. This is simply due to the fact that the NPIPE CMB maps are slightly less noisy than the PR3 ones~\cite{planck2020-LVII}. Furthermore some NPIPE low resolution patches contain a larger number of observed pixels, with respect to the PR3 dataset, whose mask contains also the contribution of half mission missing pixels (see top panels of Fig.~\ref{fig:mask}). Finally, the darker regions in error maps, which represent patches where the error on the CB angle is lower, are well correlated with the \Planck\ scanning strategy \cite{planck2013-p03f}, that observes more deeply the ecliptic poles.\par
\begin{table}[h]
    \centering
    \begin{tabular}{c|c}\hline
       \textbf{case}  & $\bm{\alpha\ \mathrm{[deg]}}$ \\\hline
        PR3 \commander\ & $\mathrm{0.27 \pm 0.05\ (stat) \pm 0.28\ (syst)}$ \\
        PR3 \nilc\      & $\mathrm{0.26 \pm 0.05\ (stat) \pm 0.28\ (syst)}$ \\
        PR3 \sevem\     & $\mathrm{0.27 \pm 0.05\ (stat) \pm 0.28\ (syst)}$ \\
        PR3 \smica\     & $\mathrm{0.24 \pm 0.05\ (stat) \pm 0.28\ (syst)}$ \\
        NPIPE \commander\ & $\mathrm{0.33 \pm 0.04\ (stat) \pm 0.28\ (syst)}$ \\
        NPIPE \sevem\ & $\mathrm{0.33 \pm 0.04\ (stat) \pm 0.28\ (syst)}$ \\\hdashline
        \cite{MinamiKomatsu2020} (PR3) & $\mathrm{0.35 \pm 0.14\ (stat)}$ \\
        \cite{Diego-Palazuelos:2022dsq} (NPIPE) &        $\mathrm{0.30 \pm 0.11\ (stat)}$ \\
        \cite{Eskilt:2022cff} (NPIPE + WMAP) &  $\mathrm{0.30^{+0.094}_{-0.091}\ (stat)}$ \\\hline
    \end{tabular}
    \caption{Monopole of the data CB angle maps produced in this work. The values found in~\cite{MinamiKomatsu2020}, \cite{Diego-Palazuelos:2022dsq}, and~\cite{Eskilt:2022cff} are also reported for reference.}
    \label{tab:monopole}
\end{table}\par
In Table~\ref{tab:monopole} we report the monopoles of the CB maps. They are obtained as a weighted average of the data maps with weights corresponding to the inverse variance of the simulations. All the monopole values are consistent with zero provided the systematic error is taken into account. Moreover they are well compatible, within the statistical uncertainty, across the different component separation methods. The statistical error contribution is computed from the standard deviation of the simulations, while the systematic error is due to the uncertainty in the orientation of \planck’s polarization-sensitive bolometers \cite{rosset2010} and is taken from Ref.~\cite{planck2014-a23}. Also here we note that, due to the lower level of the instrumental noise~\cite{planck2020-LVII} and to the larger number of non-masked pixels, the NPIPE data provide smaller statistical errors. The monopole values found in this work agree with the ones reported in literature, see e.g.~\cite{planck2014-a23,Gruppuso:2020kfy,Eskilt:2022wav,MinamiKomatsu2020,Diego-Palazuelos:2022dsq,Eskilt:2022cff}.\par
\begin{figure}[h!]
\centering
\includegraphics[width=.49\textwidth]{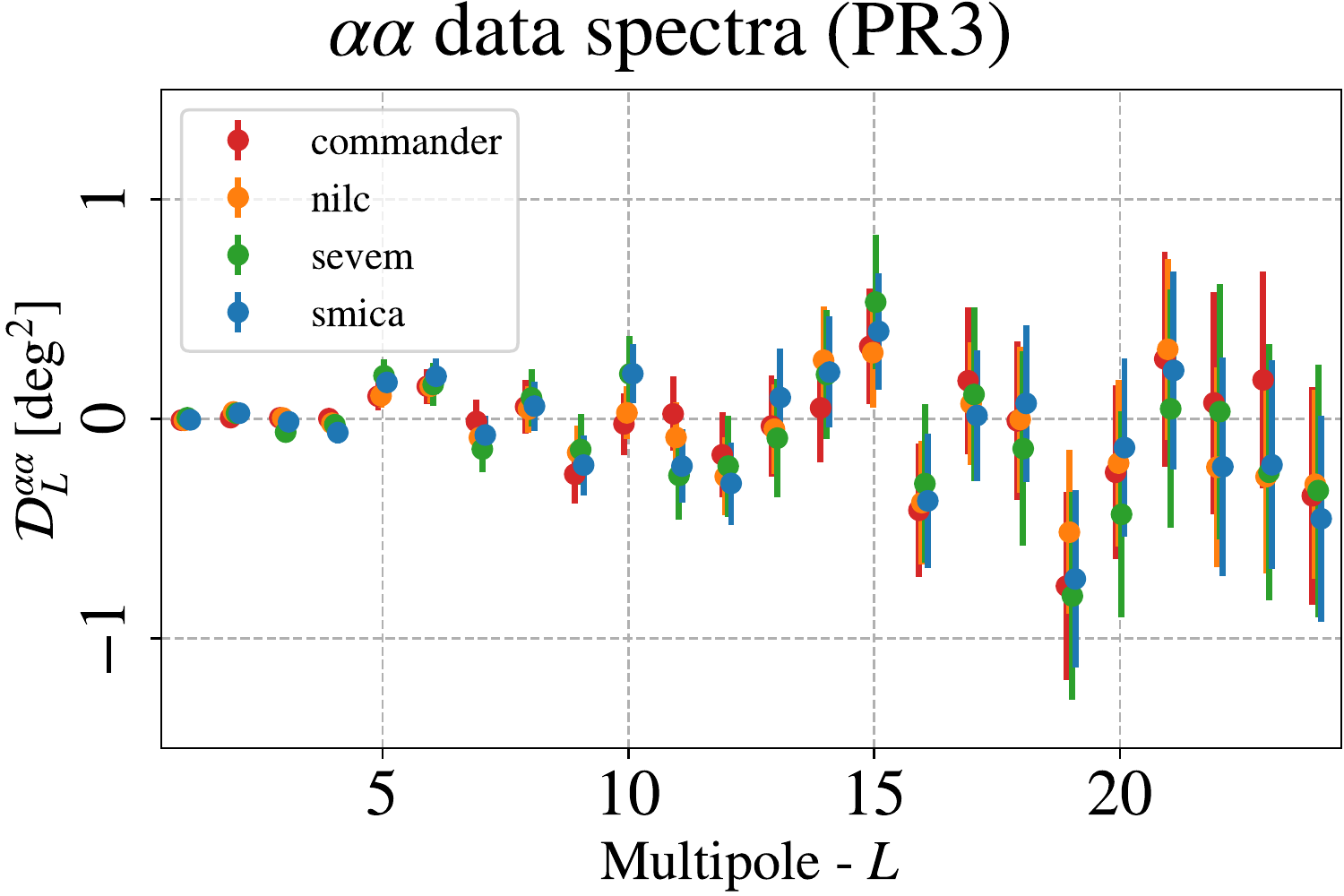}
\hfill
\includegraphics[width=.49\textwidth]{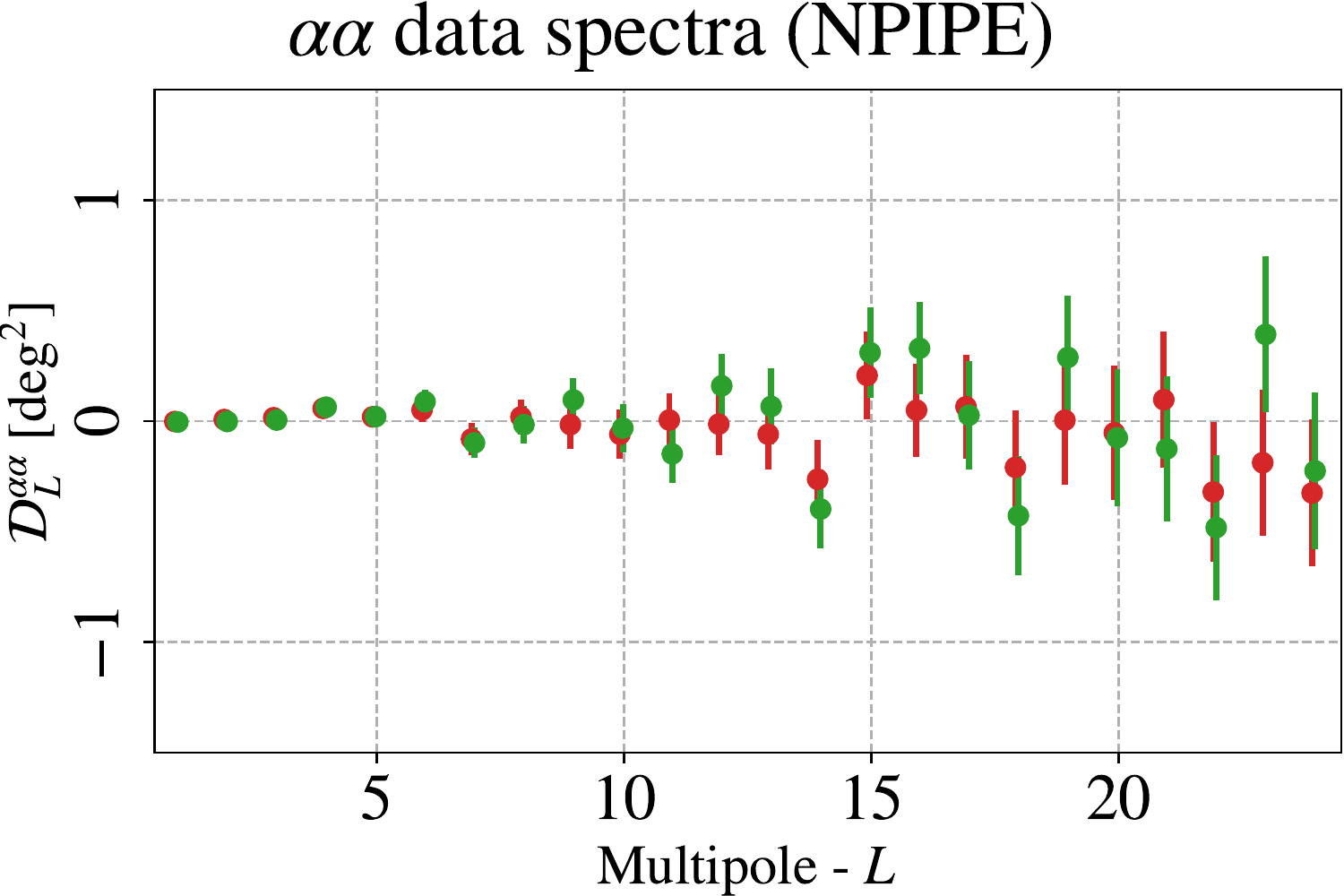}
\includegraphics[width=.49\textwidth]{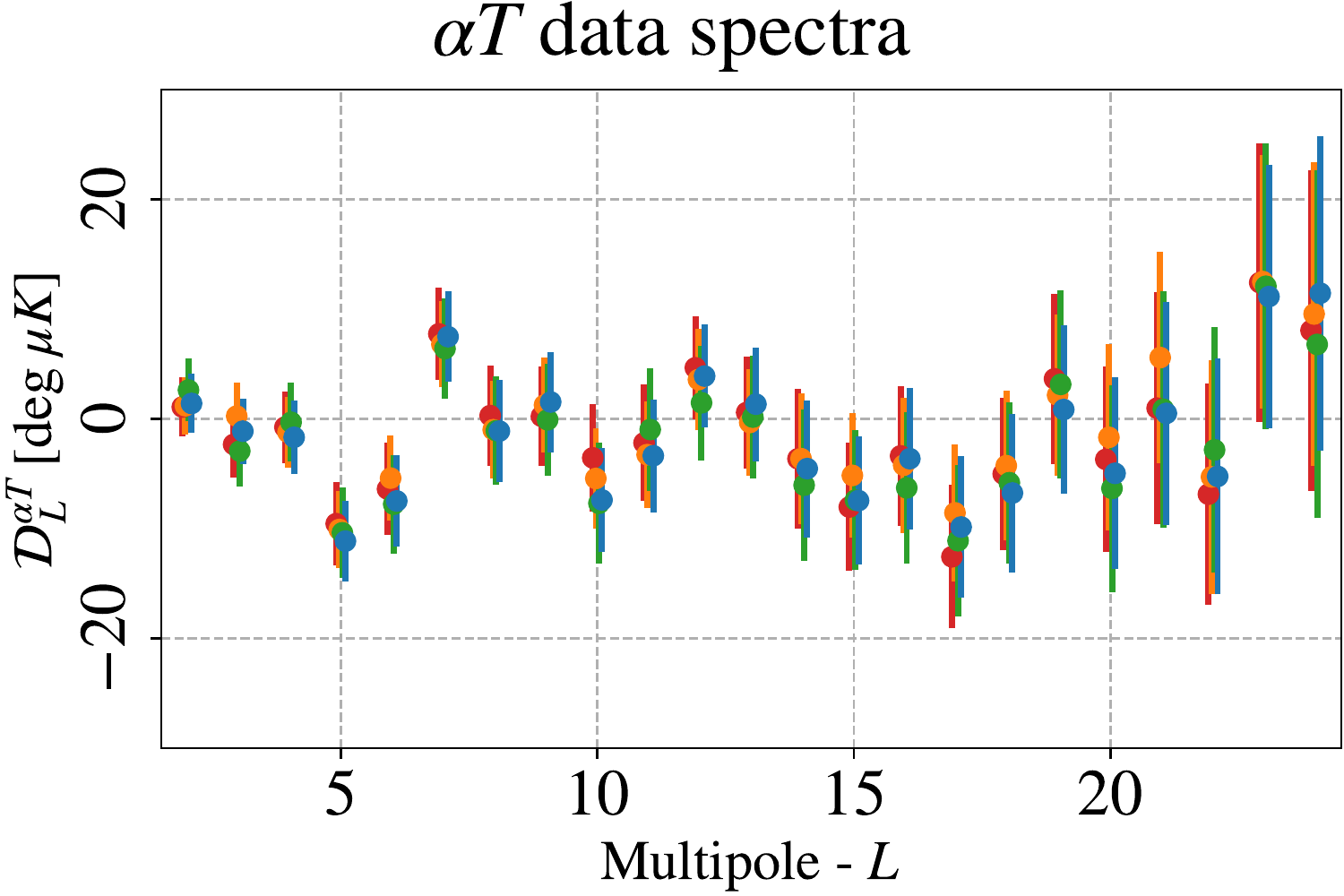}\\
\includegraphics[width=.49\textwidth]{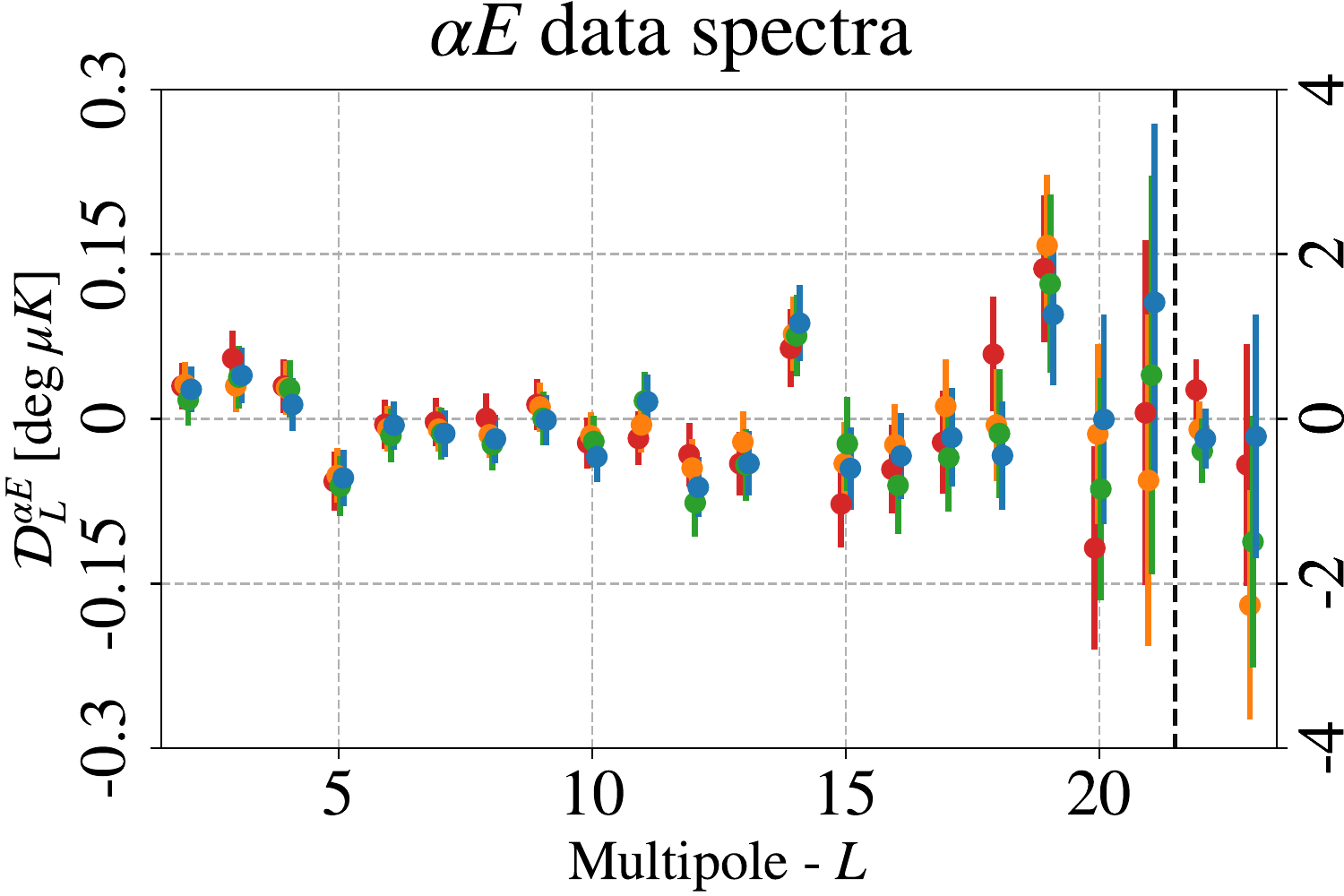}
\hfill
\includegraphics[width=.49\textwidth]{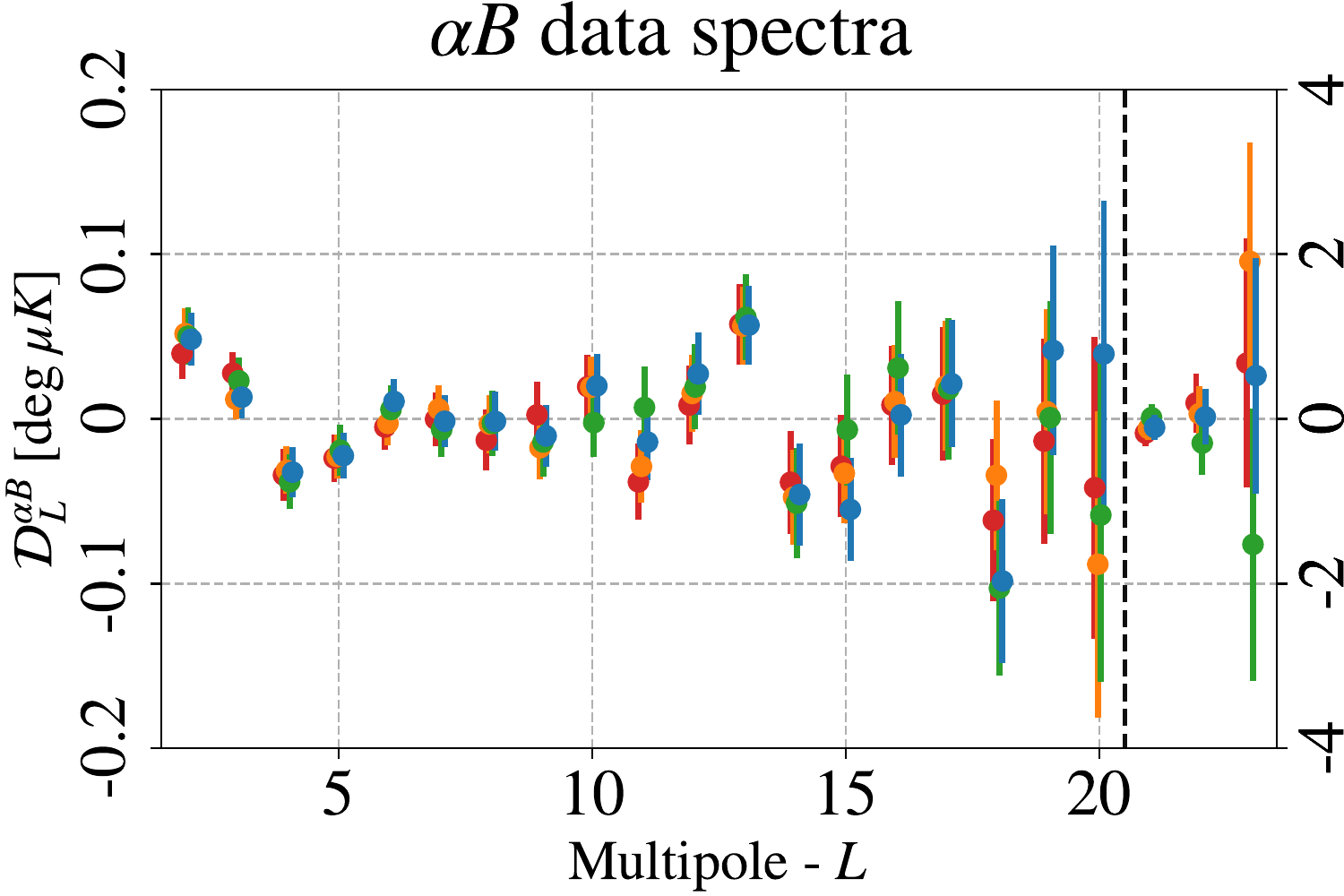}
\caption{Auto-correlation spectra of the data CB angle maps and their cross-correlation with CMB. The errors are the standard deviations of the simulations. The vertical dashed lines highlight a change in the vertical axis range. Top left panel: PR3 $\alpha\alpha$. Top right panel: NPIPE $\alpha\alpha$. Central panel: PR3 $\alpha T$. Bottom left panel: PR3 $\alpha E$. Bottom right panel: PR3 $\alpha B$.\label{fig:results_data_spectra}}
\end{figure}
\begin{figure}[h]
\centering
\includegraphics[width=.49\textwidth]{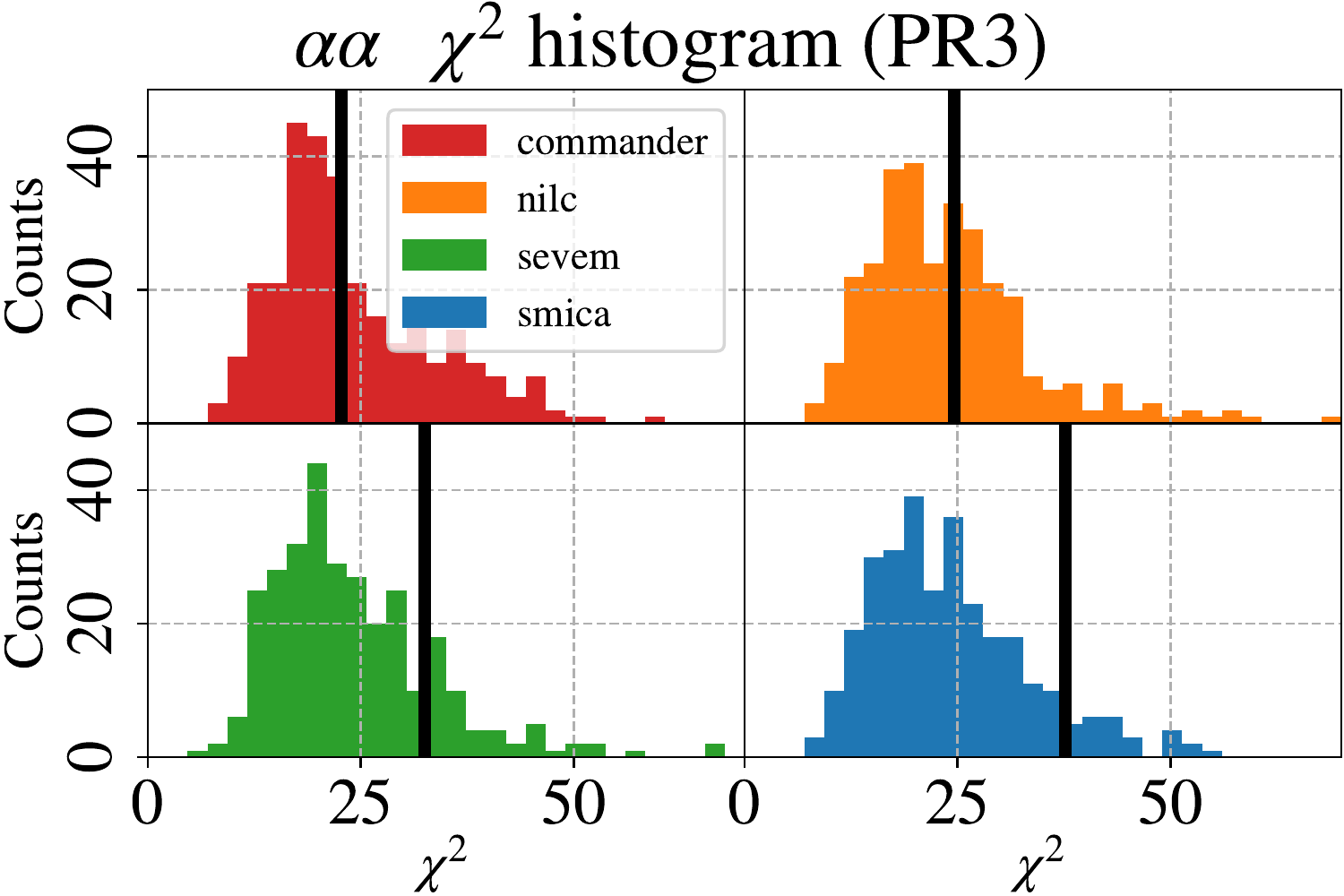}
\hfill
\includegraphics[width=.49\textwidth]{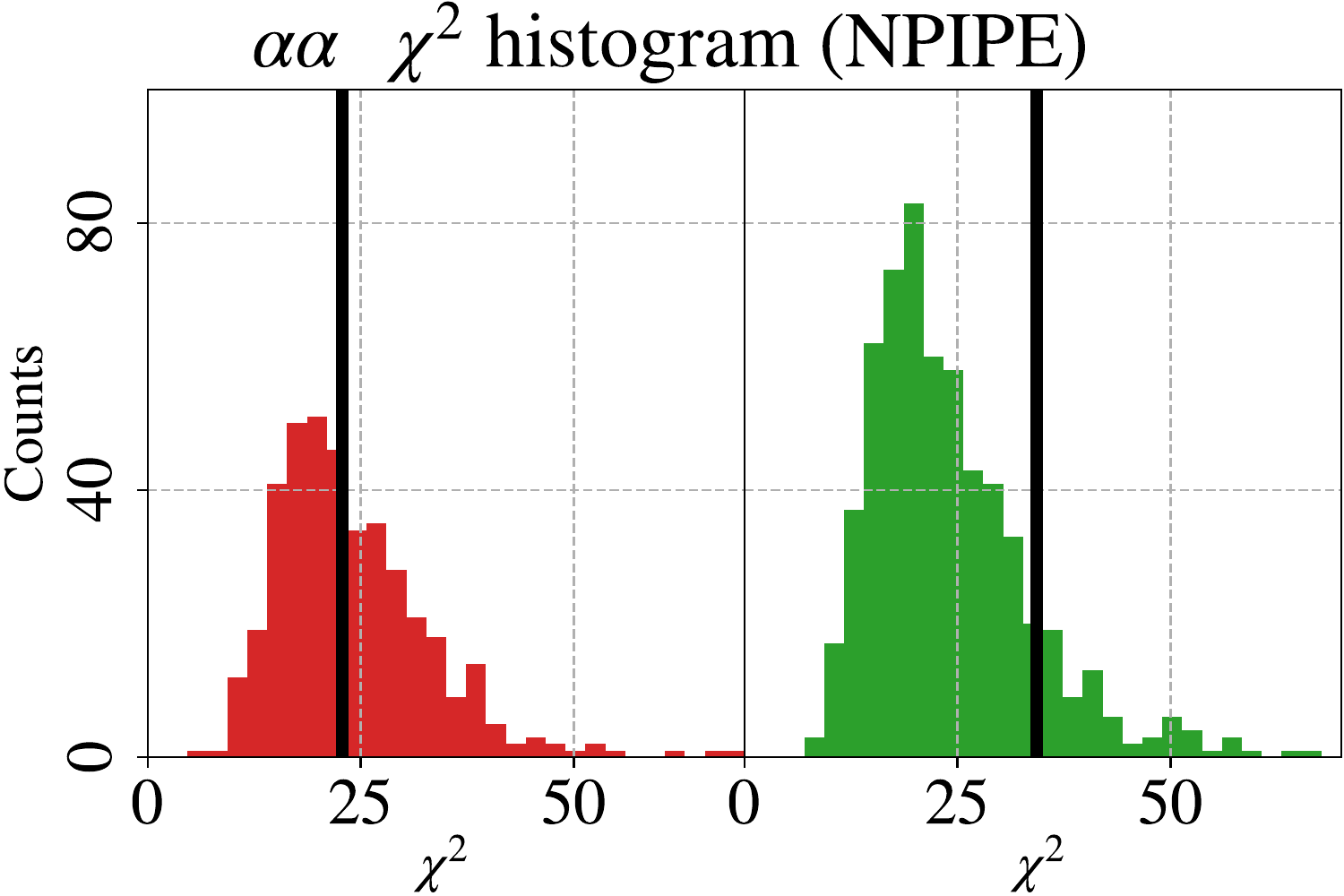}
\includegraphics[width=.49\textwidth]{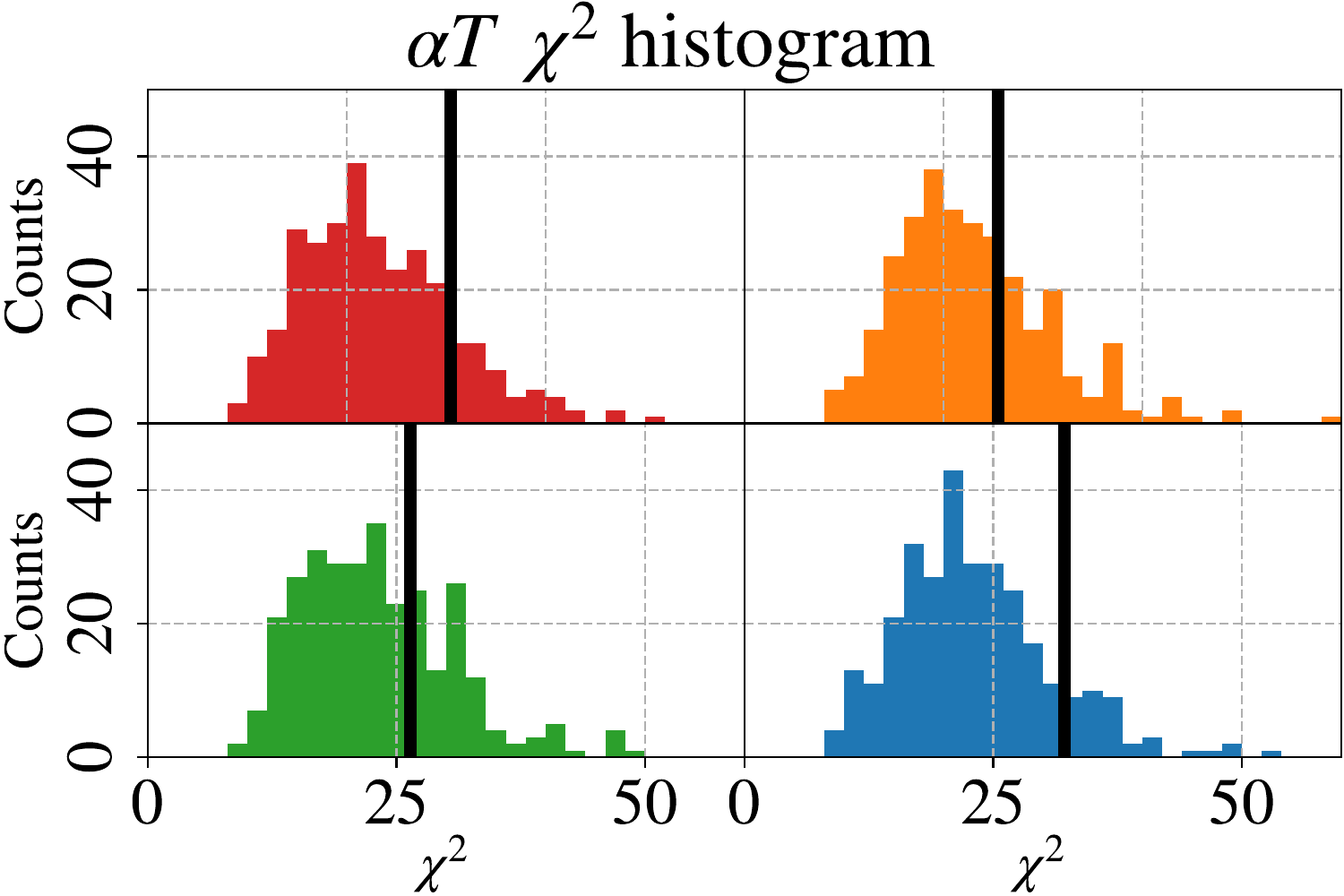}\\
\includegraphics[width=.49\textwidth]{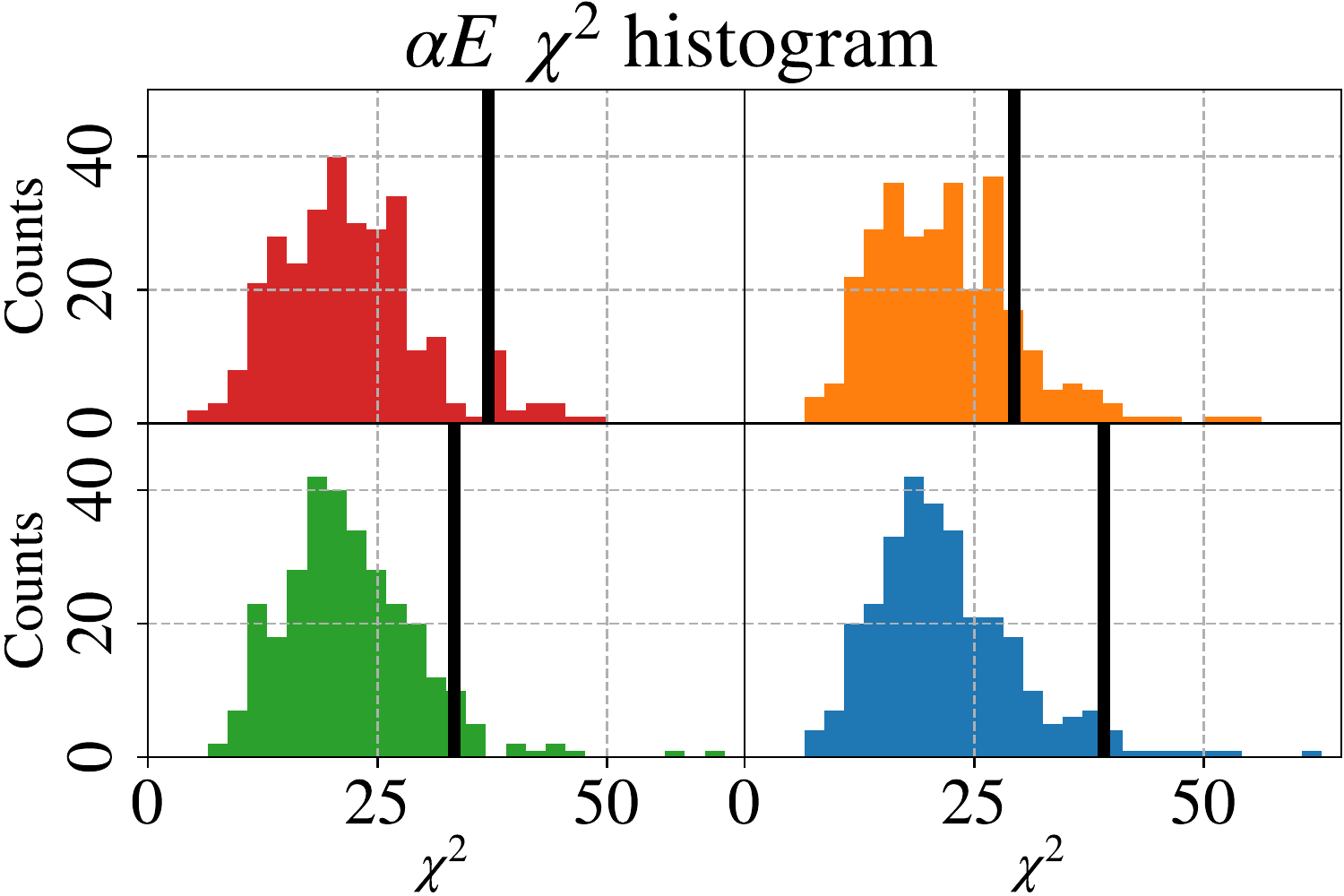}
\hfill
\includegraphics[width=.49\textwidth]{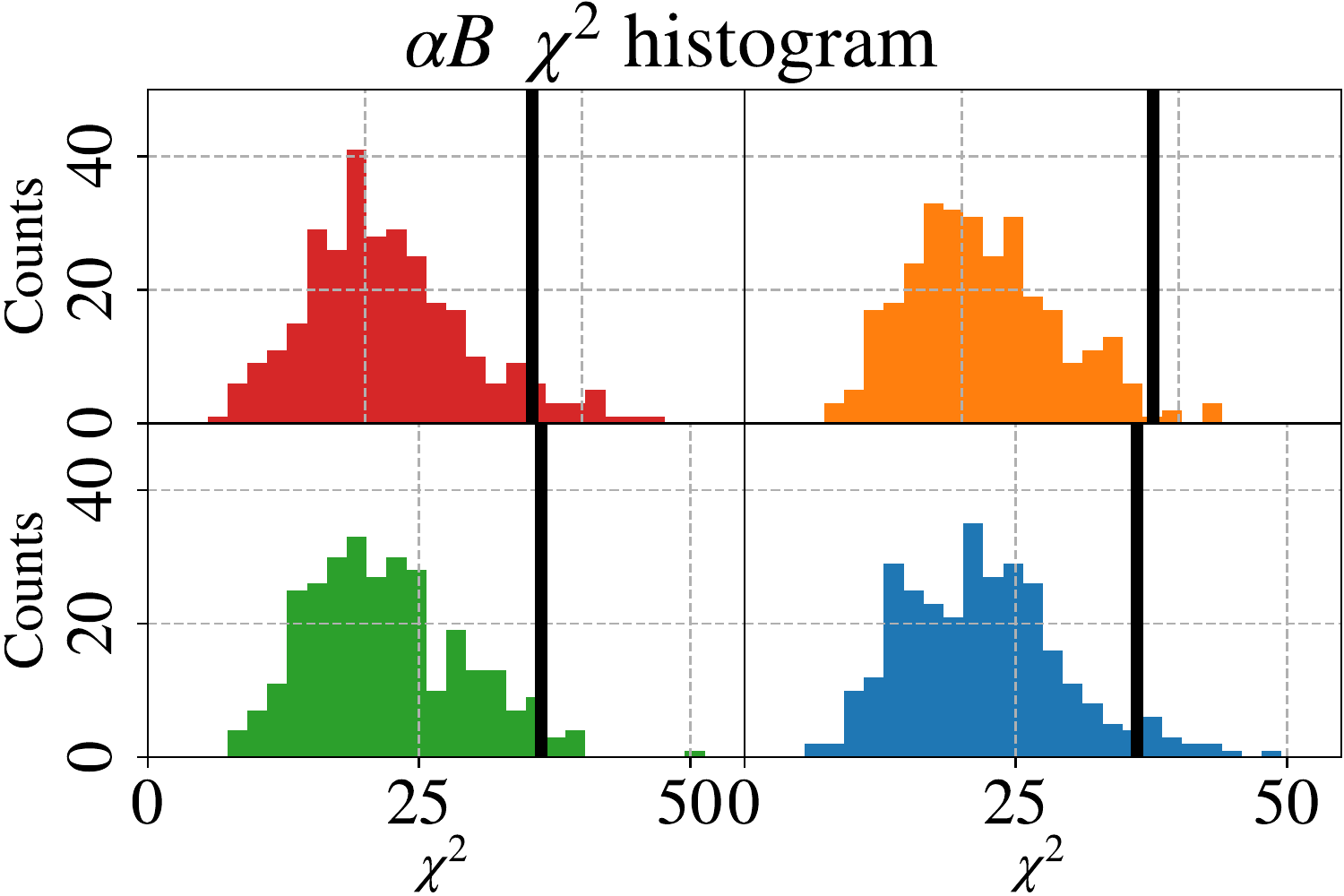}
\caption{Harmonic $\chi^2$ plots (histogram for simulations, black line for data). Top left panel: PR3 $\alpha\alpha$. Top right panel: NPIPE $\alpha\alpha$. Central panel: PR3 $\alpha T$. Bottom left panel: PR3 $\alpha E$. Bottom right panel: PR3 $\alpha B$.\label{fig:results_histo}}
\end{figure}

In Figure~\ref{fig:results_data_spectra} we show the auto-correlation of the CB maps and their cross-correlation with CMB\footnote{All the spectra shown in this paper are given in terms of bandpowers, i.e. $D_L^{\alpha X} = L(L+1)C_L^{\alpha X}/2\pi$, with $X=\alpha,T,E,B$.}. As expected the scatter of the NPIPE $\alpha\alpha$ spectra around zero and the associated errors are slightly lower with respect to PR3. 
Across the spectra, we find a general good compatibility with the null effect. 
This is assessed by comparing the harmonic $\chi^2$ of data and simulations, reported in Fig.~\ref{fig:results_histo}.
\par
The data $\chi^2$ for all the spectra lies within the histograms, suggesting that the data are compatible with the simulations, hence with the null effect. However, we note that for the $\alpha E$ and $\alpha B$ cases the $\chi^2$ of the data are closer to the right tail of the corresponding histograms, with the following Probability To Exceed (PTE): 0.9--10.6\% for $\alpha E$ and 1.4--2.5\% for $\alpha B$ (see also Tab.~\ref{tab:pte} where all the PTEs are reported).
\begin{table}[h]
    \centering
    \begin{tabular}{c|cccc}\hline
       \textbf{case} & \commander & \nilc & \sevem & \smica \\\hline
       $\alpha\alpha$ PR3   & 47.6\% & 37.1\% & 9.0\%  & 2.8\% \\
       $\alpha\alpha$ NPIPE & 47.1\% & -      & 6.2\%  & -     \\
       $\alpha T$ PR3       & 10.9\% & 27.4\% & 23.6\% & 7.5\% \\
       $\alpha E$ PR3       & 1.6\%  & 10.6\% & 4.2\%  & 0.9\% \\
       $\alpha B$ PR3       & 2.5\%  & 1.4\%  & 2.1\%  & 2.1\% \\\hline
    \end{tabular}
    \caption{Probability To Exceed obtained from the harmonic $\chi^2$ reported in Fig.~\ref{fig:results_histo}.}
    \label{tab:pte}
\end{table}
Despite the overall good compatibility with the null effect, some single multipole fluctuations deserve a closer look. Considering the $\alpha T$ spectra, $L=5$ shows a PTE between 1\% and 2.7\% for the four component separation methods, i.e. $\sim2.5\sigma$ fluctuation. For the $\alpha E$ spectra, the multipoles $L=5, 12, 14, 19$ have the largest deviation from the null effect, if compared with the corresponding empirical distributions. These PTEs are below 1\% for $D_{19}^{\alpha E}$ \commander\ and $D_{12}^{\alpha E}$ \sevem, but still within $3\sigma$ fluctuation. The same happens for $D_2^{\alpha B}$ \nilc, \sevem\ and \smica, all with PTEs lower than 1\%, but within $3\sigma$. The PTE for $D_{13}^{\alpha B}$ is between 2\% and 2.7\%, which corresponds to $\sim 2.3 \sigma$ fluctuation. All the spectra obtained with the \smica\ component separation method are reported in Tab.~\ref{tab:smica_all_spectra} along with the statistical uncertainty at $1\sigma$ level for each multipole\footnote{The other CB spectra, along with the CB angle maps, are made publicly available at \href{https://github.com/marcobortolami/AnisotropicBirefringence_patches.git}{https://github.com/marcobortolami/AnisotropicBirefringence\_patches.git}.}. The averaged spectra of the simulations are shown in App.~\ref{app:results on simulations}. They all show a nice compatibility with the null-hypothesis, meaning that the residual systematic effects included in the simulations do not have a significant impact on the spectra considered.\par
\begin{figure}[h]
\centering
\includegraphics[width=.49\textwidth]{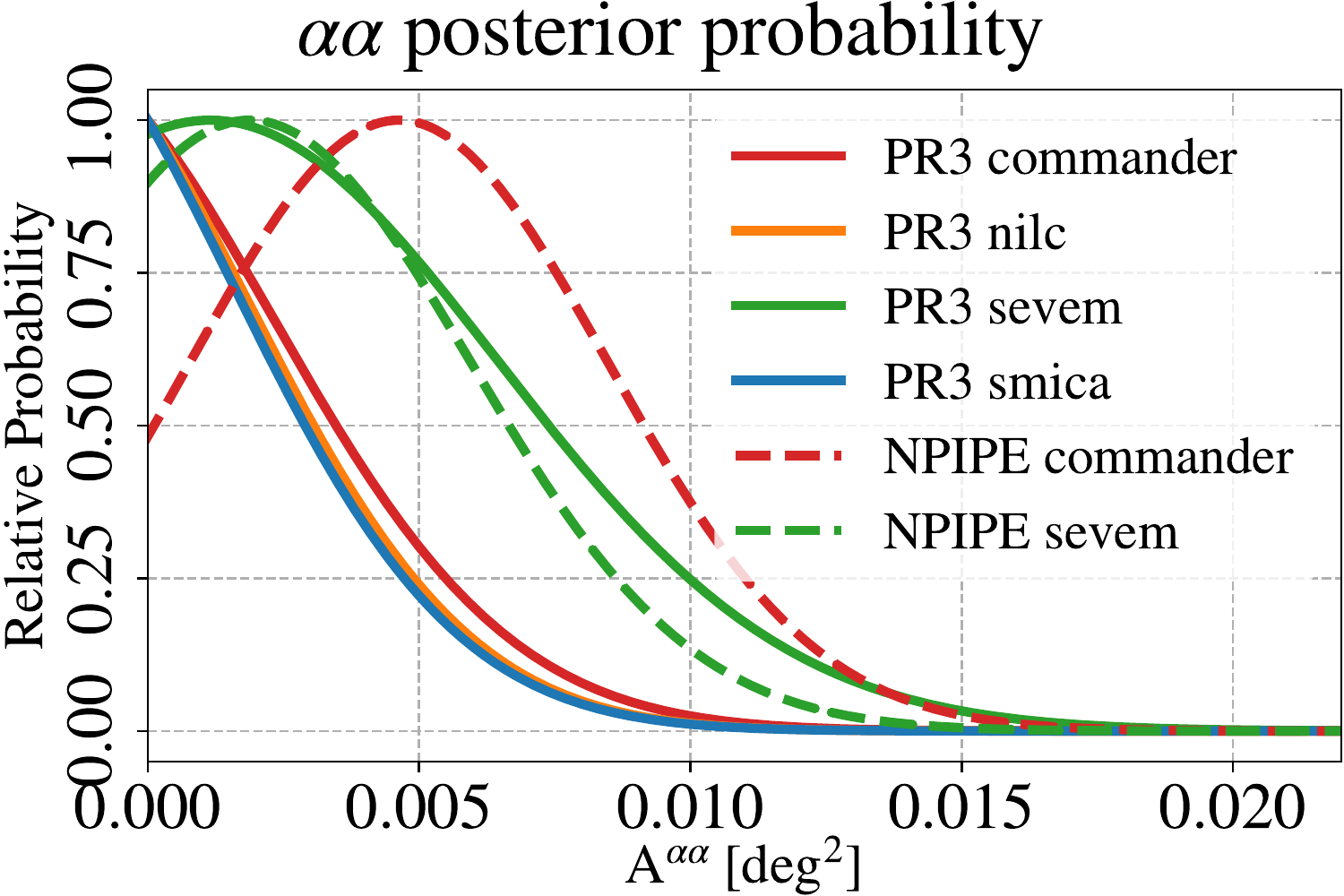}
\hfill
\includegraphics[width=.49\textwidth]{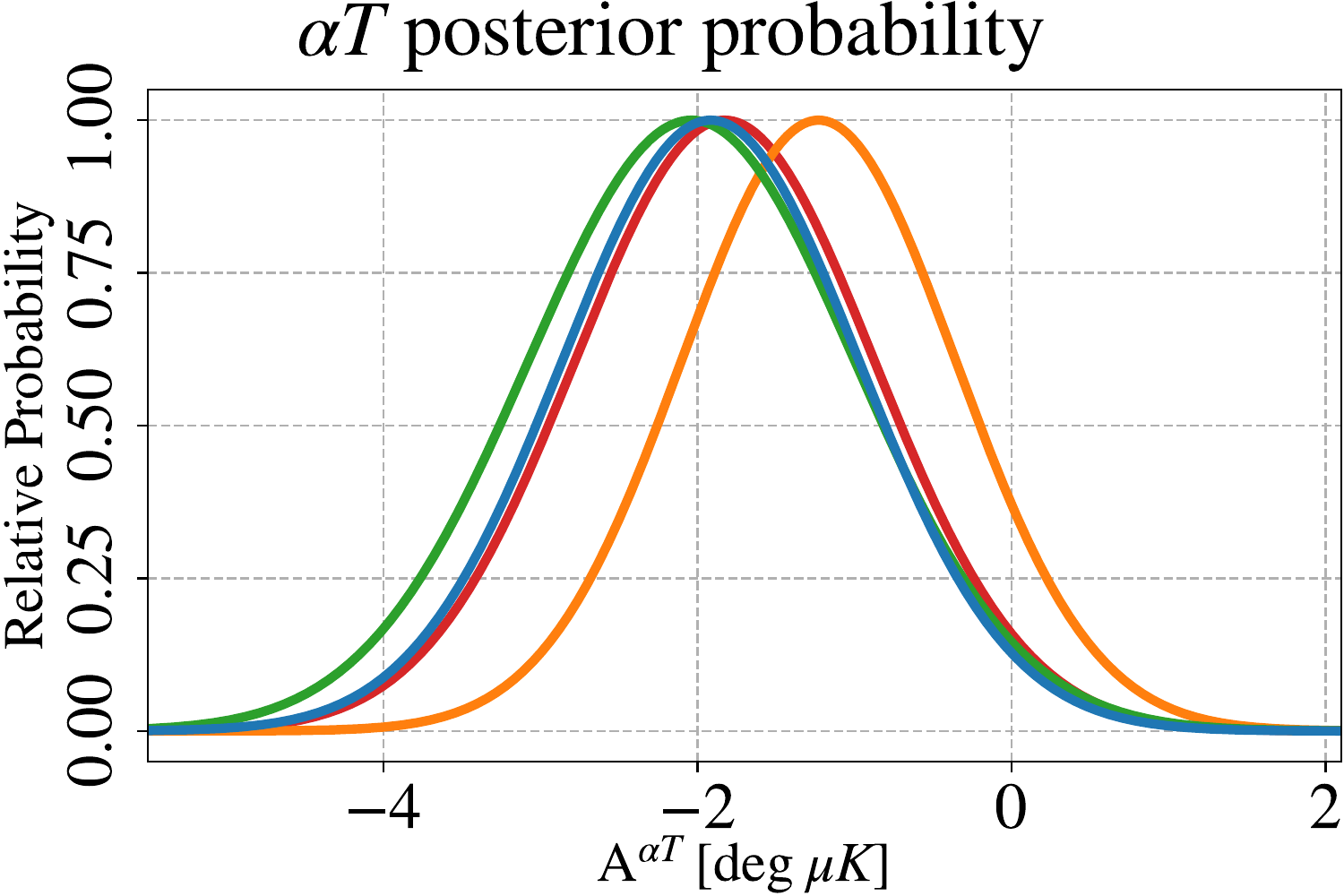}
\includegraphics[width=.49\textwidth]{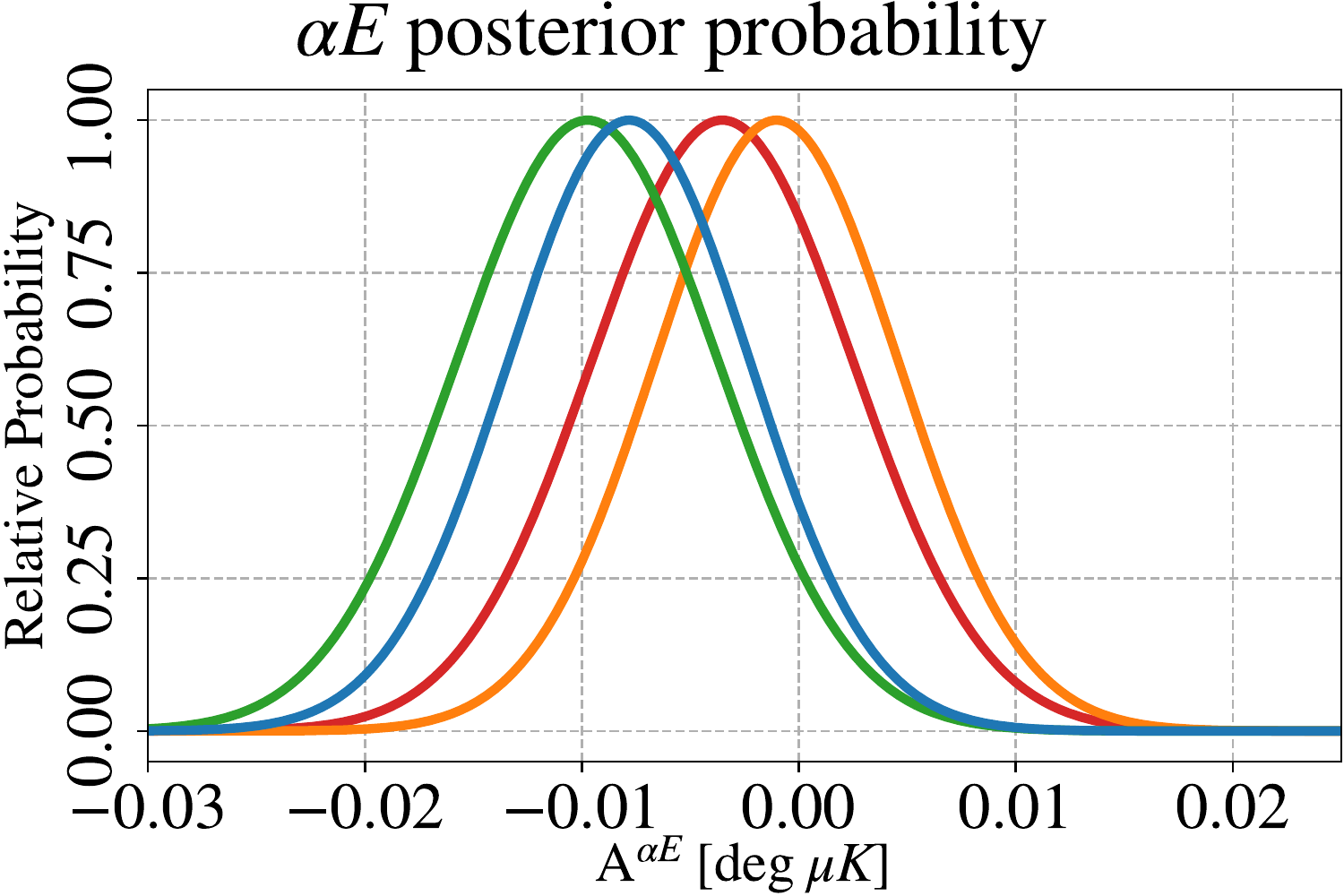}
\hfill
\includegraphics[width=.49\textwidth]{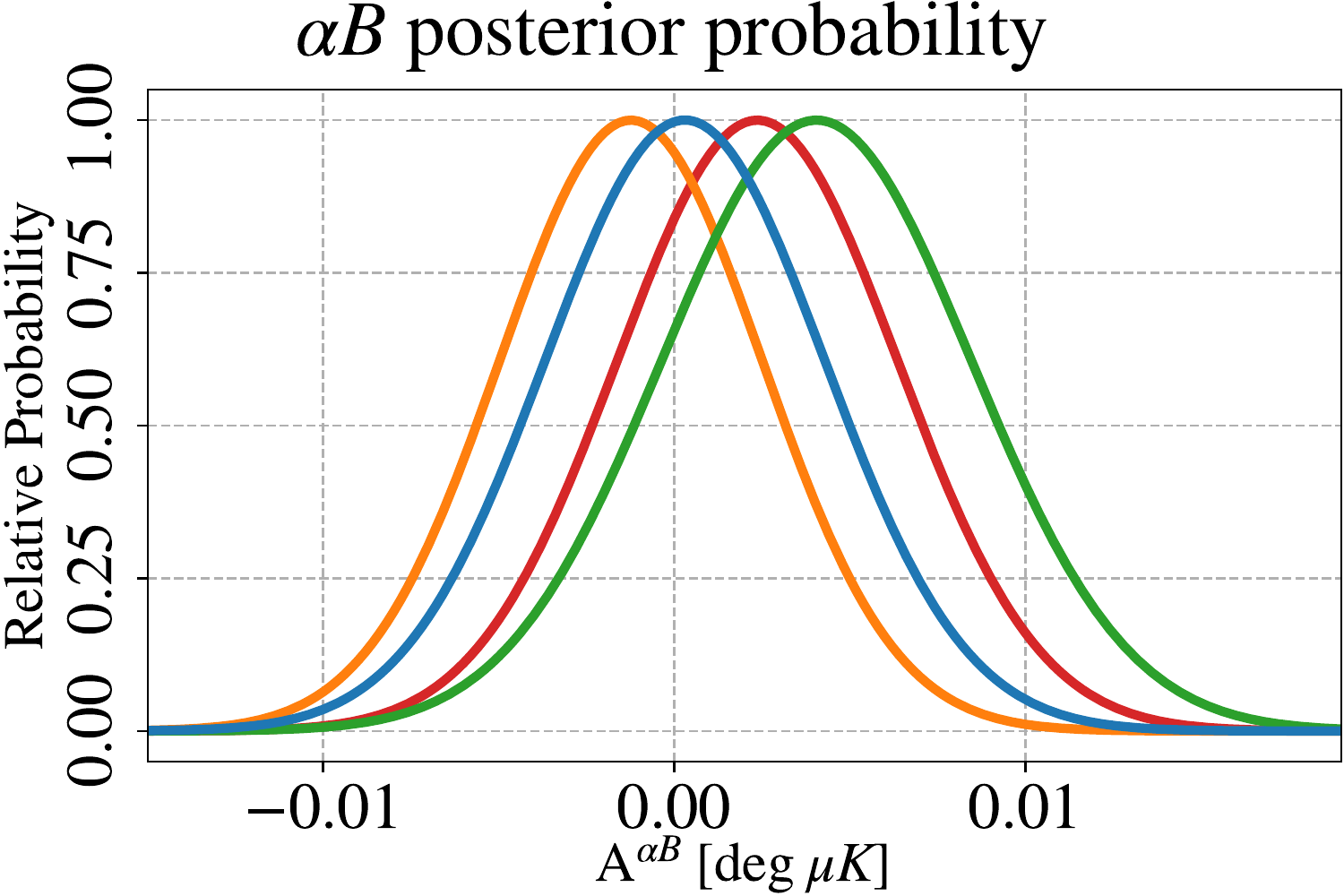}
\caption{Posteriors for the scale invariant spectrum fit. Top left panel: PR3 and NPIPE $\alpha\alpha$. Top right panel: PR3 $\alpha T$. Bottom left panel: PR3 $\alpha E$. Bottom right panel: PR3 $\alpha B$.\label{fig:results_posteriors}}
\end{figure}

\begin{table}[h]
    \centering
    \begin{tabular}{c|cccc}
    \hline
$\bm{L}$ & $\bm{D_L^{\alpha\alpha} [deg^2]}$ & $\bm{D_L^{\alpha T} [\mu K\ deg]}$ & $\bm{D_L^{\alpha E} [\mu K\ deg]}$ & $\bm{D_L^{\alpha B} [\mu K\ deg]}$ \\ \hline
 1 & $-0.004 \pm 0.006$ & - & - & - \\
 2 & $\phantom{-}0.026 \pm 0.019$ & $ \phantom{-}1.404  \pm  2.683 $ & $ \phantom{-}0.027  \pm  0.021 $ & $ \phantom{-}0.048  \pm  0.016 $ \\
 3 & $-0.013 \pm 0.028$ & $ -1.123  \pm  2.996 $ & $ \phantom{-}0.040  \pm  0.025 $ & $ \phantom{-}0.013  \pm  0.013 $ \\
 4 & $-0.063 \pm 0.049$ & $ -1.678  \pm  3.351 $ & $ \phantom{-}0.013  \pm  0.024 $ & $ -0.032  \pm  0.015 $ \\
 5 & $\phantom{-}0.166 \pm 0.060$ & $ -11.140  \pm  3.668 $ & $ -0.054  \pm  0.026 $ & $ -0.022  \pm  0.014 $ \\
 6 & $\phantom{-}0.194 \pm 0.083$ & $ -7.479  \pm  4.151 $ & $ -0.006  \pm  0.022 $ & $ \phantom{-}0.011  \pm  0.014 $ \\
 7 & $-0.074 \pm 0.090$ & $ \phantom{-}7.492  \pm  4.141 $ & $ -0.013  \pm  0.021 $ &  $ -0.001  \pm  0.016 $ \\
 8 & $\phantom{-}0.058 \pm 0.112$ & $ -1.114  \pm  4.663 $ & $ -0.018  \pm  0.022 $ & $ -0.002  \pm  0.018 $ \\
 9 & $-0.211 \pm 0.137$ & $ \phantom{-}1.536  \pm  4.577 $  & $ -0.001  \pm  0.023 $ & $ -0.010  \pm  0.019 $ \\
 10 & $\phantom{-}0.206 \pm 0.135$ & $ -7.389  \pm  4.740 $ & $ -0.035  \pm  0.023 $ & $ \phantom{-}0.020  \pm  0.019 $ \\
 11 & $-0.215 \pm 0.169$ & $ -3.359  \pm  5.139 $ & $ \phantom{-}0.016  \pm  0.025 $ & $ -0.014  \pm  0.023 $ \\
 12 & $-0.294 \pm 0.188$ & $ \phantom{-}3.900  \pm  4.707 $ & $ -0.062  \pm  0.028 $ & $ \phantom{-}0.027  \pm  0.025 $ \\
 13 & $\phantom{-}0.097 \pm 0.224$ & $ \phantom{-}1.340  \pm  5.185 $ & $ -0.041  \pm  0.029 $ & $ \phantom{-}0.057  \pm  0.024 $ \\
 14 & $\phantom{-}0.213 \pm 0.253$ & $ -4.532  \pm  6.245 $ & $ \phantom{-}0.087  \pm  0.034 $ & $ -0.046  \pm  0.031 $ \\
 15 & $\phantom{-}0.399 \pm 0.266$ & $ -7.444  \pm  5.834 $ & $ -0.045  \pm  0.038 $ & $ -0.055  \pm  0.031 $ \\
 16 & $-0.373 \pm 0.307$ & $ -3.625  \pm  6.447 $ & $ -0.034  \pm  0.039 $ & $ \phantom{-}0.002  \pm  0.037 $ \\
 17 & $\phantom{-}0.015 \pm 0.298$ & $ -9.853  \pm  6.462 $ & $ -0.017  \pm  0.045 $ & $ \phantom{-}0.021  \pm  0.039 $ \\
 18 & $\phantom{-}0.071 \pm 0.357$ & $ -6.756  \pm  7.213 $ & $ -0.033  \pm  0.049 $ & $ -0.098  \pm  0.050 $ \\
 19 & $-0.729 \pm 0.403$ & $ \phantom{-}0.848  \pm  7.662 $ & $ \phantom{-}0.095  \pm  0.065 $ & $\phantom{-} 0.042  \pm  0.064 $ \\
 20 & $-0.131 \pm 0.406$ & $ -4.962  \pm  8.737 $ & $ -0.000  \pm  0.095 $ & $ \phantom{-}0.039  \pm  0.093 $ \\
 21 & $\phantom{-}0.221 \pm 0.450$ & $ \phantom{-}0.500 \pm  10.144 $ & $ \phantom{-}0.106  \pm  \phantom{-}0.163 $ & $ -0.104  \pm  0.149 $ \\
 22 & $-0.218 \pm 0.497$ & $ -5.240  \pm  10.748 $ & $ -0.241  \pm  0.365 $ & $ \phantom{-}0.024  \pm  0.338 $ \\
 23 & $-0.209 \pm 0.476$ & $\phantom{-}11.141  \pm  11.989 $ & $ -0.213  \pm  1.477 $ & $ \phantom{-}0.526  \pm  1.429 $ \\
 24 & $-0.455 \pm 0.470$ & $\phantom{-} 11.452  \pm  14.336 $ & - & - \\ \hline
    \end{tabular}
    \caption{CB spectra for the PR3 \smica\ case. The $\alpha E$ and $\alpha B$ spectra are calculated up to $L = 23$ as the $coswin$ smoothing (and thus the spectra) are null for higher multipoles, as discussed in Sec.~\ref{sec:Analysis pipeline}.}
    \label{tab:smica_all_spectra}
\end{table}
In Figure~\ref{fig:results_posteriors} we display the posterior distributions for the scale invariant amplitude, $A^{\alpha X}$, computed using Eq.~(\ref{eq:chi2_A}). In Table~\ref{tab:results_constraints} we report the upper limits at 95\% C.L., for $A^{\alpha\alpha}$, and the constraints at 68\% C.L., for $A^{\alpha T}$, $A^{\alpha E}$ and $A^{\alpha B}$. As before, also for these parameters, we find a good compatibility among the different component separation methods and with the null effect. Furthermore, for $A^{\alpha\alpha}$, there is good compatibility between the PR3 and the NPIPE results. The NPIPE upper limit using the \commander\ component separation method is looser with respect to the PR3 one because the posterior peak is slightly shifted to the right. Even if a small shift is present also for \sevem, the lower width of the posterior of NPIPE results in a tighter constraint on $A^{\alpha\alpha}$, when compared with \sevem\ PR3.
The $\alpha T$ constraints are compatible with, but tighter than~\cite{Gruppuso:2020kfy}, where they used the same approach adopted in this work but at the lower resolution $N_{side}=4$. Thus, a higher resolution provides more constraining power for $\alpha T$.
\begin{table}[h]
    \centering
    \begin{tabular}{c|cccc}\hline
       \textbf{parameter} & \commander & \nilc & \sevem & \smica \\\hline
       $\mathrm{A^{\alpha\alpha}\ [deg^2]\ PR3}$ & $< 0.007$ & $< 0.007$ & $< 0.010$ & $< 0.007$ \\
       $\mathrm{A^{\alpha\alpha}\ [deg^2]\ NPIPE}$ & $< 0.010$ & - & $< 0.009$ & - \\
       $\mathrm{A^{\alpha T}\ [\mu K\ deg]\ PR3}$ & $-1.827 \pm 0.953$ & $-1.229 \pm 0.873$ & $-2.037 \pm 1.038$ & $-1.916 \pm 0.945$ \\\hdashline
       $\mathrm{A^{\alpha E}\ [nK\ deg]\ PR3}$ & $-3.5 \pm 6.0$ & $-1.0 \pm 5.6$ & $-9.7 \pm 6.0$ & $-7.8 \pm 5.6$ \\
       $\mathrm{A^{\alpha B}\ [nK\ deg]\ PR3}$ & $\hphantom{-}2.4 \pm 4.0$ & $-1.2 \pm 3.7$ & $\hphantom{-}4.0 \pm 4.4$ & $\hphantom{-}0.3 \pm 4.0$ \\\hline
    \end{tabular}
    \caption{Constraints on the scale invariant power spectrum $A^{\alpha X}$, with $X=\alpha,T,E,B$, set in this work.}
    \label{tab:results_constraints}
\end{table}

\section{Conclusions}\label{sec:Conclusions}
In this work we build CB maps at angular scales larger than $\sim 7$ deg, exploiting \planck\ CMB component separated maps from both PR3 and NPIPE releases. From these CB maps we estimate the monopoles (see Tab.~\ref{tab:monopole}), i.e. the isotropic birefringence angle, finding a very good compatibility with previous results~\cite{planck2014-a23,Gruppuso:2020kfy,Eskilt:2022wav,MinamiKomatsu2020,Diego-Palazuelos:2022dsq}. We also compute their spectra and the cross-correlation with the CMB temperature map. Moreover, we provide for the first time the spectra of the cross-correlation of the CB field with the CMB E and B fields. The data CB angle maps and spectra, for both auto- and cross-correlations with CMB, are made publicly available\footnote{\href{https://github.com/marcobortolami/AnisotropicBirefringence_patches.git}{https://github.com/marcobortolami/AnisotropicBirefringence\_patches.git}}.\par
We quantify the compatibility of all the aforementioned spectra with null effect through an harmonic $\chi^2$ analysis, see Fig.~\ref{fig:results_histo}, and through a scale invariant amplitude $A^{\alpha X}$, with $X=\alpha,T,E,B$, see Fig.~\ref{fig:results_posteriors}. We find no significant evidence of deviation from the null effect. The constraints on $A^{\alpha X}$ are summarized in Tab.~\ref{tab:results_constraints}. The latter are obtained through a $\chi^2$ minimization assuming null CB, which is supported by data. We also constrain jointly $A^{\alpha \alpha}$ and $A^{\alpha T}$ with a pixel-based likelihood which naturally takes into account the effect of CB-induced cosmic variance, see App.~\ref{app:2D plots}.\par
In this work we extend to $L=24$ the multipole range of the $\alpha\alpha$ and $\alpha T$ spectra covered in~\cite{Gruppuso:2020kfy}, which adopts the same technique for the construction of the CB maps. Note that for the PR3 $\alpha T$ case we obtain constraints compatible with, but tighter than, \cite{Gruppuso:2020kfy}. In addition, we provide for the first time estimates of the \Planck\ NPIPE $\alpha\alpha$ spectra. However, the main novelty of this work is represented by the estimates of the \Planck\ PR3 $\alpha E$ and $\alpha B$ spectra, which are presented here for the first time. The latter might be fruitfully considered to constrain models of anisotropic birefringence that predict correlations with the CMB fields.

\acknowledgments
We acknowledge the financial support from the INFN InDark initiative and from the COSMOS network (www.cosmosnet.it) through the ASI (Italian Space Agency) Grants 2016-24-H.0 and 2016-24-H.1-2018, as well as 2020-9-HH.0 (participation in LiteBIRD phase A). We acknowledge the use of the \texttt{healpy}~\cite{Zonca:2019vzt,Gorski:2004by}, \texttt{NaMaster}~\cite{Alonso:2018jzx}, \texttt{numpy} \citep{Harris:2020xlr}, \texttt{matplotlib} \citep{Hunter:2007ouj} software packages, and the use of computing facilities at CINECA.  

\bibliographystyle{JHEP}
\bibliography{bibliography,Planck_bib}

\appendix

\section{E and B modes purification for CMB APS extraction}\label{E and B modes purification for CMB APS extraction}
In each patch we estimate the CMB power spectra using Pymaster~\cite{Alonso:2018jzx}, which is a pseudo-$C_\ell$ estimator. This technique is known to suffer from E/B leakage when applied on a masked sky. This effect can be alleviated considering a purification process, which removes the misinterpreted modes, at a cost of information loss. However, for our case, i.e. the \planck\ noise and a sky fraction of the size of the patches, the standard master algorithm provides unbiased spectra and the purification process just worsens the uncertainty of the estimated spectra. For this reason, we decide not to apply any purification. In order to prove the former statement, we compare the estimated spectra for all the possible purification cases: master (i.e. no purification), pure E (i.e. purification of the E modes only), pure B (i.e. purification of the B modes only) and pure EB (i.e. purification of both the E and B modes). We employ 100 CMB simulations with a \planck-like noise level, i.e. the polarization noise is $\sigma_P$ = 57.7 $\mu$K arcmin, and a sky fraction of $f_{sky,patch} \simeq 0.13\%$.\par
\begin{figure}[h]
    \centering
    \includegraphics[width=1.\textwidth]{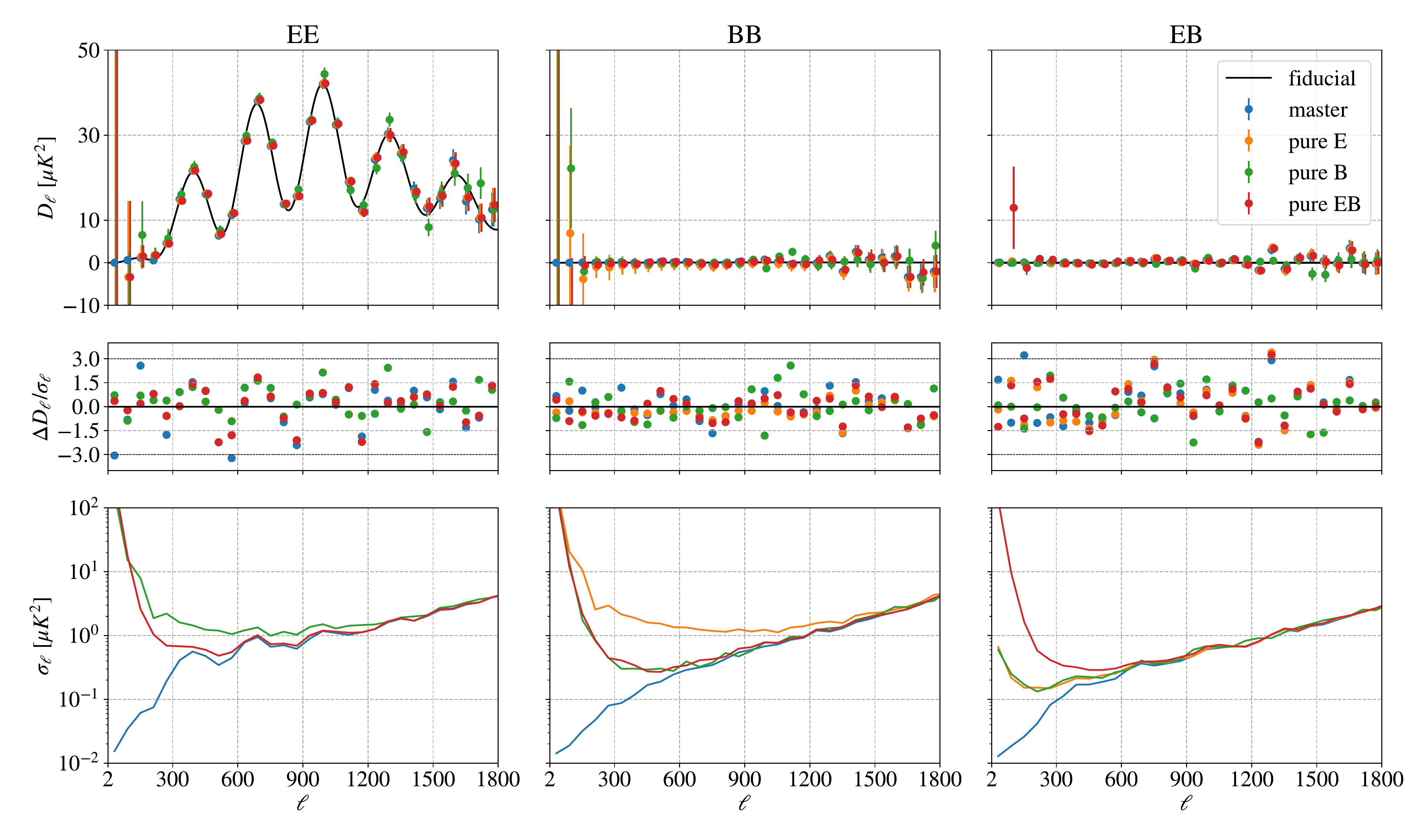}
    \caption{Top: EE (left), BB (center) and EB (right) CMB power spectra for the four purification cases and for 100 \Planck-like simulations. The black line represents the fiducial, while the colored points are the simulation average spectra. The error bars represent the error of the mean. Center: EE, BB and EB difference between the simulation average and the fiducial spectra, divided by the error of the mean. Bottom: EE, BB and EB errors of the mean.}
    \label{fig:purification_study}
\end{figure}

In Figure~\ref{fig:purification_study} we show the results of this test.
The spectra are validated for all the four cases. The error on the spectra is similar for high multipoles, while it is smaller for the master case at low multipoles. Thus, for the analysis conditions considered in this work, the purification process increases the uncertainty of the estimated spectra and turns out to be not useful.

\section{Effective beam calculation}\label{app:effective beam calculation}
Strictly speaking, both CMB data and simulations contain a non-Gaussian beam and a non-ideal pixel window function (see Figure F.1. of~\cite{planck2013-p03c}). Thus, we employ an effective ``smoothing function'' $b_{\ell,eff}^{2}$ obtained by taking these steps:
\begin{enumerate}
    \item From 999 signal-only \planck\ PR3 simulated $a_{\ell m}^i$, not smoothed, we compute the corresponding $C_\ell^i$ for each simulation $i$.
    \item From 999 signal-only \planck\ PR3 simulated maps at $N_{side} = 2048$, that are smoothed with a realistic beam, get the corresponding $C_{\ell,sm}^i$ in full sky for each simulation $i$.
    \item Divide the two power spectra to obtain $\left(b_{\ell,eff}^{2}\right)^i = C_{\ell,sm}^i / C_\ell^i$ for each simulation $i$.
    \item Average the smoothing function over the simulation index $i$.
\end{enumerate}
The smoothing function obtained contains both an effective beam window function and an effective pixel window function. This procedure is followed for the PR3 simulations of the 4 component separation methods maps separately, obtaining 4 different effective smoothing functions for \commander, \nilc, \sevem\ and \smica. For NPIPE \commander\ and \sevem\ maps we use the same smoothing function computed for PR3. For each component separation method, the same smoothing function is used for each patch.

\section{CB auto- and cross-correlation results on simulations}\label{app:results on simulations}
\begin{figure}[h]
\centering
\includegraphics[width=.49\textwidth]{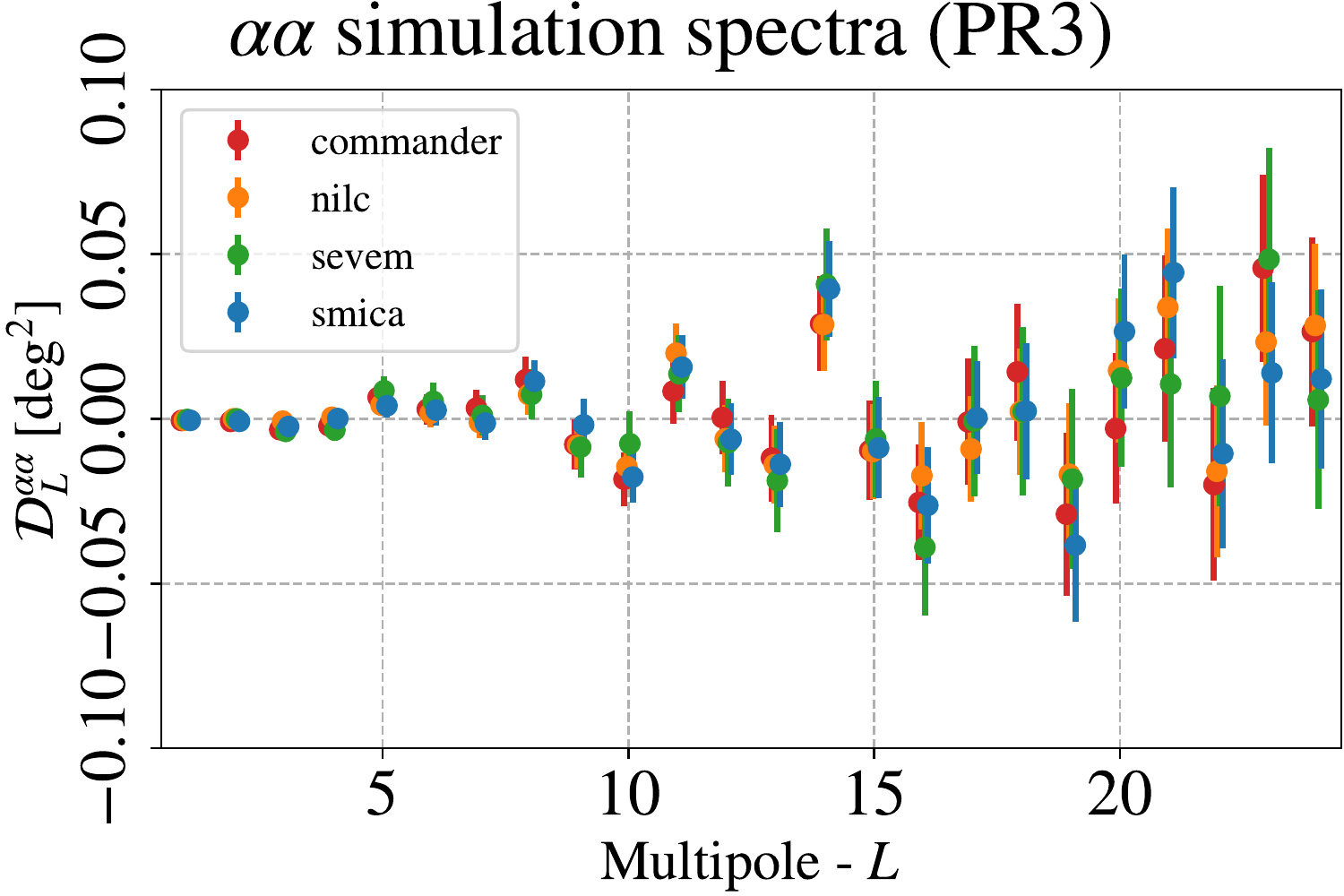}
\hfill
\includegraphics[width=.49\textwidth]{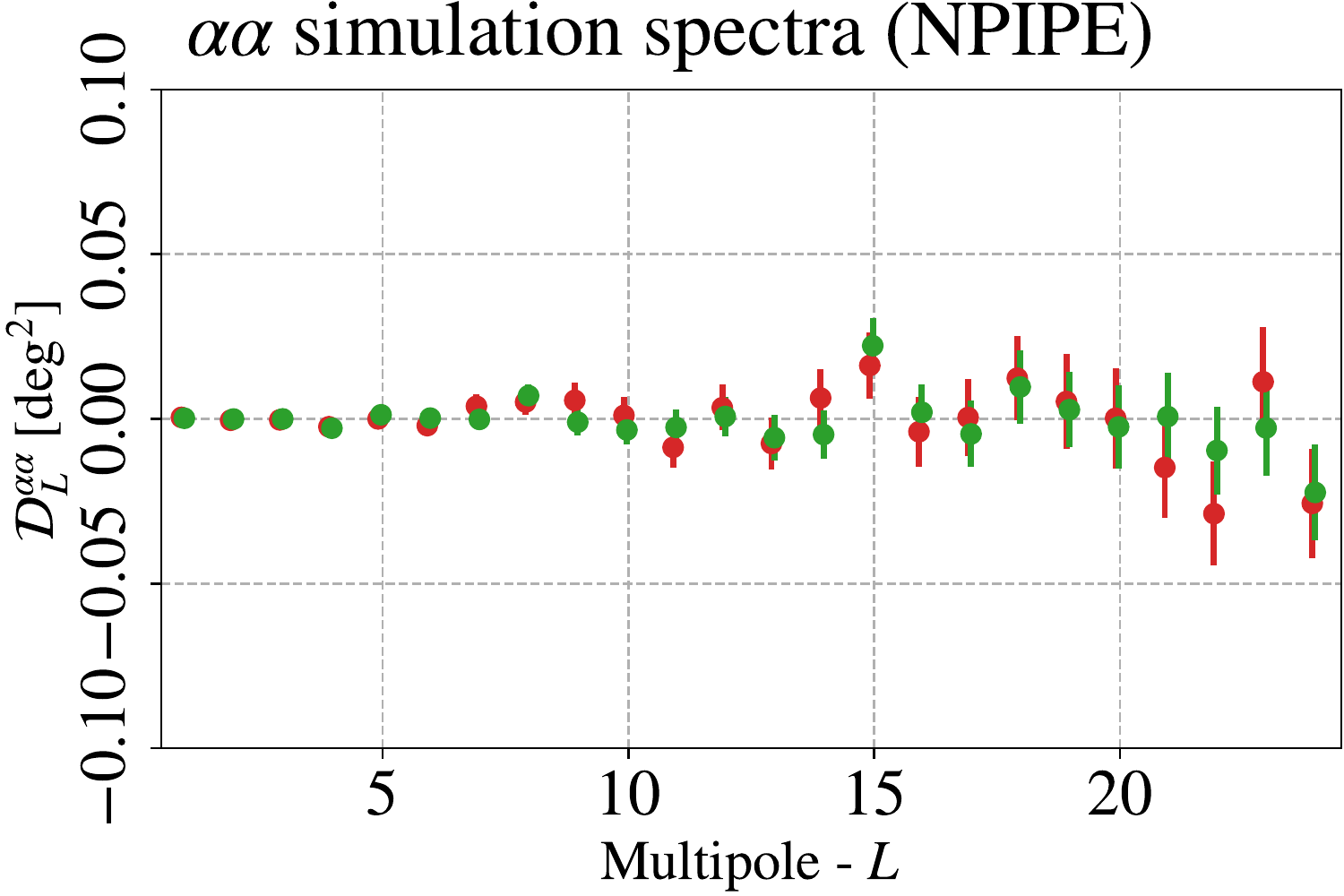}
\includegraphics[width=.49\textwidth]{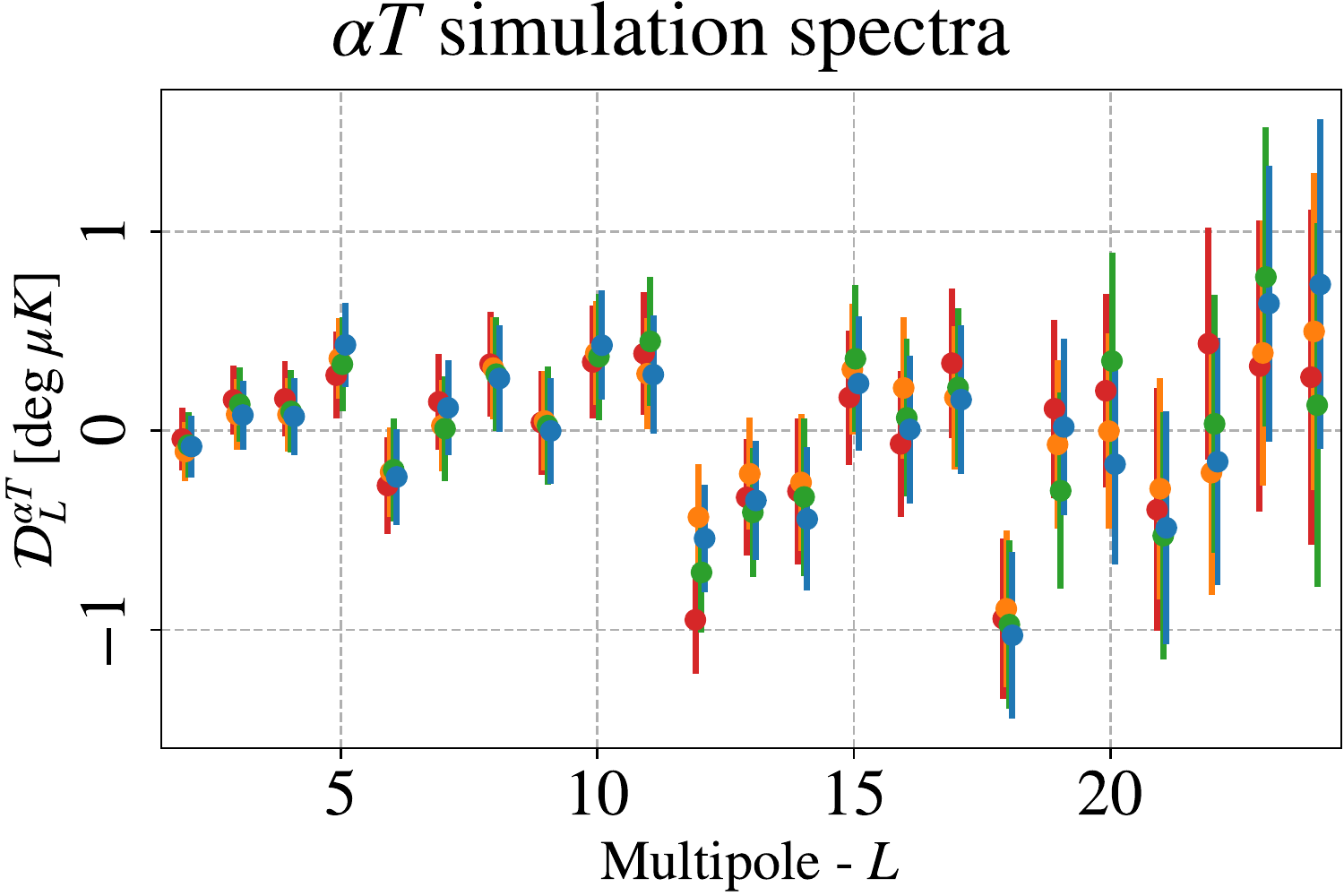}\\
\includegraphics[width=.49\textwidth]{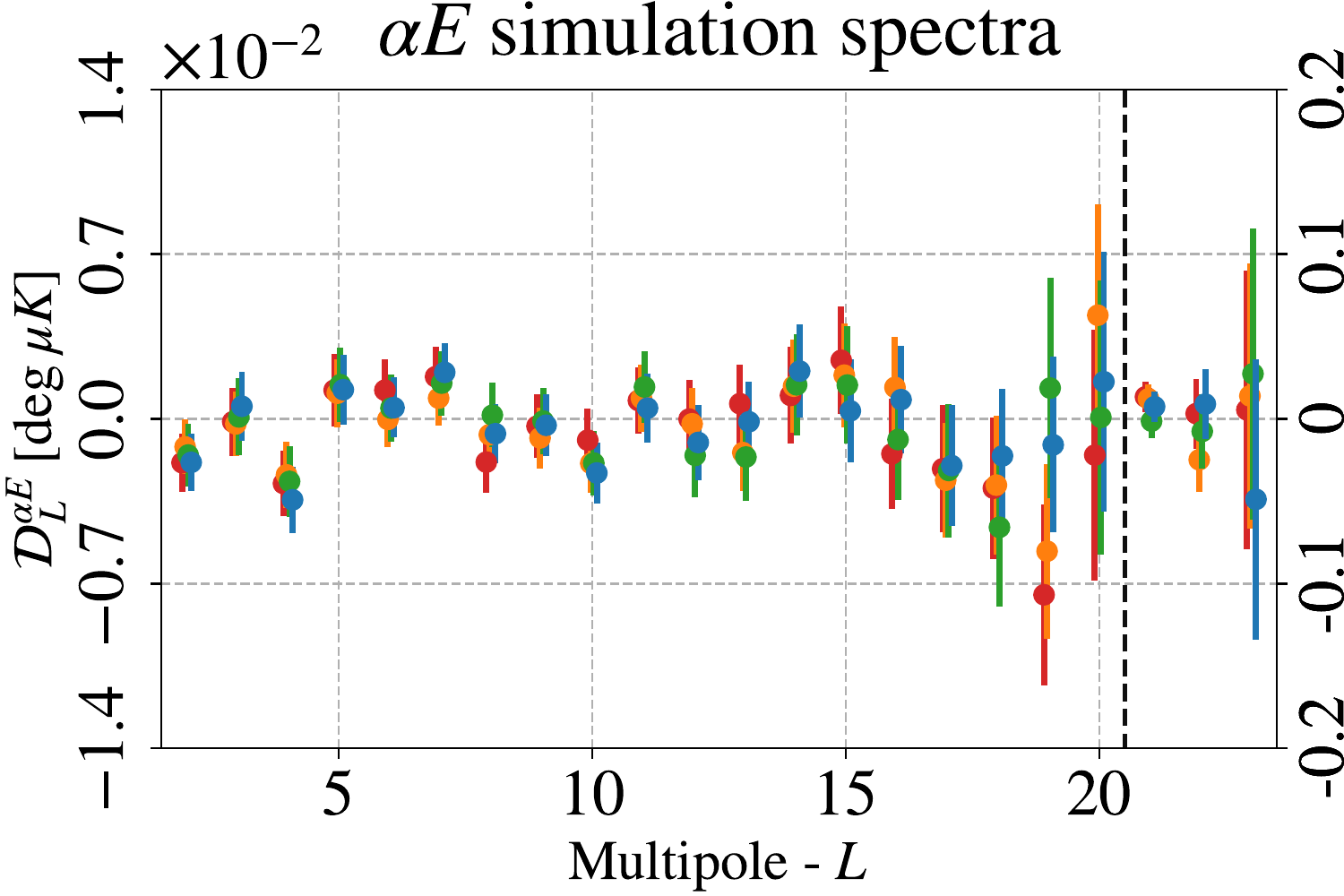}
\hfill
\includegraphics[width=.49\textwidth]{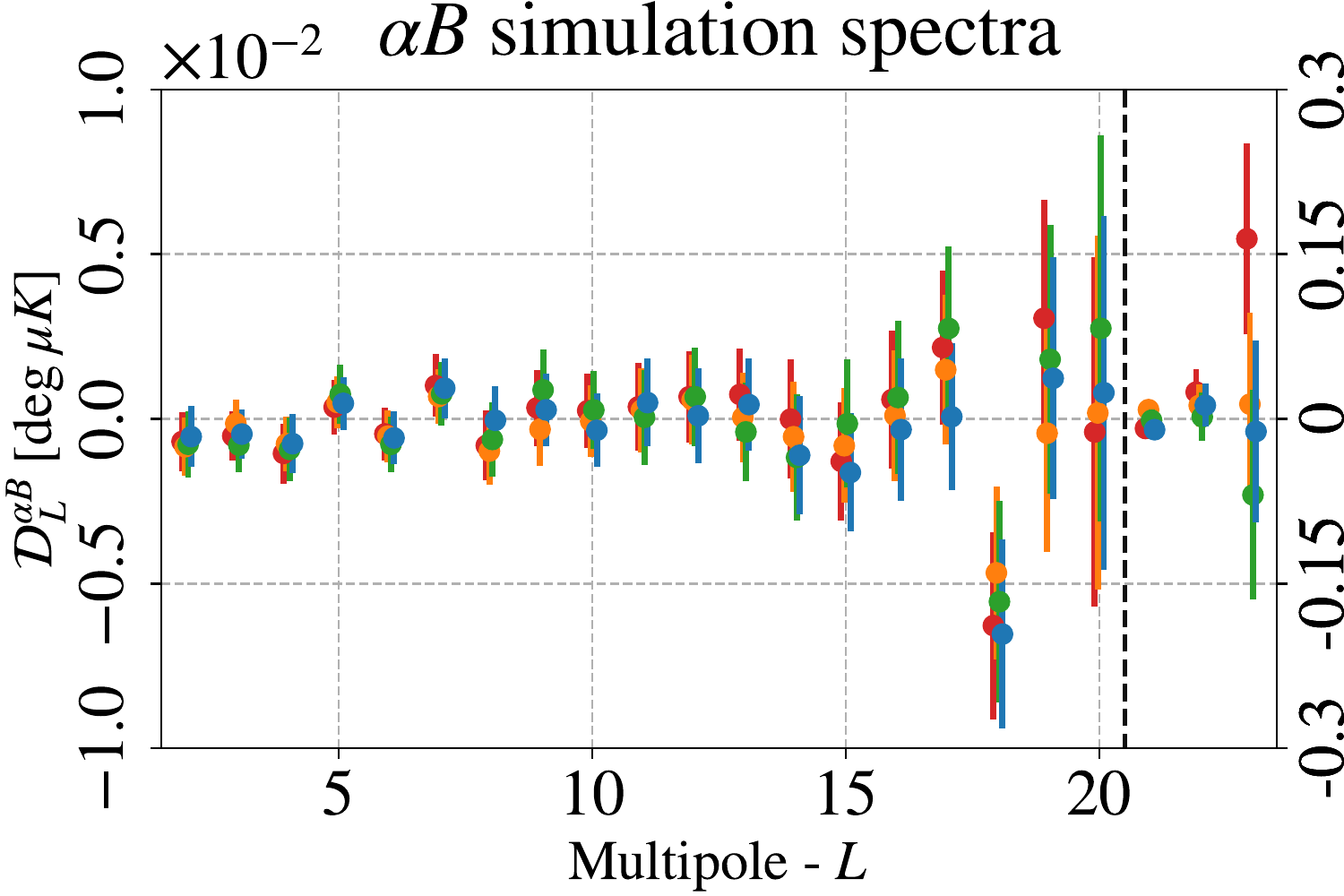}
\caption{Auto-correlation spectra of the simulation CB angle maps and their cross-correlation with CMB. The dots represent the average of the simulation spectra, while the errors are the errors of the average. The vertical dashed lines highlight a change in the vertical axis range. Top left panel: PR3 $\alpha\alpha$. Top right panel: NPIPE $\alpha\alpha$. Central panel: PR3 $\alpha T$. Bottom left panel: PR3 $\alpha E$. Bottom right panel: PR3 $\alpha B$.\label{fig:results_sims_spectra}}
\end{figure}

We report in Fig.~\ref{fig:results_sims_spectra} the auto-correlation of the CB angle maps and their cross-correlation with CMB temperature and polarization fields obtained on simulations. All the spectra are given in terms of bandpowers, as in Sec.~\ref{sec:Results}. Figure~\ref{fig:results_sims_spectra} shows that the systematic effects included in the simulations have a negligible impact because all the spectra are well compatible with zero, as the deviations from zero are less than $3\sigma$ for all the multipoles considered. As for the data spectra, the scatter around zero of the simulation NPIPE $\alpha\alpha$ spectra and the errors are lower with respect to the PR3 $\alpha\alpha$ case. This is again due to the lower levels of noise present in the NPIPE maps.

\section{Analysis robustness tests}\label{app:Analysis robustness tests}
In order to test the robustness of our results, we modify some analysis settings of our pipeline. Since we find good stability for all the considered spectra, for the sake of brevity we present here robustness tests only for the cross-correlation between the CB and the CMB temperature field. This choice is mainly due to the fact that, even if statistically well compatible, the most likely value for $A^{\alpha T}$ is found here negative while in~\cite{Gruppuso:2020kfy} it is positive. Since the analysis in~\cite{Gruppuso:2020kfy} was performed at \healpix\ resolution $N_{side} = 4$ (while our analysis is performed at $N_{side} = 8$) we rerun our codes at lower resolution for the \commander\ component separation method. The maximum multipole for the $\alpha T$ spectra at this resolution is 12. We find that data and simulations spectra are compatible with zero, as at $N_{side} = 8$. However, the scale-invariant amplitude moves towards positive values, as we obtain $\mathrm{A^{\alpha T} = (0.192 \pm 1.102)\ \mu K\ deg}$. The constraint is worse than the one at $N_{side} = 8$ due to the lower resolution of the patch maps. In addition, if we use a more aggressive mask by extending our mask by the size of one patch as done in~\cite{Gruppuso:2020kfy} ($\approx 15^\circ$ at $N_{side}=4$), we obtain $\mathrm{A^{\alpha T} = (1.890 \pm 1.297)\ \mu K\ deg}$. Our constraints at $N_{side}=4$ are well compatible (even for the sign of the most likely value) with~\cite{Gruppuso:2020kfy} and with null cross-correlation between the CB and the CMB temperature fields. To test if the change in resolution can explain the shift of the data spectra to more negative values, we calculate the difference between the two data spectra for the first 12 multipoles at the two different resolutions and we compare it with the same quantity calculated on simulations. We find that the shift in the bandpowers is compatible with the scatter of the simulations.\par
In order to see if the negative sign of the $\mathrm{A^{\alpha T}}$ parameter is given by a bias in the simulations, we subtract the average of the CB simulation maps from all the CB simulations and data maps. We then run the QML estimator for the $D_L^{\alpha T}$ spectra and fit with a scale-invariant spectrum for the $\mathrm{A^{\alpha T}}$ parameter. The posteriors move towards slightly less negative values of $\mathrm{A^{\alpha T}}$ and the errors are slightly lower, but this shift cannot explain the change in sign of $\mathrm{A^{\alpha T}}$. The constraints in the debias case are reported in Tab.~\ref{tab:robustness_tests_constraints}.
\begin{table}[h]
    \centering
    \begin{tabular}{c|cccc}\hline
       \textbf{case}    & \commander & \nilc & \sevem & \smica \\\hline
       reference        & $-1.827 \pm 0.953$ & $-1.229 \pm 0.873$ & $-2.037 \pm 1.038$ & $-1.916 \pm 0.945$ \\
       debias           & $-1.766 \pm 0.948$ & $-1.192 \pm 0.868$ & $-1.997 \pm 1.034$ & $-1.886 \pm 0.941$ \\
       extended         & $-1.383 \pm 1.098$ & $-0.724 \pm 0.993$ & $-1.636 \pm 1.164$ & $-1.354 \pm 1.066$ \\
       $\ell_{min}=302$ & $-1.931 \pm 0.995$ & $-1.439 \pm 0.899$ & $-2.468 \pm 1.092$ & $-2.122 \pm 0.987$ \\\hline
    \end{tabular}
    \caption{Constraints on the scale invariant power spectrum $A^{\alpha T}$, given in $\mu K\ deg$. The reference case is discussed in Sec.~\ref{sec:Analysis pipeline}. The debias, extended and $\ell_{min}=302$ cases are explained in this Appendix.}
    \label{tab:robustness_tests_constraints}
\end{table}\par
To study the effect of possible foregrounds not excluded by the mask applied to the CB angle and CMB temperature maps, we try to use a more aggressive mask. We thus extend the mask shown in the left panel of Fig.~\ref{fig:CB_spectra_masks} by the size of one patch, i.e. $\approx 8^\circ$, leaving $\approx$ 60\% of the sky non-masked. The scatter of the data spectra around zero is slightly larger than for the non extended mask due to the lower sky fraction, but the compatibility with the null effect is still obtained. The posteriors move towards less negative values of the scale invariant spectrum, but the errors on the parameter increase, again due to the reduced sky fraction. The constraints with the extended mask are reported in Tab.~\ref{tab:robustness_tests_constraints}.\par
Finally, we change the CMB multipole range used for the minimization of the $\chi^2$ to obtain the CB angles in each patch. We increase the minimum multipole considered from 62 to 302 and then produce the CB angle maps and get the final results. The posteriors move towards more negative values of $\mathrm{A^{\alpha T}}$, but the errors on this parameter increase and there is again good compatibility with null cross-correlation between CB and CMB temperature fields. The constraints with $\ell_{min}=302$ case are reported in Tab.~\ref{tab:robustness_tests_constraints}.

\section{$\alpha\alpha$ and $\alpha T$ joint constraints}\label{app:2D plots}
\begin{figure}[h]
\centering
\includegraphics[width=.49\textwidth]{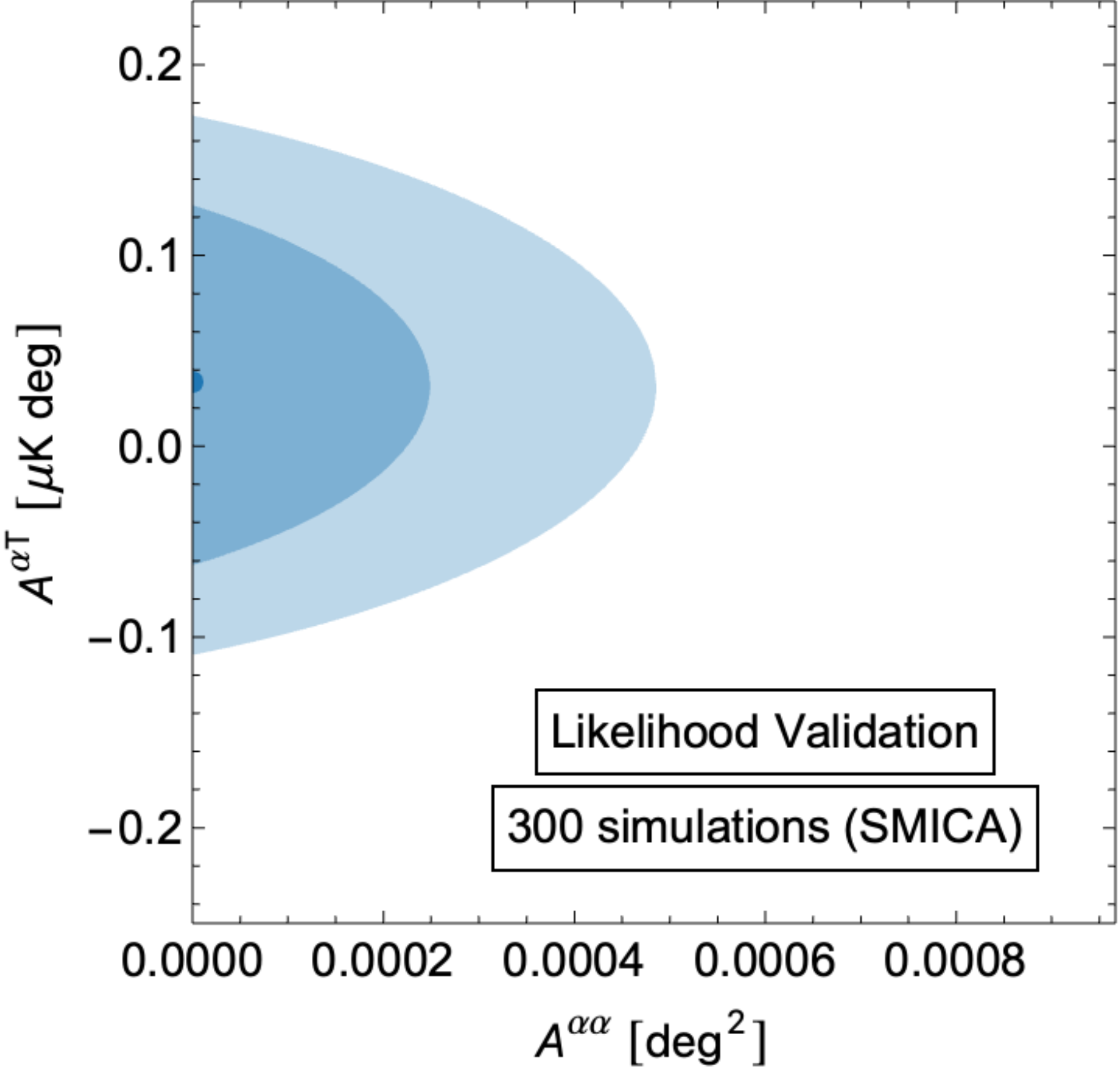}
\caption{2D contour plots for the $\mathrm{A^{\alpha\alpha}}$ and $\mathrm{A^{\alpha T}}$ parameters obtained with the \smica\ simulations. The darker (lighter) area represents the 68\% (95\%) confidence region.}\label{fig:PR3_2D_validation}
\end{figure}

We set 2D constraints on $A^{\alpha \alpha}$ and $A^{\alpha T}$ following the methodology explained in~\cite{Gruppuso:2020kfy}. We report in Fig.~\ref{fig:PR3_2D_validation} the validation of the 2D analysis for the \smica\ case. The posterior shown in Fig.~\ref{fig:PR3_2D_validation} is the product of the posteriors obtained from the single simulations. Since the simulations do not contain the CB effect and the posterior is consistent with zero, the analysis is validated. We obtain similar results for the other component separation methods.
\par
\begin{figure}[h]
\centering
\includegraphics[width=.49\textwidth]{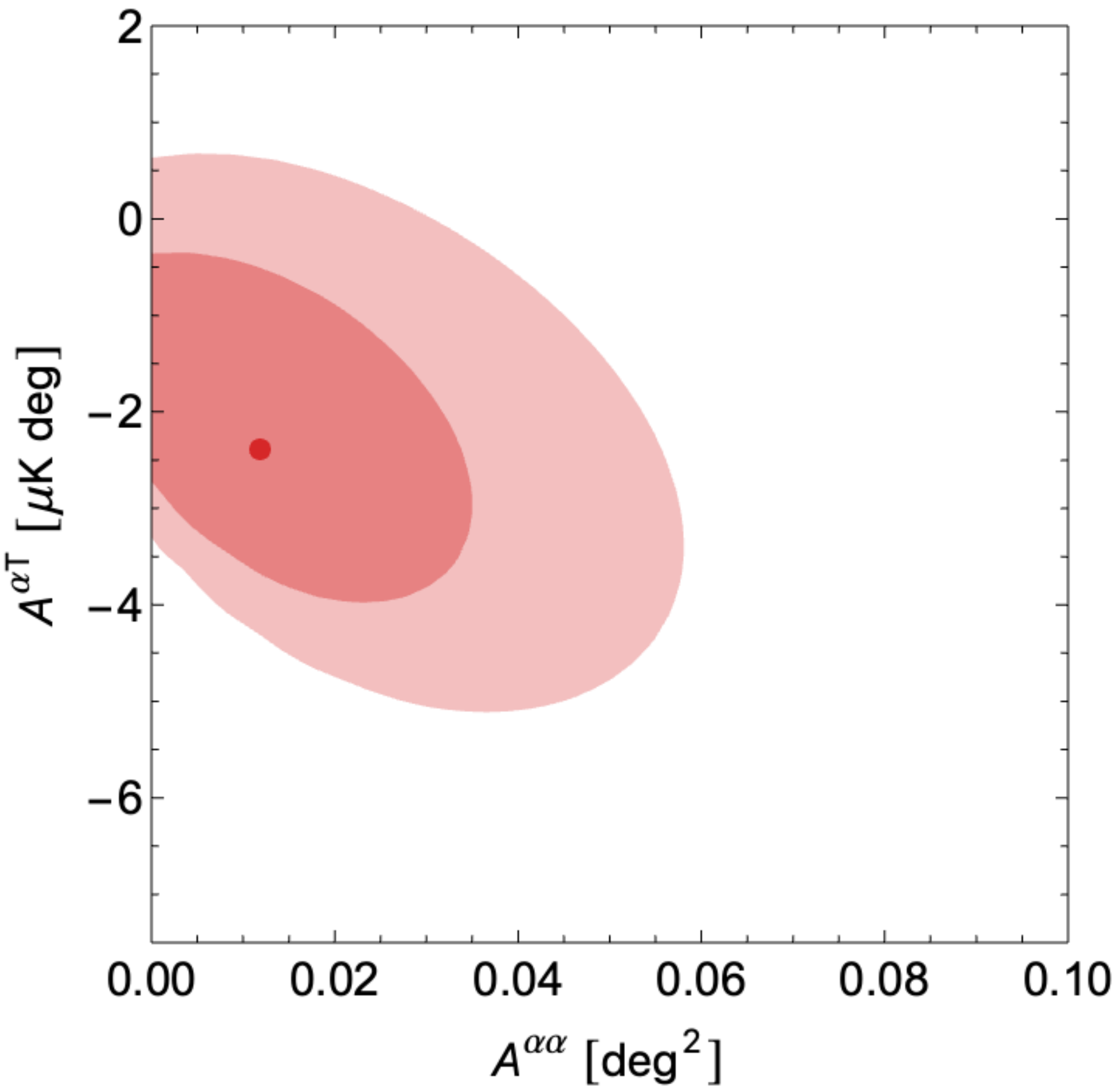}
\hfill
\includegraphics[width=.49\textwidth]{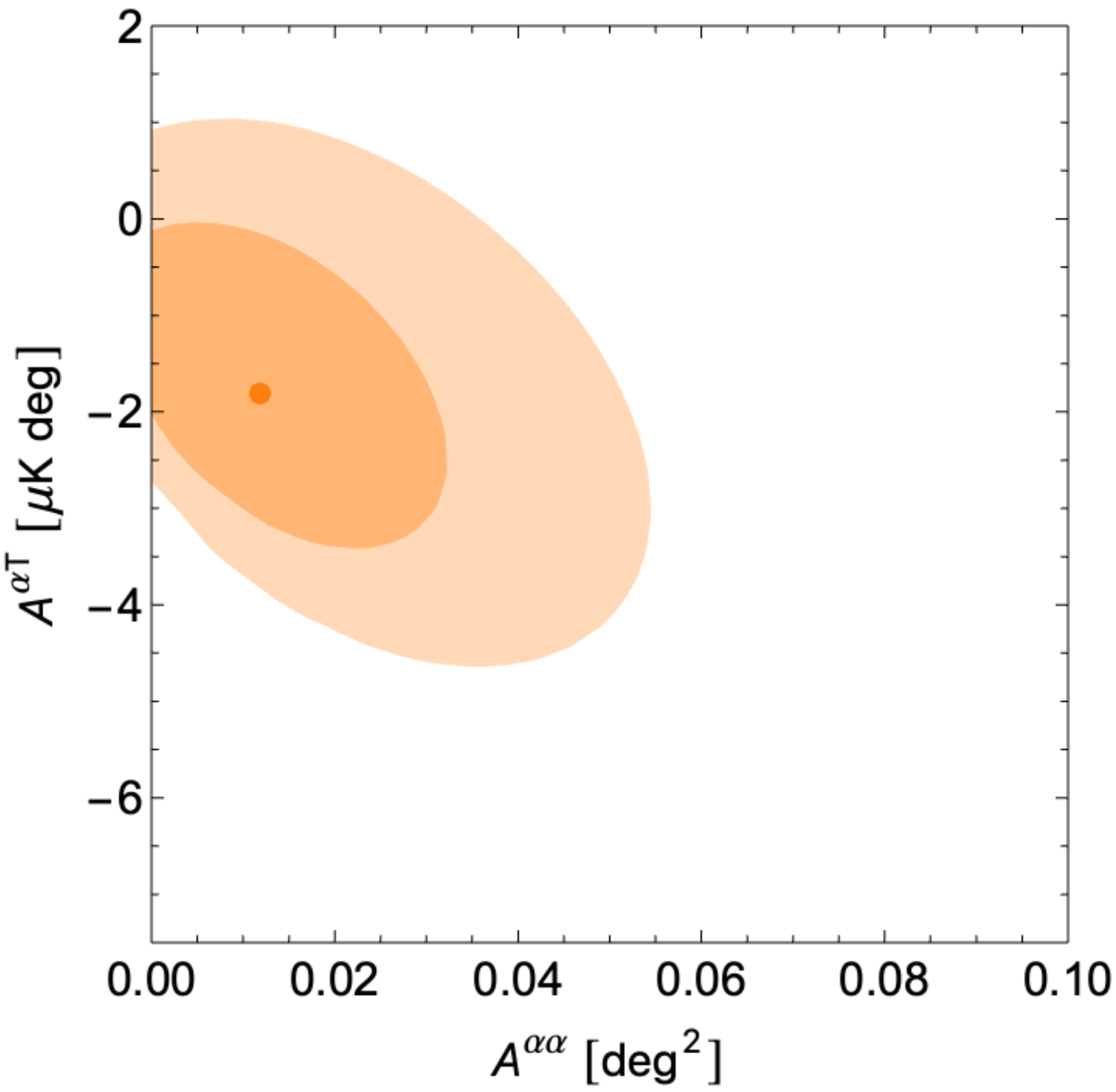}
\includegraphics[width=.49\textwidth]{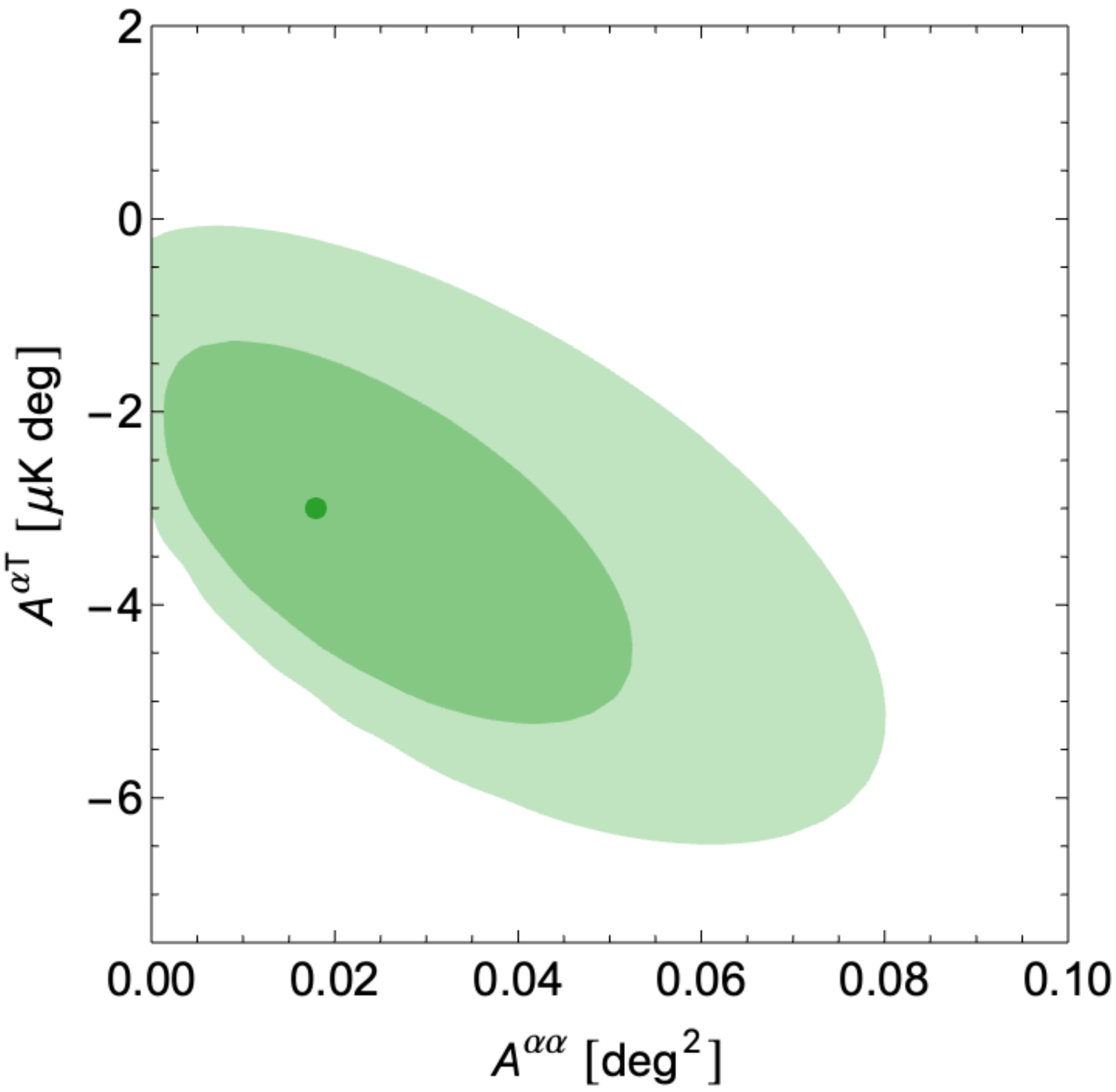}
\hfill
\includegraphics[width=.49\textwidth]{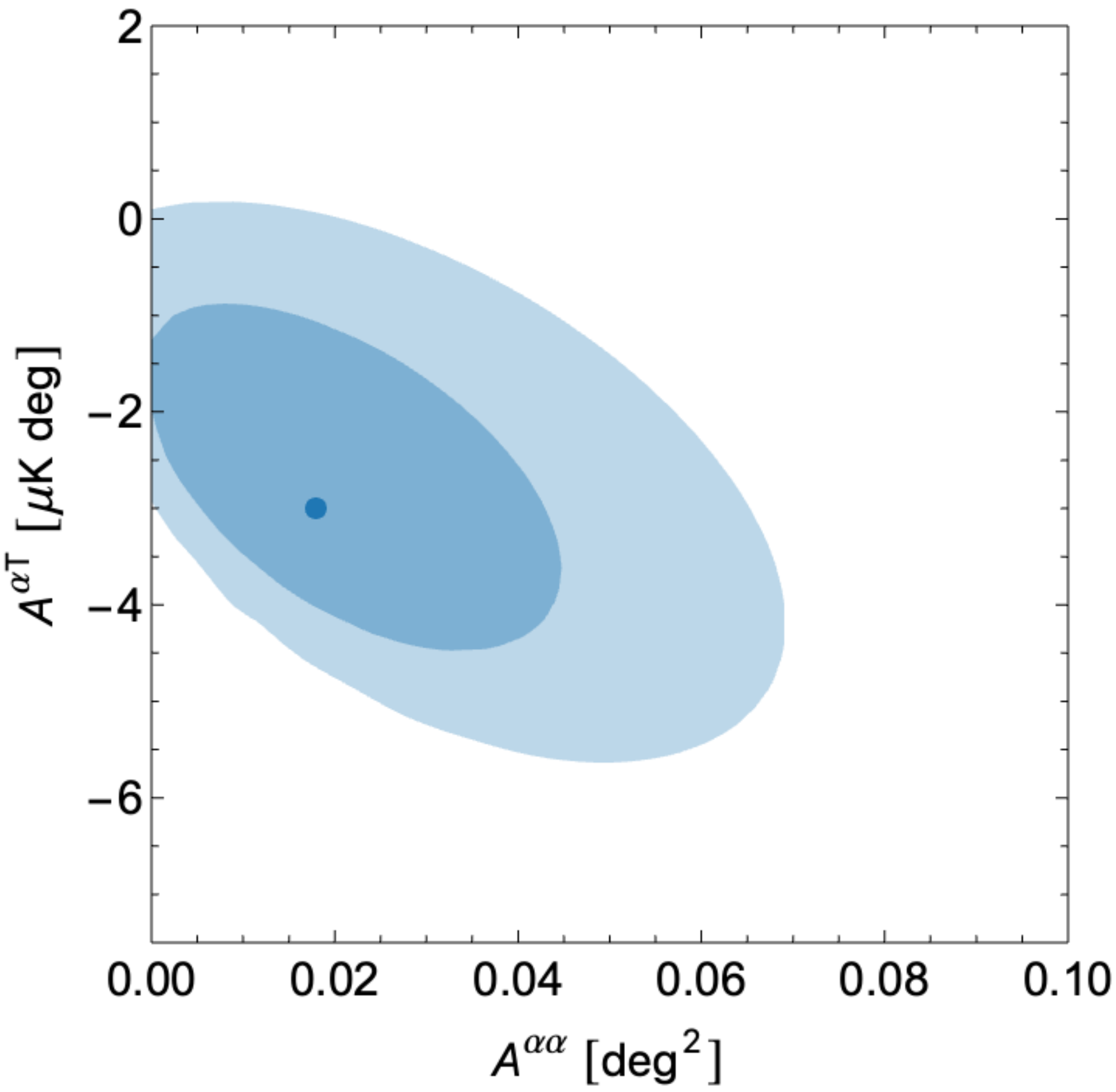}
\caption{2-dimensional contour plots for the $\mathrm{A^{\alpha\alpha}}$ and $\mathrm{A^{\alpha T}}$ parameters obtained with the \commander\ (top left panel), \nilc\ (top right panel), \sevem\ (bottom left panel) and \smica\ (bottom right panel) data. The darker (lighter) area represents the 68\% (95\%) confidence region.}\label{fig:PR3_2D}
\end{figure}

In Figure~\ref{fig:PR3_2D} we show the 2D contour plots for the $\mathrm{A^{\alpha\alpha}}$ and $\mathrm{A^{\alpha T}}$ parameters obtained with the \commander, \nilc, \sevem\ and \smica\ CB data maps for the PR3 case and with the \commander\ CMB temperature map. We use two different masks for the CB angles and CMB temperature maps in order to exploit the largest amount of data: the former has a sky fraction of 74\%, while the latter of 84\%. The covariance matrices are built analytically. For this motivation, the uncertainties in the 2D case are larger with respect to the 1D analysis described in Sec.~\ref{sec:Analysis pipeline}, as in the latter the CB cosmic variance is not taken into account. The four component separation methods provide similar constraints. The \sevem\ component separation method is the one having the lowest compatibility with null CB. However, the deviation is not larger than 3 standard deviations. Thus, we find no evidence for the CB effect.\par
\begin{figure}[h]
\centering
\includegraphics[width=.42\textwidth]{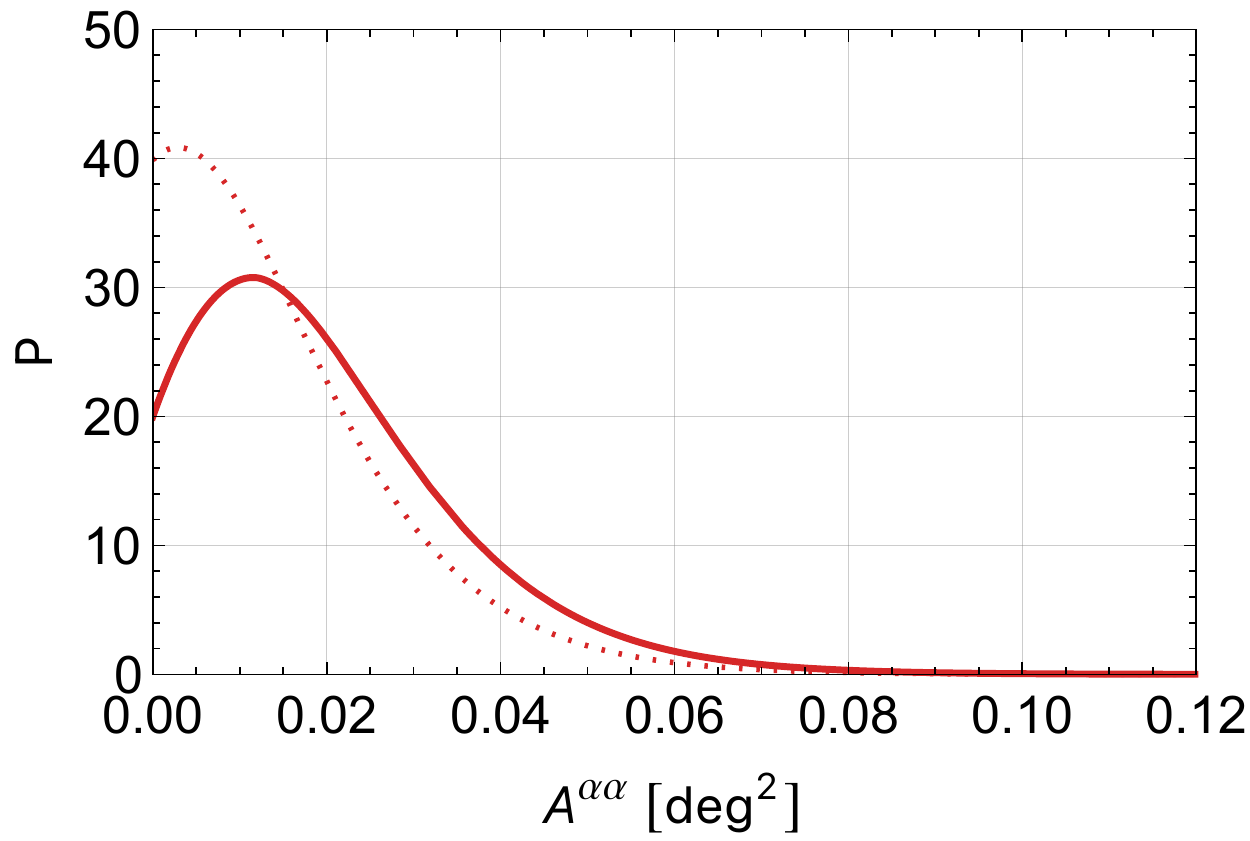}
\hfill
\includegraphics[width=.42\textwidth]{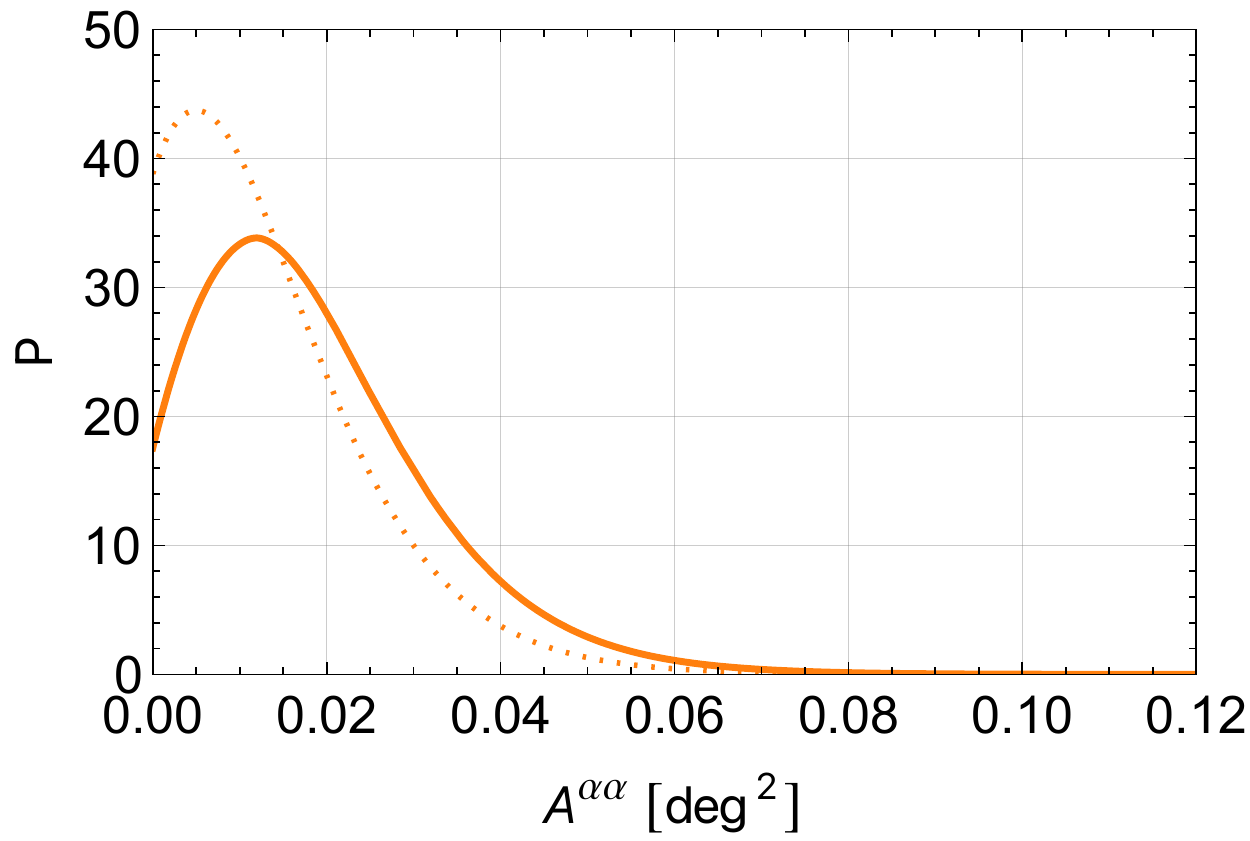}
\includegraphics[width=.42\textwidth]{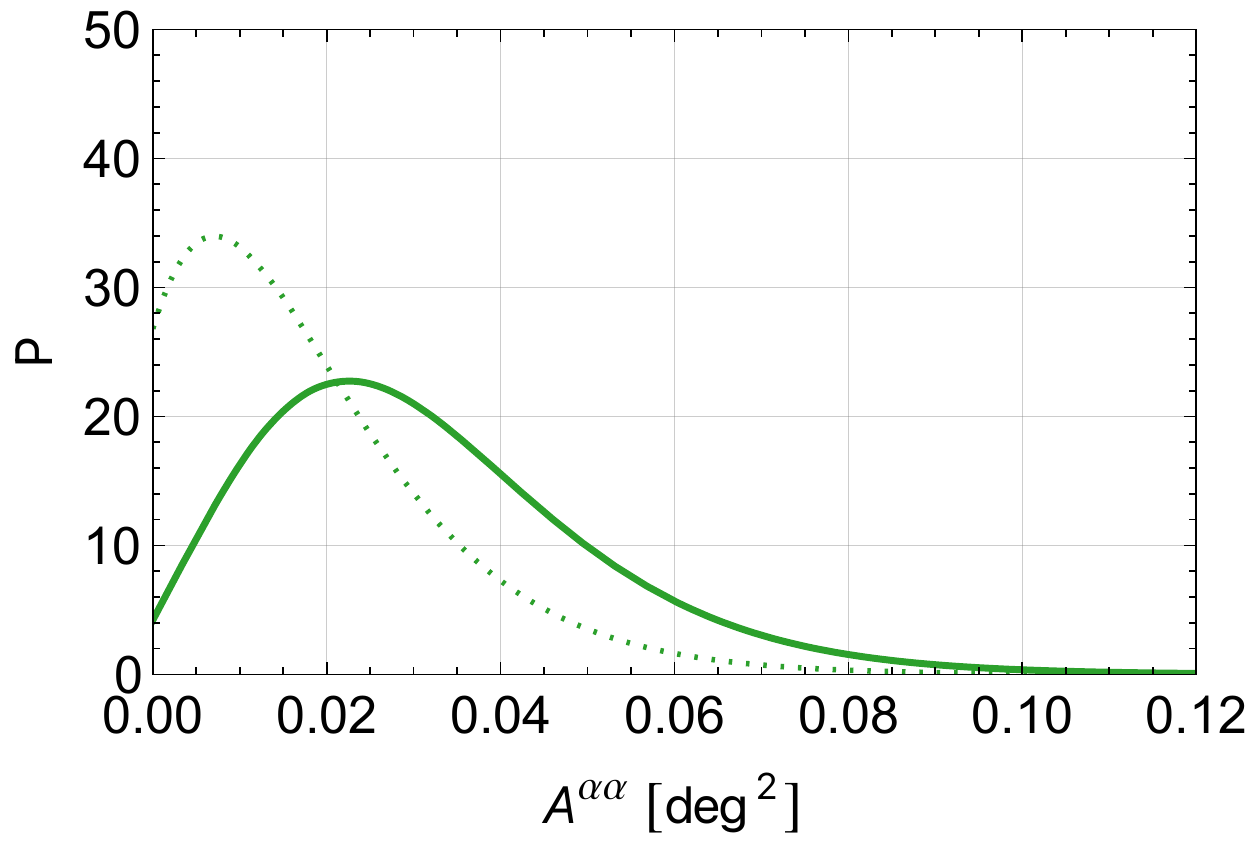}
\hfill
\includegraphics[width=.42\textwidth]{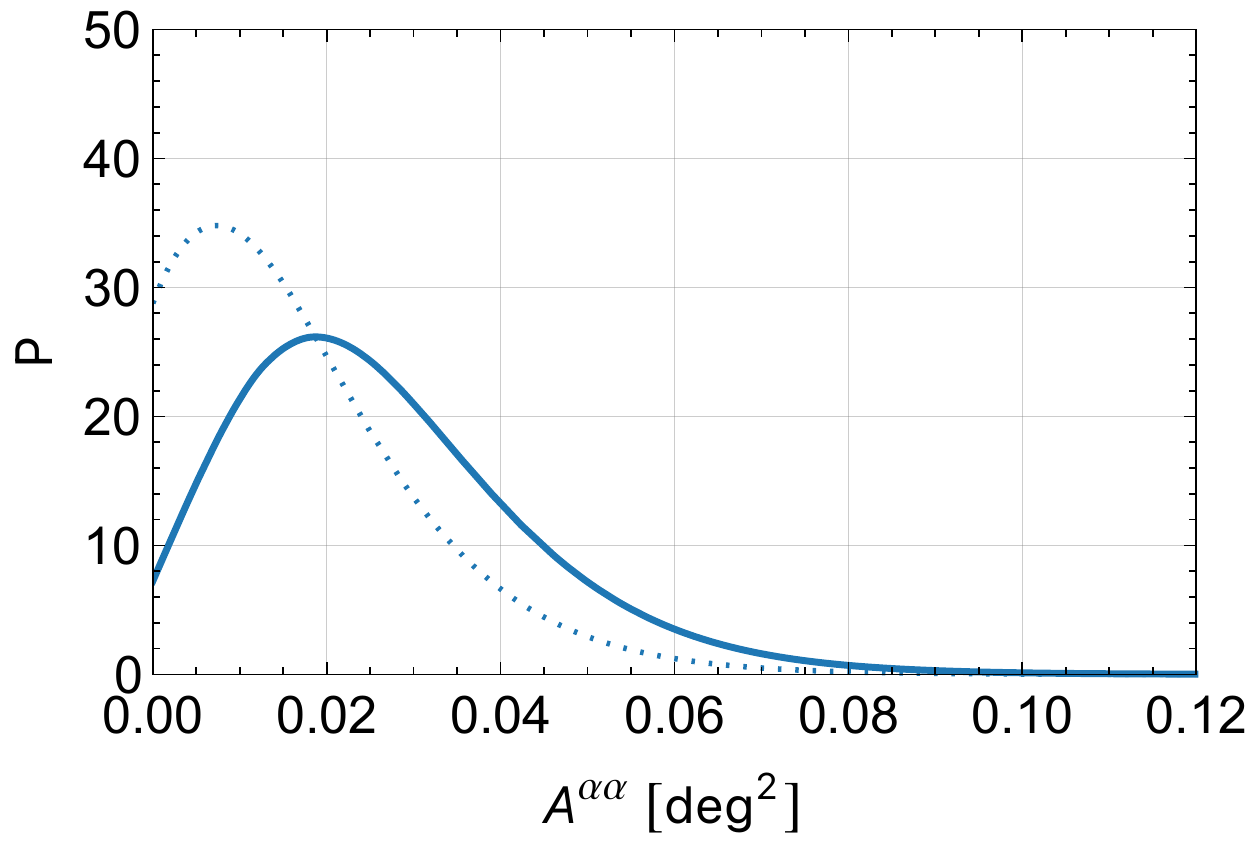}
\caption{Posterior distribution functions for the $\mathrm{A^{\alpha\alpha}}$ parameter obtained from \commander\ (top left panel), \nilc\ (top right panel), \sevem\ (bottom left panel) and \smica\ (bottom right panel) data. The solid curves are computed marginalising the 2D distribution functions over $\mathrm{A^{\alpha T}}$ and the dashed ones slicing the 2D distribution functions at $\mathrm{A^{\alpha T}} = 0$.}\label{fig:PR3_2D_marginaliseT}
\end{figure}

We then marginalize the 2D posterior distribution functions over $\mathrm{A^{\alpha T}}$, obtaining the posterior distribution functions of $\mathrm{A^{\alpha\alpha}}$ reported in Fig.~\ref{fig:PR3_2D_marginaliseT} (solid curves). We also slice the 2D probability at $\mathrm{A^{\alpha T}} = 0$, obtaining the probability distribution of $\mathrm{A^{\alpha\alpha}}$ shown in Fig.~\ref{fig:PR3_2D_marginaliseT} (dashed curves). The posterior is closer to zero for the sliced case, providing more stringent constraints than the marginalized posterior. The four component separation methods agree among themselves, even if the \sevem\ distribution is larger than the other three, reflecting what discussed for Fig.~\ref{fig:PR3_2D}.\par
\begin{figure}[h]
\centering
\includegraphics[width=.42\textwidth]{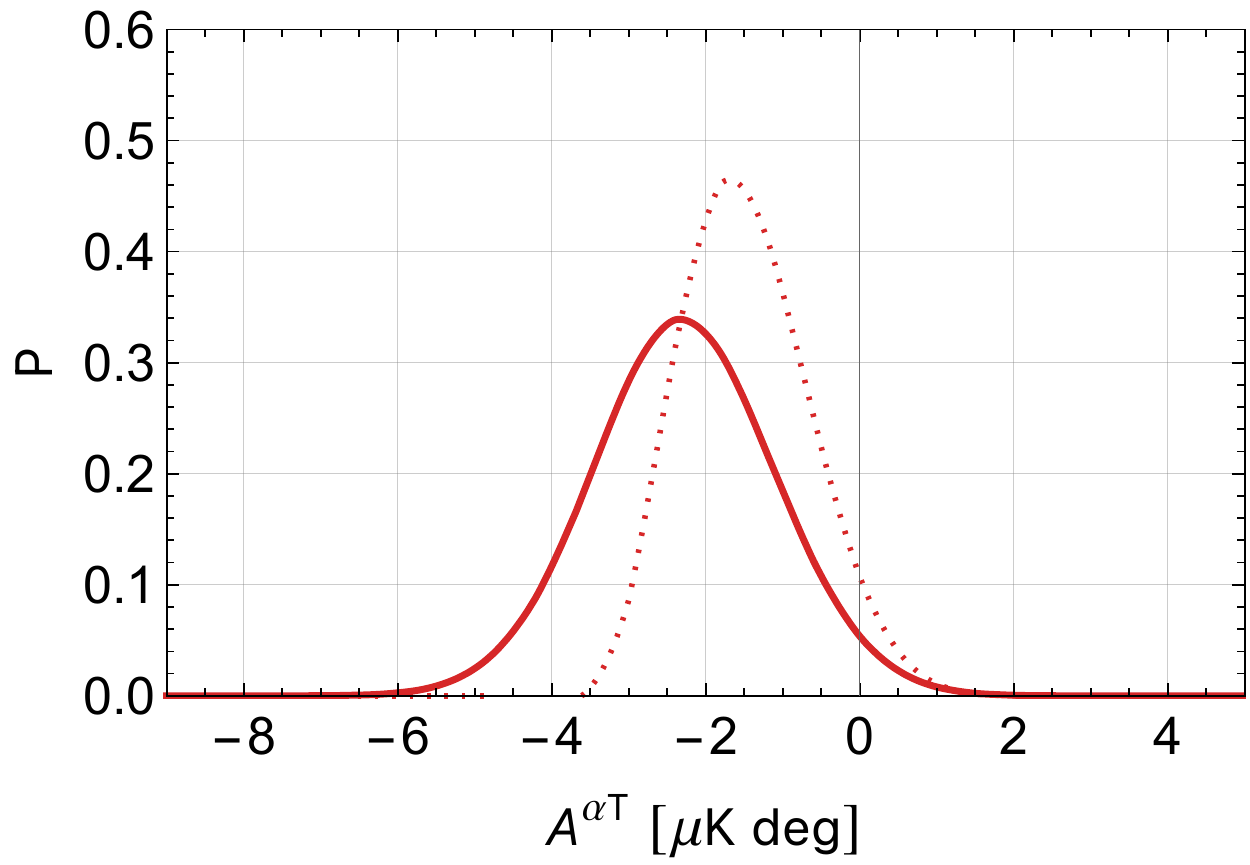}
\hfill
\includegraphics[width=.42\textwidth]{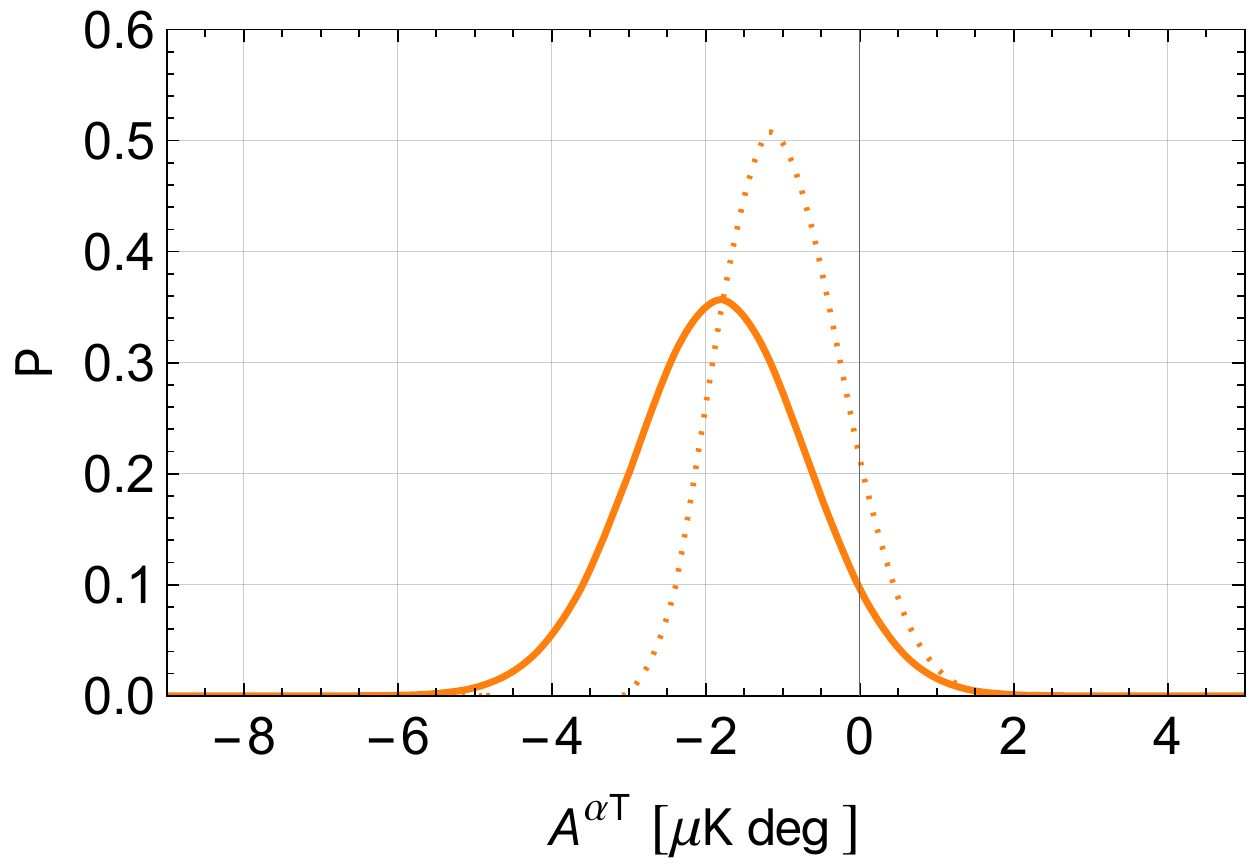}
\includegraphics[width=.42\textwidth]{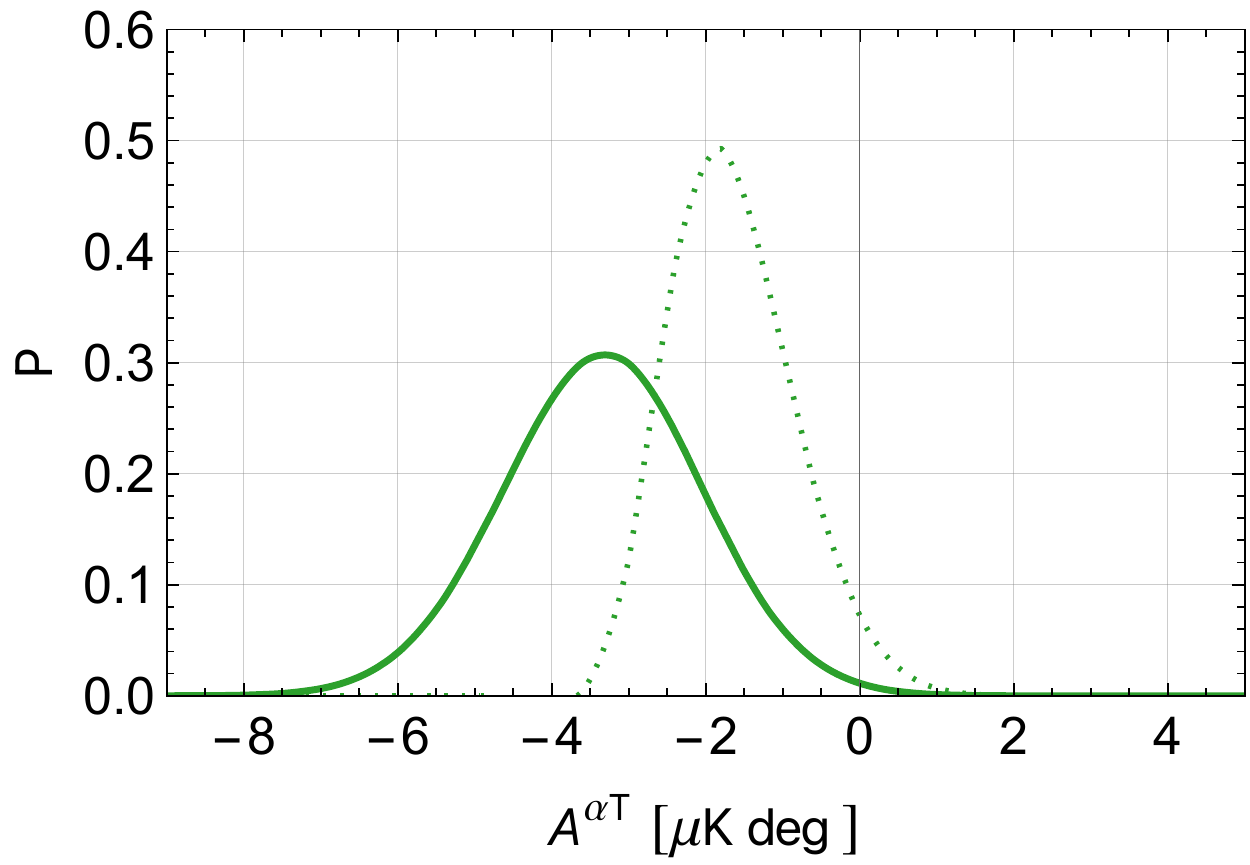}
\hfill
\includegraphics[width=.42\textwidth]{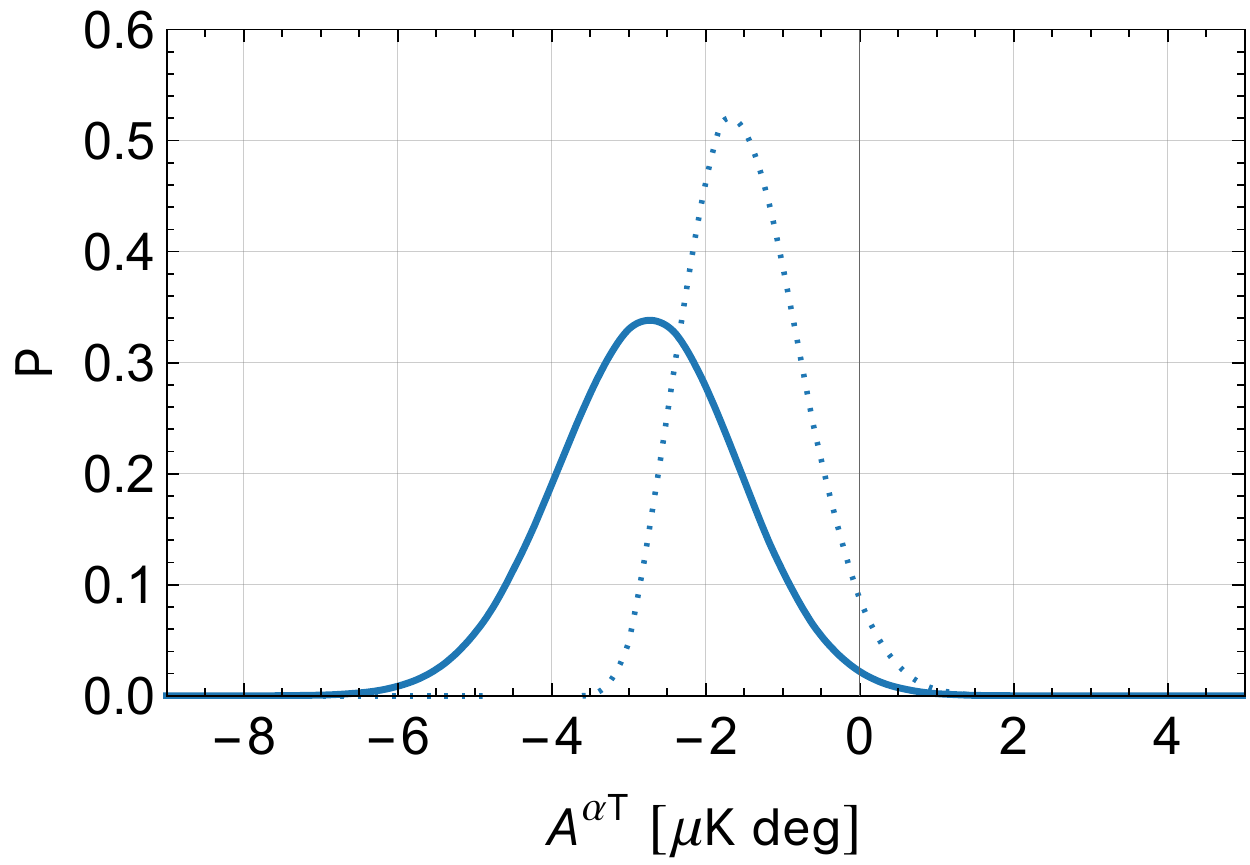}
\caption{Posterior distribution functions for the $\mathrm{A^{\alpha T}}$ parameter obtained from \commander\ (top left panel), \nilc\ (top right panel), \sevem\ (bottom left panel) and \smica\ (bottom right panel) data. The solid curves are computed marginalising the 2D distribution functions over $\mathrm{A^{\alpha \alpha}}$ and the
dashed ones slicing the 2D distribution functions at $\mathrm{A^{\alpha \alpha}} = 0$.}\label{fig:PR3_2D_marginalisealpha}
\end{figure}

If we instead marginalize over $\mathrm{A^{\alpha\alpha}}$, we obtain the posterior distribution functions for $\mathrm{A^{\alpha T}}$ reported in Fig.~\ref{fig:PR3_2D_marginalisealpha} (solid curves). Slicing the 2D contours at $\mathrm{A^{\alpha\alpha}} = 0$, we obtain the probability distribution of $\mathrm{A^{\alpha T}}$ shown in Fig.~\ref{fig:PR3_2D_marginalisealpha} (dashed curves). Also here the sliced case provides tighter posteriors than for the marginalized one, as expected. All these results are compatible with null CB.
\end{document}